\documentclass[12pt,letterpaper]{article}

\usepackage{graphicx}
\usepackage{amsmath}
\usepackage[psamsfonts]{amssymb}
\usepackage{amsthm}
\usepackage{fullpage}
\usepackage{indentfirst}
\usepackage{xspace}

\numberwithin{equation}{section}

\newtheoremstyle{example}{\topsep}{\topsep}%
{}%         Body font
{}%         Indent amount (empty = no indent, \parindent = para indent)
{\bfseries}% Thm head font
{}%        Punctuation after thm head
{\newline}%     Space after thm head (\newline = linebreak)
{\thmname{#1}\thmnumber{ #2}: \thmnote{#3}}%         Thm head spec

\theoremstyle{example}
\newtheorem{example}{Example}[section]

\newcommand{\R}{\ensuremath{\mathbb R}}
\newcommand{\C}{\ensuremath{\mathbb C}}
\newcommand{\PP}{\ensuremath{\mathbb P}}
\newcommand{\Z}{\ensuremath{\mathbb Z}}

\newcommand{\half}{\ensuremath{\frac{1}{2}}}

\newcommand{\N}{{\mathcal N}}

\newcommand{\abs}[1]{\lvert#1\rvert}

\newcommand{\IP}[1]{\langle#1\rangle}
\newcommand{\dwrt}[1]{\frac{\partial}{\partial#1}}
\newcommand{\eps}{\epsilon}

\newcommand{\ti}[1]{\textit{#1}}

\newcommand{\gl}{{\mathfrak g}}
\newcommand{\kahler}{{K\"ahler}\xspace}
\newcommand{\p}{{\partial}}
\newcommand{\OO}{{\mathcal O}}
\newcommand{\vol}{{\mathrm{vol}}}
\newcommand{\de}{{\mathrm d}}
\newcommand{\De}{{\mathrm D}}
\newcommand{\Tr}{{\mathrm{Tr}}\,}
\newcommand{\I}{{\mathrm i}}
\newcommand{\bp}{\overline{\p}}
\newcommand{\M}{{\mathcal M}}
\newcommand{\F}{{\mathcal F}}
\newcommand{\W}{{\mathcal W}}
\newcommand{\re}{{\mathrm{Re}}\,}
\newcommand{\im}{{\mathrm{Im}}\,}
\newcommand{\LL}{{\mathcal L}}
\newcommand{\threehalf}{\frac{3}{2}}

\newcommand{\fig}[2]{
\begin{figure}[t]
\begin{center}
\includegraphics{#1}
\end{center}
\caption{#2}
\label{#1}
\end{figure}}

\newcommand{\figscaled}[3]{
\begin{figure}[t]
\begin{center}
\includegraphics[#3]{#1}
\end{center}
\caption{#2}
\label{#1}
\end{figure}}

\begin{document}

\bibliographystyle{utphys}

\setcounter{page}{1}
\pagestyle{plain}

\begin{titlepage}

\begin{center}
\hfill HUTP-04/A039\\
\hfill hep-th/0410178

\vskip 1.5 cm
{\huge \bf Topological strings and their physical applications}
\vskip 1.3 cm
{\large Andrew Neitzke and Cumrun Vafa}\\
\vskip 0.5 cm
{Jefferson Physical Laboratory,
Harvard University,\\
Cambridge, MA 02138, USA}
\vskip 0.3cm
{
{\tt neitzke@fas.harvard.edu}\\
{\tt vafa@physics.harvard.edu}
}
\end{center}

\vskip 0.5 cm
\begin{abstract}
We give an introductory review of topological strings and their
application to various aspects of superstrings and supersymmetric
gauge theories.
This review includes developing the necessary mathematical background for topological
strings, such as the notions of Calabi-Yau manifold and toric geometry,
as well as physical methods developed for solving them, such as mirror symmetry,
large $N$ dualities, the topological vertex and quantum foam.
In addition, we discuss applications of topological strings to ${\cal N}=1,2$ supersymmetric
gauge theories in 4 dimensions as well as to BPS black hole entropy in 
$4$ and $5$ dimensions.
(These are notes from lectures given by the 
second author at the 2004 Simons Workshop in Mathematics
and Physics.) 
\end{abstract}

\end{titlepage}

\renewcommand{\baselinestretch}{1.2}
\small\normalsize

\tableofcontents

\section{Introduction}

The topological string grew out of attempts to extend computations which occurred in 
the physical string theory.  Since then it has developed in many interesting directions
in its own right.  
Furthermore, the study of the topological string yielded 
an unanticipated but very exciting bonus:  
it has turned out that the topological string has many physical applications 
far beyond
those that motivated its original construction!

In a sense, the topological string is a natural locus where mathematics and physics meet.
Unfortunately, though, the topological string is not very well-known
among physicists; and conversely, although mathematicians are able to understand what the
topological string is mathematically, they are generally less aware of its physical content.
These lectures are intended as a short overview of the topological string, hopefully
accessible to both groups, as a place to begin.  When we have the choice, we mostly
focus on specific examples rather than the general theory.
In general, we make no pretense at being 
complete; for more details on any of the subjects we treat, one should consult the references.

These lectures are organized as follows; for a more detailed overview of the individual 
sections, see the beginning of each section.  We begin by introducing Calabi-Yau spaces, 
which are the geometric setting within which the topological string lives.
In Section \ref{sec-cy}, we define these spaces, give some examples, and briefly explain
why they are relevant for the physical string.  Next, in Section \ref{sec-toric-geometry}, we 
discuss a particularly important class
of Calabi-Yaus which can be described by ``toric geometry''; as we 
explain, toric geometry is convenient
mathematically and also admits an enlightening physical realization, which has been
particularly important for making progress in the topological string.  

With this background
out of the way, we can then move on to the topological string itself, which we introduce in 
Section \ref{sec-top-string}.  There we give the definition of the topological string, and discuss
its geometric meaning, with particular emphasis on the ``simple'' case of genus 
zero.  Having defined
the topological string the next question is how to compute its amplitudes, and in Section
\ref{sec-computing} we describe a variety of methods for computing topological
string amplitudes at all genera, including
mirror symmetry, large $N$ dualities and direct target space analysis.

Having computed all these amplitudes one would like to use them for something;
in Section \ref{sec-physical-applications}, we consider the physical applications
of the topological string.  We consider applications
to $\N=1,2$ supersymmetric gauge
theories as well as to BPS black hole counting in four and five dimensions.

Finally, in Section \ref{sec-top-mtheory} we briefly describe some speculations
on a ``topological M-theory'' which could give a nonperturbative definition and
unification of the two topological string theories.

\section{Calabi-Yau spaces} \label{sec-cy}

Before defining the topological string, we need some basic geometric background.
In this section we introduce the notion of ``Calabi-Yau space.''  We begin
with the mathematical definition and a short discussion of the reason why
Calabi-Yau spaces are relevant for physics.  Next we give some representative
examples of Calabi-Yau spaces in dimensions $1$, $2$ and $3$, both compact
and non-compact.
We end the section with a short overview of a particularly important non-compact Calabi-Yau
threefold, namely the conifold, and the topology changing transition between its
``deformed'' and ``resolved'' versions.

\subsection{Definition of Calabi-Yau space} \label{sec-cy-definition}

We begin with a review of the notion of ``Calabi-Yau space.''  There are many
definitions of Calabi-Yau spaces, which are not quite equivalent to one another; but here
we will not be too concerned about such subtleties, and all the spaces
we will consider are Calabi-Yau under any reasonable definition.  For us
a Calabi-Yau space is a manifold $X$ with a Riemannian metric $g$, satisfying
three conditions:

\begin{itemize}
\item {\bf I. $X$ is a complex manifold}.  This means $X$ looks locally
like $\C^n$ for some $n$, in the sense that it can be covered by patches 
admitting local complex coordinates 
\begin{equation} \label{complex-coords}
z_1, \dots, z_n,
\end{equation}
and the transition functions between patches are holomorphic.
In particular, the real dimension of $X$ is $2n$, so it is always even.
Furthermore the metric $g$ should be Hermitian with respect to the complex structure,
which means
\begin{equation}
g_{ij} = g_{\overline{i} \overline{j}} = 0,
\end{equation}
so the only nonzero components are $g_{i \overline{j}}$.

\item {\bf II. $X$ is \kahler}.  This means that
locally on $X$ there is a real function $K$ such that
\begin{equation} \label{kahler-potential}
g_{i \overline{j}} = \p_i \p_{\overline{j}} K.
\end{equation}
Given a Hermitian metric $g$ one can define its associated \kahler
form, which is of type $(1,1)$,
\begin{equation} \label{kahler-form}
k = g_{i \overline{j}} \de z_i \wedge \de \overline{z_j}.
\end{equation}
Then the \kahler condition is $\de k = 0$.

\item {\bf III. $X$ admits a global nonvanishing holomorphic $n$-form.}  In each local
coordinate patch of $X$ one
can write many such forms, 
\begin{equation}
\Omega = f(z_1, \dots, z_n) \de z_1 \wedge \dots \wedge \de z_n,
\end{equation}
for an arbitrary holomorphic function $f$.  The condition is that such an $\Omega$ exists globally on $X$.
For compact $X$ there is always at most one such $\Omega$ up to an \ti{overall} scalar rescaling; its existence is equivalent to the topological condition
\begin{equation}
c_1(TX) = 0,
\end{equation}
where $TX$ is the tangent bundle of $X$.

\end{itemize}

If conditions I, II, and III are satisfied there is an important consequence.  Namely, according
to Yau's Theorem \cite{yau-thm}, $X$ admits a metric $g$ for which the Ricci curvature vanishes:
\begin{equation}
R_{i \overline{j}} = 0.
\end{equation}
Except in the simplest examples, it is difficult to determine the Ricci-flat \kahler metrics on 
Calabi-Yau spaces.  Nevertheless it is important and useful to know that such a metric \ti{exists},
even if we cannot construct it explicitly.
One thing we can construct explicitly is the \ti{volume form} of the Ricci-flat metric;
it is (up to a scalar multiple)
\begin{equation}
\vol = \Omega \wedge \overline{\Omega}.
\end{equation}

Strictly speaking Yau's Theorem as stated above applies to compact $X$, and has to be supplemented
by suitable boundary conditions at infinity for the holomorphic $n$-form $\Omega$ when $X$ is
non-compact.  For physical applications we
do not require that $X$ be compact; in fact, as we will see, many topological string computations
simplify in the non-compact case, and this is also the case which is directly relevant for the 
connections to gauge theory.

\subsection{Why Calabi-Yau?} \label{sec-why-cy}

Before turning to examples, let us
briefly explain the role that the Calabi-Yau conditions play in superstring theory.
First, why are we interested in Riemannian manifolds at all?
The reason is that they provide a class of candidate backgrounds on which the strings could propagate.  The requirement that the background $X$ be complex and \kahler turns 
out to have a rather direct consequence for the physics of observers living in the
target space:  namely, it implies that these observers will see supersymmetric physics.  
Since supersymmetry is interesting phenomenologically, this
is a natural condition to impose.  Finally, the requirement that $X$ be Ricci-flat is even more fundamental:
string theory would not even make sense without it, as we will sketch in Section \ref{sec-top-string}.

In addition to these motivations from the physical superstring, 
once one specializes to the topological string, one finds other reasons to be interested in Calabi-Yau spaces
and particularly Calabi-Yau threefolds; so we will revisit the question ``why Calabi-Yau?'' in 
Section \ref{sec-why-cy3}.  
Although the Calabi-Yau conditions can be relaxed to give
``generalized Calabi-Yau spaces,'' with correspondingly more general
notions of topological string, the examples which have played the biggest role in the development
of the theory so far are honest Calabi-Yaus.  Therefore, in this review we focus on the honest Calabi-Yau
case.

\subsection{Examples of Calabi-Yau spaces} \label{sec-examples-cy}

\subsubsection{Dimension $1$} \label{sec-examples-cy-d1}

We begin with the case of complex dimension $n=1$.  In this case one can easily list all
the Calabi-Yau spaces.  

\begin{example}[The complex plane] 
The simplest example is just the complex plane $\C$, with a single complex 
coordinate $z$, and the usual flat metric
\begin{equation}
g_{z \overline{z}} = - 2\I.
\end{equation}
In this case the holomorphic 1-form is simply
\begin{equation}
\Omega = \de z.
\end{equation}
\end{example}

\begin{example}[The punctured complex plane, aka the cylinder]
The next simplest example is $\C^\times = \C \setminus \{ 0 \}$, with its cylinder metric
\begin{equation}
g_{z \overline{z}} = - 2\I / \abs{z}^2,
\end{equation}
and holomorphic 1-form
\begin{equation}
\Omega = \de z/z.
\end{equation}
\end{example}

\begin{example}[The 2-torus] \label{example-torus}
Finally there is one compact example, namely the torus $T^2 = S^1 \times S^1$.  We can picture
it as a rectangle which we have glued together at the boundaries, as shown in Figure
\ref{fig-rectangular-torus}.

\fig{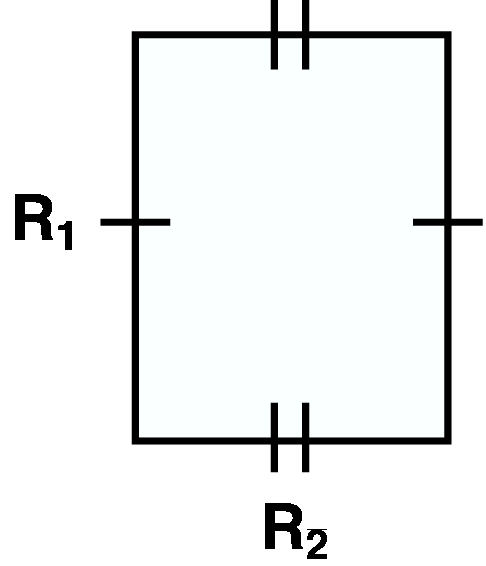}{A rectangular torus; the top and bottom sides are identified,
as are the left and right sides.}

This torus has an obvious flat metric, namely the metric of the page;
this metric depends on two parameters $R_1$, $R_2$ which are the lengths of the sides, so
we say we have a two-dimensional ``moduli space'' of Calabi-Yau metrics on $T^2$, parameterized by
the pair $(R_1, R_2)$.
It is convenient to repackage the moduli of $T^2$ into
\begin{align} \label{torus-moduli}
A &= \I R_1 R_2, \\
\tau &= \I R_2 / R_1.
\end{align}
Then $A$ describes the overall area of the torus, or its ``size,'' while $\tau$ describes its complex structure,
or its ``shape.''  
A remarkable fact about string theory
is that it is in fact invariant under the exchange of size and shape,
\begin{equation} \label{mirror-symmetry-torus}
A \leftrightarrow \tau.
\end{equation}
This is the
simplest example of ``mirror symmetry,'' which we will discuss further in Section \ref{sec-mirror-symmetry}. 
Here we just note that the symmetry \eqref{mirror-symmetry-torus} is
quite unexpected from the viewpoint of classical geometry; for example, when combined with the obvious
geometric symmetry $R_1 \leftrightarrow R_2$, it implies that string theory is invariant under 
$A \leftrightarrow 1 / A$!

\fig{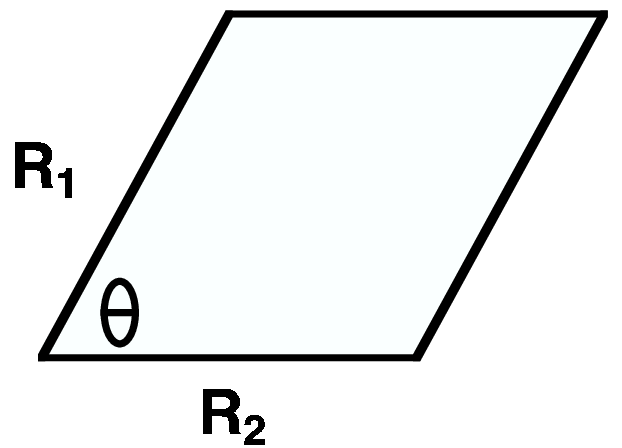}{A 2-torus with a more general metric; again, opposite sides of the figure
are identified.}

We could also consider a more general 2-torus, as shown in Figure \ref{fig-tilted-torus}, again with
the flat metric inherited from the plane.
This is still a Calabi-Yau space.
It is natural to include such tori in our moduli space by letting 
the parameter $\tau$ have a real part as well as an imaginary part:  namely, one can define
the torus to be the quotient $\C / (\Z \oplus \tau \Z)$, equipped with the \kahler metric 
inherited from $\C$.
But then in order for the symmetry
\eqref{mirror-symmetry-torus} to make sense, $A$ should also be allowed to have a real part;
in string theory this real part is naturally provided by an extra field, known as the ``$B$ field.'' 
For general $X$ this $B$ field is a
class in $H^2(X,\R)$, which should be considered as part of the moduli of the Calabi-Yau space 
along with the metric; it naturally combines with $k$ to give the complex 2-form $k+\I B$.
In our case $X = T^2$, $H^2(X,\R)$ is 1-dimensional, and it exactly
provides the missing real part of $A$.

Finally, let us introduce some terminology which will recur repeatedly throughout this review.
We call $\tau$ a ``complex modulus'' of $T^2$ because changing $\tau$ changes the complex structure
of the torus.  In contrast, we can change $A$ just by changing the (complexified) 
\kahler metric without changing the complex structure, so we
call $A$ a ``\kahler modulus.''
\end{example}

\subsubsection{Dimension $2$}

Now let us move to Calabi-Yau spaces of complex dimension $2$.  Here the supply of examples
is somewhat richer.  First there is a trivial example:

\begin{example}[Cartesian products]
One can obtain Calabi-Yau spaces of dimension $2$ by taking Cartesian products of the 
ones we had in dimension $1$, e.g. $\C^2, \C \times \C^\times, \C \times T^2$.
\end{example}

Next we move on to the nontrivial compact examples.  
Up to diffeomorphism there are only two, namely the four-torus $T^4$
and the ``K3 surface.''  We focus here on K3.

\begin{example}[K3] \label{example-k3}
The fastest way to construct a K3 surface 
is to obtain it as a quotient $T^4 / \Z_2$, using the $\Z_2$ identification
\begin{equation} \label{z2-quotient}
(x_1, x_2, x_3, x_4) \sim (-x_1, -x_2, -x_3, -x_4),
\end{equation}
where the $x_i$ are coordinates on $T^4$ (so they are periodically identified.)
Strictly speaking, this quotient gives a singular K3 surface, with 16 singular points which are
the fixed points of \eqref{z2-quotient}.  The singular points can be ``blown up''
(this roughly means replacing them by embedded 2-spheres, see e.g. \cite{MR95d:14001}) 
to obtain a smooth K3 surface.  In string
theory both singular K3 surfaces and smooth K3 surfaces are allowed; the singular ones
correspond to a particular sublocus of the moduli space of K3 surfaces.

One can also define the K3 surface directly by means of algebraic equations.  To begin
with we introduce an auxiliary space $\C\PP^n$, which is also important in its own right:  

\begin{example}[Complex projective space] \label{example-cpn}
$\C\PP^n$ consists of all
$(n+1)$-tuples $(z_1, \dots, z_{n+1}) \in \C^{n+1}$, excluding the point $(0, 0, \dots, 0)$,
modulo the identification
\begin{equation} \label{projective-identification}
(z_1, \dots, z_{n+1}) \sim (\lambda z_1, \dots, \lambda z_{n+1}),
\end{equation}
for all $\lambda \in \C^\times$.
Then $\C\PP^n$ is an $n$-dimensional complex manifold, roughly because we can use the identification
\eqref{projective-identification} to eliminate one coordinate.  $\C\PP^n$ is not Ricci-flat, so it is not
a Calabi-Yau space.

A useful special case to remember is $\C\PP^1$, which is simply the 
Riemann sphere $S^2$.  The same is not true in higher dimensions, though --- e.g. $\C\PP^2$ is not topologically the
same as $S^4$ (the latter is not even a complex manifold.)
\end{example}

Having introduced complex projective space, now we return to the job of constructing 
K3.  We consider the equation
\begin{equation} \label{k3-equation}
P_4(z_1, \dots, z_4) = 0,
\end{equation}
where $P_4$ is some homogeneous polynomial of degree $4$.  Then we define K3 to be the set of solutions
to \eqref{k3-equation} inside the complex projective space 
$\C\PP^3$.  Since $\C\PP^3$ is 3-dimensional and \eqref{k3-equation} is
1 complex equation, K3 so defined will be 2-dimensional.
(Note that in order for this definition to make sense it is
important that $P_4$ is a \ti{homogeneous} polynomial --- otherwise the condition \eqref{k3-equation}
would not be well-defined after the identification \eqref{projective-identification}.)

Different choices for the polynomial $P_4$ give rise to different K3 surfaces, in the sense that they
have different complex structures, although they are all diffeomorphic.  $P_4$ has 
$20$ complex coefficients,
but the equation \eqref{k3-equation} is obviously independent of the overall scaling of $P_4$, so this
rescaling does not affect the complex structure of the resulting K3; all the other coefficients do affect
the complex structure, so one gets a $19$-parameter family of K3 surfaces from this construction. These $19$ parameters are the analog of the single parameter $\tau$ in Example \ref{example-torus}.\footnote{These 
are not quite all the complex moduli of K3 --- there is one more complex deformation possible,
for a total of $20$, but after making this deformation one gets a surface which cannot be realized 
by algebraic equations inside $\C\PP^3$.}

So far we have only discussed K3 as a complex manifold, but it is indeed a Calabi-Yau space,
as we now explain.  It is easy
to see that it is \kahler since it inherits a \kahler metric from $\C\PP^4$.  To see that it has a Ricci-flat
\kahler metric one can invoke Yau's Theorem, as we mentioned in Section \ref{sec-cy-definition}; that 
reduces the task to showing that K3 satisfies the
topological condition $c_1 = 0$.  By using the ``adjunction formula'' from algebraic 
geometry \cite{MR95d:14001} one finds that given a polynomial equation of degree $d$ inside $\C\PP^{k-1}$, the
resulting hypersurface $X$ has
\begin{equation} \label{adjunction-c1}
c_1(X) \sim (d - k) c_1(\C\PP^{k-1}).
\end{equation}
In this case we took $d = k = 4$, so $c_1(X) = 0$ as desired.  This shows the existence of the desired Calabi-Yau
metric.  However, the explicit form of the metric is not known, except at special points in the moduli space.
\end{example}

\begin{example}[ALE spaces] \label{example-ale}
The ``asymptotically locally Euclidean,'' or ``ALE,'' 
spaces form an important class of non-compact Calabi-Yaus of complex dimension 2.
Roughly speaking, these spaces are are obtained as $\C^2 / G$, where $G$ is a finite subgroup of
$SU(2)$ acting linearly on $\C^2$.  (The condition that $G \subset SU(2)$ implies that it preserves
the holomorphic 2-form on $\C^2$, so that it descends to a holomorphic 2-form on $\C^2 / G$, which
is therefore a Calabi-Yau.)
More precisely, the ALE space is not quite $\C^2 / G$; that quotient
has a singularity at the origin, because that point is fixed by the linear action of $G$.
One obtains the ALE space by a local modification near the origin known as ``resolving'' the singularity. 
This resolution replaces the singularity by 
a number of $\C\PP^1$'s localized near the origin.  The number of $\C\PP^1$'s which one gets and
their intersection numbers with one another are determined by the group $G$; for example, if $G = \Z_n$
one gets $n-1$ such $\C\PP^1$'s $C_j$, $j=1, \dots, n-1$, with intersection numbers
\begin{align}
C_i \cap C_i &= -2, \\
C_i \cap C_j &= 1 \quad \textrm{if}\ \abs{i - j} = 1, \\
C_i \cap C_j &= 0 \quad \textrm{if}\ \abs{i - j} > 1.
\end{align}
These intersection numbers are exactly the Cartan matrix of the Lie algebra
$A_{n-1} = su(n)$.  So the curves $C_i$ are playing the role of the simple roots of $A_{n-1}$.
This ``coincidence'' also extends to other choices for $G \subset SU(2)$.  
One possibility is that $G$ can be a double cover of the dihedral group on
$n$ elements; in this case resolving the singularity gives the simple roots of $D_{n-1} = so(2n-2)$.  The other
possibilities for $G$ are the ``exceptional subgroups'' of $SU(2)$, namely double covers of 
the tetrahedral, octahedral and dodecahedral groups, and these give the simple roots of $E_6$, $E_7$, $E_8$ respectively.
This relation between singularities $\C^2 / G$ and simply-laced Lie algebras is known as an ``ADE classification.''
The meaning of the Lie algebras which appear here will become more clear in Section \ref{sec-geom-eng} where
they will be related to gauge symmetries.

After resolving the singularity of $\C^2 / G$, one obtains the ALE space,
which admits a Calabi-Yau metric.  In fact, as with our other examples, it has a whole
moduli space of such metrics:  in particular, for each of the curves $C_i$ obtained by resolving the 
singularity, there is a \kahler modulus 
$t_i = \int_{C_i} k + \I B$ determining its size.  In the limit $t_i \to 0$ the metric reduces
to that of the singular space $\C^2 / G$.  In this sense one can think of the singularity 
of $\C^2 / G$ as containing a number of ``zero size $\C\PP^1$'s.''
\end{example}

\subsubsection{Dimension $3$}

Now we move to the case which is most interesting for topological string theory.  In $d=3$ the problem
of classifying Calabi-Yau spaces is far more complicated, 
even if we restrict to compact Calabi-Yaus; while in $d=1$ we had just $T^2$, and in $d=2$ just
$T^4$ and K3, in $d=3$ it is not even known whether the number of compact Calabi-Yau spaces up
to diffeomorphism is finite.
So we content ourselves with a few examples.

\begin{example}[The quintic threefold]  \label{example-quintic}
The quintic threefold is defined similarly to our algebraic construction of K3 in Example \ref{example-k3}; namely
we consider the equation
\begin{equation} \label{quintic-equation}
P_5(z_1, \dots, z_5) = 0,
\end{equation}
where $P_5$ is homogeneous of degree $5$.  The solutions of \eqref{quintic-equation} inside $\C\PP^4$ give a 3-dimensional space which we call the ``quintic threefold.''  It is a Calabi-Yau space again using \eqref{adjunction-c1}
just as we did for K3.

The quintic threefold has $101$ complex moduli, and is in some sense the simplest compact Calabi-Yau
threefold.  As such it has been extensively studied, e.g. as the first
example of full-fledged mirror symmetry.
\end{example}

\begin{example}[Local $\C\PP^2$.]
One non-compact Calabi-Yau can be obtained by starting with four complex coordinates 
$(x, z_1, z_2, z_3)$, subject to the condition $(z_1, z_2, z_3) \neq (0, 0, 0)$, and making the identification
\begin{equation} \label{o3-identification}
(x, z_1, z_2, z_3) \sim (\lambda^{-3} x, \lambda z_1, \lambda z_2, \lambda z_3)
\end{equation}
for all $\lambda \in \C^\times$.  Mathematically, this space is known as the total space of the line 
bundle $\OO(-3) \to \C\PP^2$; we can think of it as obtained by starting with the $\C\PP^2$ spanned
by $z_1, z_2, z_3$ and adjoining the extra coordinate $x$.  See Figure \ref{fig-local-o3}.
Locally on $\C\PP^2$, our space has the structure of $\C\PP^2 \times \C$.
In this sense it has ``4 compact directions'' and ``2 non-compact directions.''

\figscaled{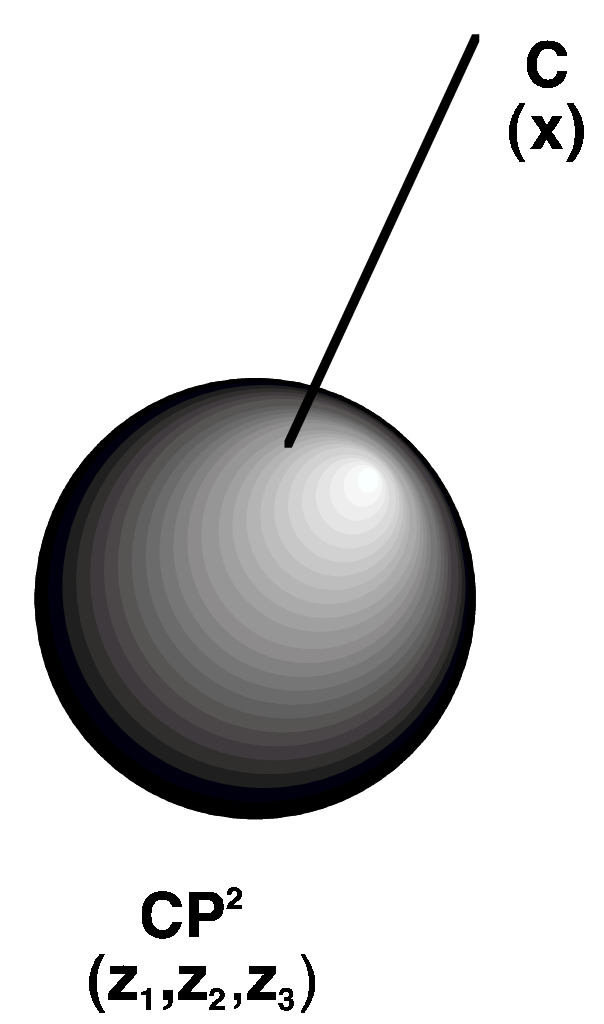}{A crude representation of the local $\C\PP^2$ geometry, $\OO(-3) \to \C\PP^2$.}
{height=3in}
The rule \eqref{o3-identification} 
characterizes the behavior of $x$ under rescalings of the homogeneous coordinates on $\C\PP^2$,
or equivalently, it determines how $x$ transforms as one moves between different 
coordinate patches on $\C\PP^2$.  

Although the local $\C\PP^2$ 
geometry is non-compact, it can arise naturally even if we start with a compact Calabi-Yau --- namely,
it describes the geometry of a Calabi-Yau space containing a $\C\PP^2$, in the limit where
we focus on the immediate neighborhood of the $\C\PP^2$.
\end{example}

\begin{example}[Local $\C\PP^1$.] \label{example-local-cp1}
Similar to the last example, we can start with four complex coordinates 
$(x_1, x_2, z_1, z_2)$, subject to the condition $(z_1, z_2) \neq (0, 0)$, and make the identification
\begin{equation} \label{o1o1-identification}
(x_1, x_2, z_1, z_2) \sim (\lambda^{-1} x_1, \lambda^{-1} x_2, \lambda z_1, \lambda z_2)
\end{equation}
for all $\lambda \in \C^\times$.  This gives the total space of the line bundle $\OO(-1) \oplus \OO(-1)
\to \C\PP^1$.  Similarly to the previous example, it is obtained by starting with $\C\PP^1$, which has
``2 compact directions,'' and then adjoining the coordinates $x_1, x_2$, which contribute ``4
non-compact directions.''  See Figure \ref{fig-local-o1-o1}.

\figscaled{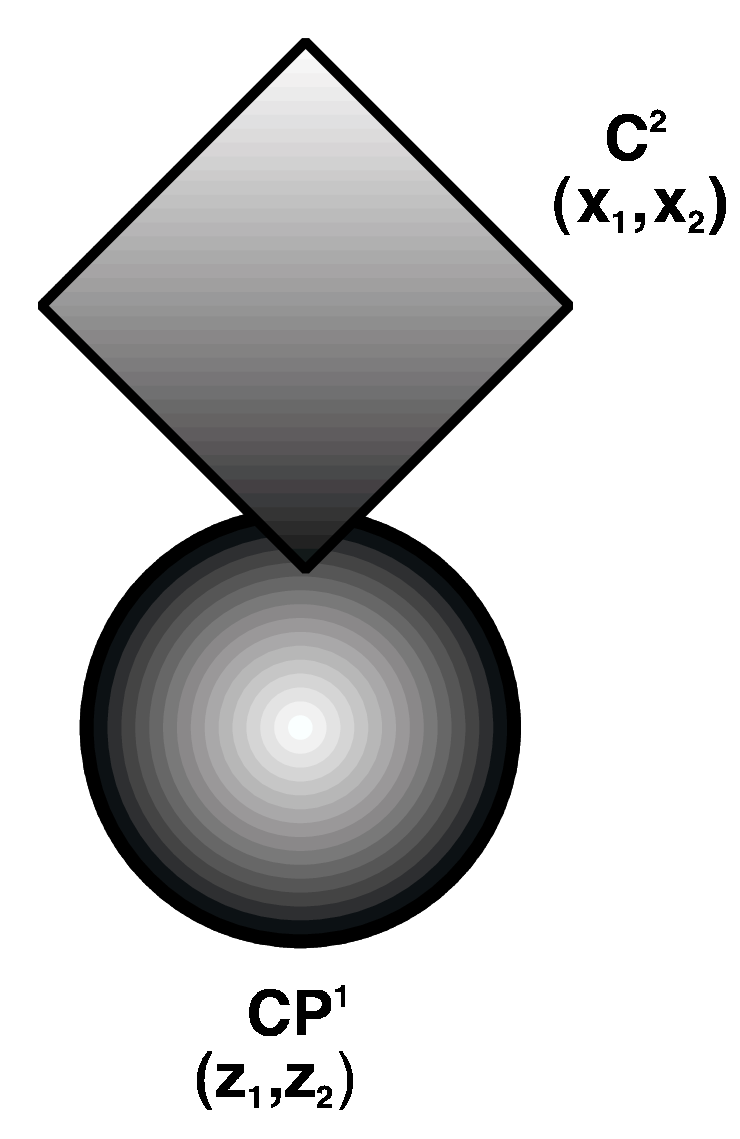}{A crude representation of the local $\C\PP^1$ geometry, $\OO(-1) \oplus \OO(-1)
\to \C\PP^1$.}{height=3in}

This example is also known as the ``resolved conifold,'' a name to which we will return in 
Section \ref{sec-conifolds}.
\end{example}

\begin{example}[Local $\C\PP^1 \times \C\PP^1$.]  \label{example-local-cp1-cp1}
Another standard example comes by starting with five complex
coordinates $(x, y_1, y_2, z_1, z_2)$, with $(y_1, y_2) \neq (0,0)$ and $(z_1,z_2) \neq (0,0)$, and
making the identification
\begin{equation} \label{o22-identification}
(x, y_1, y_2, z_1, z_2) \sim (\lambda^{-2} \mu^{-2} x, \lambda y_1, \lambda y_2, \mu z_1, \mu z_2)
\end{equation}
for all $\lambda, \mu \in \C^\times$.  This gives the total space of the line bundle $\OO(-2,-2)
\to \C\PP^1 \times \C\PP^1$.  It has four compact directions and two non-compact directions.
\end{example}

\begin{example}[Deformed conifold.] All the local examples we discussed so far were ``rigid,''
in other words, they had no deformations of their complex structure.\footnote{Strictly speaking, this is
a delicate statement in the non-compact case 
since we should specify what kind of boundary conditions we are imposing at
infinity.  When we say that these local examples are rigid we essentially mean that the compact part, 
$\C\PP^1$ or $\C\PP^2$, has no complex deformations.}  Now let us consider an example which is 
not rigid.  Starting with the complex coordinates $(x,y,z,t) \in \C^4$, this time without any 
projective identification, we look at the space of solutions to
\begin{equation}
xy - zt = \mu.
\end{equation}
This gives a Calabi-Yau 3-fold for any value $\mu \in \C$, so $\mu$ spans 
the 1-dimensional moduli space of complex structures.  If $\mu = 0$ then the Calabi-Yau
has a singularity at $(x,y,z,t) = (0,0,0,0)$, known as the ``conifold singularity.''  
For finite $\mu$ it is smooth.  Since we obtain
the smooth Calabi-Yau from the singular one just by varying the parameter $\mu$, which
deforms the complex structure, we call the smooth $\mu \neq 0$ version the ``deformed conifold.''
We will discuss it in more detail in Section \ref{sec-conifolds}.
\end{example}

\subsection{Conifolds} \label{sec-conifolds}

In the last section we introduced the singular conifold
\begin{equation} \label{singular-conifold}
xy - zt = 0,
\end{equation}
and the deformed conifold
\begin{equation} \label{deformed-conifold}
xy - zt = \mu.
\end{equation}
Since the deformed conifold is such an important example it will be useful to describe it in another way. 
Namely, by a change of variables we can rewrite \eqref{deformed-conifold} as
\begin{equation}
x_1^2 + x_2^2 + x_3^2 + x_4^2 = r.
\end{equation}
Describing it this way it is easy to see that there is an $S^3$ in the geometry, namely,
just look at the locus where all $x_i \in \R$.  The full geometry where we include also the 
imaginary parts of $x_i$ is in fact diffeomorphic to the cotangent bundle, $T^* S^3$.

This space is familiar to physicists as the phase space of a particle which moves on $S^3$; it has three
``position'' variables labeling a point $x \in S^3$ and three ``momenta'' spanning the cotangent space
at $x$.  Now we want to describe its geometry ``near infinity,'' i.e. at large distances,
similar to how we might describe the infinity of Euclidean $\R^3$ as looking like a large $S^2$.
In the case of $T^* S^3$ the position coordinates are bounded, so 
looking near infinity means choosing large values for the momenta,
which gives a large $S^2$ in the cotangent space $\R^3$.  Therefore the infinity of $T^* S^3$
should look like some $S^2$ bundle over the position space $S^3$, 
i.e. locally on $S^3$ it should look like $S^2 \times S^3$.  It turns out that this is enough to imply 
that it is even \ti{globally} $S^2 \times S^3$.  

So at infinity the deformed conifold has the geometry of $S^2 \times S^3$.  As we move from 
infinity toward the origin both $S^2$ and $S^3$ shrink, 
until the $S^2$ disappears altogether, leaving just an $S^3$ with radius $r$, which
is the core of the $T^* S^3$ geometry (the zero section of the cotangent bundle.)  This is
depicted on the left side of Figure \ref{fig-three-conifolds}.

Now let us describe another way of smoothing the conifold singularity.
First rewrite \eqref{singular-conifold} as
\begin{equation} \label{conifold-determinant}
\det \begin{pmatrix} x & z \\ t & y \end{pmatrix} = 0.
\end{equation}
This equation is equivalent to the existence of nontrivial solutions to
\begin{equation} \label{conifold-matrixeq}
\begin{pmatrix} x & z \\ t & y \end{pmatrix} \begin{pmatrix} \xi_1 \\ \xi_2 \end{pmatrix} = 0.
\end{equation}
Indeed, away from $(x,y,z,t) = (0,0,0,0)$, \eqref{conifold-determinant} just states that the
matrix has rank $1$, so $(\xi_1, \xi_2)$ solving \eqref{conifold-matrixeq} are unique up to
an overall rescaling.  So away from $(x,y,z,t) = (0,0,0,0)$ one could describe the singular 
conifold as the space of solutions to \eqref{conifold-matrixeq}, with $(\xi_1, \xi_2) \neq (0,0)$,
and with the identification
\begin{equation} \label{resolved-conifold-identification}
(\xi_1, \xi_2) \sim (\lambda \xi_1, \lambda \xi_2)
\end{equation}
where $\lambda \in \C^\times$.  But at $(x,y,z,t) = (0,0,0,0)$ something new happens:
any pair $(\xi_1, \xi_2)$ now solves \eqref{conifold-matrixeq}.  Taking into account
\eqref{resolved-conifold-identification}, $(\xi_1, \xi_2)$ parameterize a
$\C\PP^1$ of solutions.  In summary, \eqref{singular-conifold} and \eqref{conifold-matrixeq},\eqref{resolved-conifold-identification} are
equivalent, except that $(x,y,z,t) = (0,0,0,0)$ describes a single point in \eqref{singular-conifold},
but a whole $\C\PP^1$ in \eqref{conifold-matrixeq},\eqref{resolved-conifold-identification}.  We refer to the space described by \eqref{conifold-matrixeq},\eqref{resolved-conifold-identification} 
as the ``resolved conifold.''  (In fact, it is isomorphic to the local $\C\PP^1$ geometry of Example \ref{example-local-cp1}.)

Mathematically this discussion would be summarized by saying that the resolved conifold is 
obtained by making a ``small resolution'' of the conifold singularity.  We emphasize, however,
that physically it is natural to consider this as a continuous process, contrary to the usual
mathematical description in which it seems to be a discrete jump.  
This is because physically we consider the full Calabi-Yau metric rather
than just the complex structure.
Namely, the resolved conifold has a single \kahler modulus for its Calabi-Yau metric,\footnote{Once
again, we are here considering only variations of the metric which preserve suitable boundary
conditions at infinity.} naturally parameterized by
\begin{equation}
t = \vol(\C\PP^1) = \int_{\C \PP^1} k + \I B.
\end{equation}
In the limit $t \to 0$, the $\C\PP^1$ shrinks to a point and the Calabi-Yau metric on 
the resolved conifold approaches the Calabi-Yau metric on the singular conifold.
So the resolved conifold is obtained by a \kahler deformation of the metric without
changing the complex structure.\footnote{Mathematically, the resolved conifold and the singular conifold
are not the same as complex manifolds, but they are birationally equivalent.  Physically we want
to consider birationally equivalent spaces as really having the same complex structure.}

In summary, we have two different non-compact Calabi-Yau geometries, as depicted in 
Figure \ref{fig-three-conifolds}:  
the deformed conifold, which has one complex modulus $r$ and
no \kahler moduli, and the resolved conifold, which has no complex moduli but one \kahler modulus $t$;
we can interpolate from one space to the other by passing through the
singular conifold geometry.  The deformed conifold has a single $S^3$ at its heart, whose
size is determined by $r$, while the resolved conifold has a single $S^2$, whose size is
determined by $t$.

\figscaled{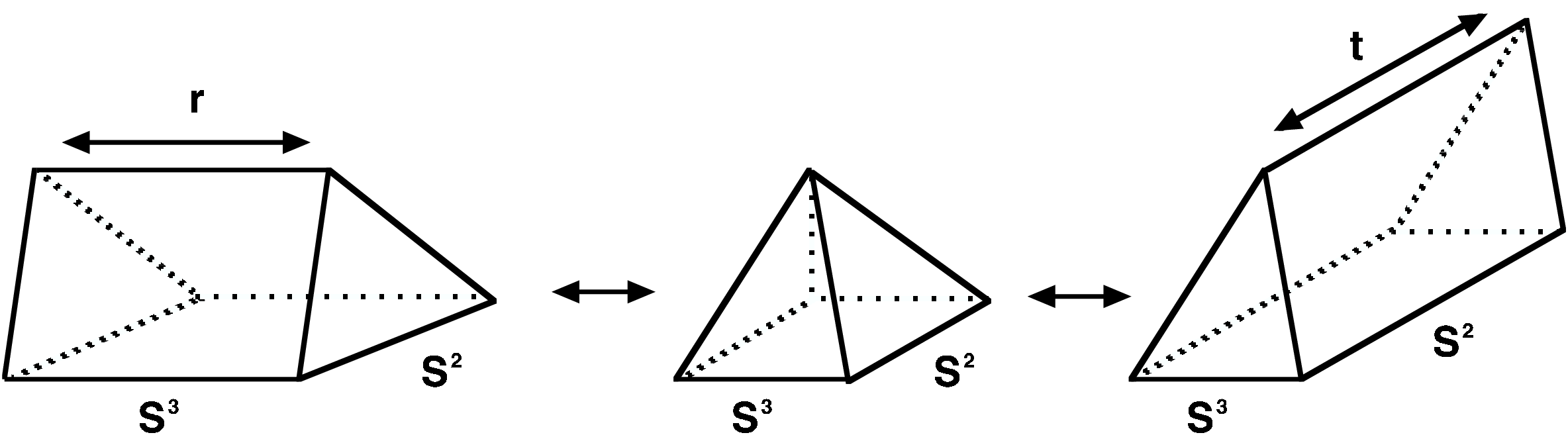}{The three conifold geometries:  from left to right, deformed, singular and resolved.  Both geometries look like $S^2 \times S^3$ near infinity (the bottom of the figure); they
are distinguished by whether the $S^2$ or the $S^3$ shrinks to zero size in the interior (the
top of the figure.)}{width=6in}

Note that from the perspective of Figure \ref{fig-three-conifolds}, 
the $S^2$ and $S^3$ which appear when we resolve the singular conifold seem very natural;
in some sense they were both in 
the game even before resolving, as we see from the $S^2 \times S^3$ at infinity.  
All three cases --- deformed, singular, and resolved --- look the same at infinity; they differ
only near the tip of the cone.  This is exactly what we expect since we were trying to study only
localized deformations.

We will return to the conifold repeatedly in later sections.  For more information about
its geometry, including the explicit Calabi-Yau metrics, see \cite{Candelas:1990js}.

\section{Toric geometry} \label{sec-toric-geometry}

Now we want to introduce a particularly convenient representation of a special class of
algebraic manifolds, which includes and generalizes some of the examples we considered above.
Mathematically this representation is called ``toric geometry''; for a more detailed review
than we present here, see e.g. \cite{MR2003030}.  As we will see, toric manifolds have two closely
related virtues:  first, they are easily described in terms of a finite amount of combinatorial
data; second, they can be concretely realized via two-dimensional field theories of a 
particularly simple type.

We begin with the simplest of all toric manifolds.

\begin{example}[$\C^n$] Consider the $n$-complex-dimensional manifold $\C^n$, with complex coordinates $(z_1, \dots, z_n)$ and the standard flat metric, and parameterize it in an idiosyncratic way:  writing
\begin{equation}
z_i = \abs{z_i} e^{\I \theta_i},
\end{equation}
choose the coordinates $((\abs{z_1}^2, \theta_1), \dots, (\abs{z_n}^2, \theta_n))$.  This coordinate
system emphasizes the symmetry $U(1)^n$ which acts on $\C^n$ by shifts of the $\theta_i$.  It is 
also well suited to describing the symplectic structure given by the \kahler form $k$:
\begin{equation}
k = \sum_i \de z_i \wedge \de \overline{z_i} = \sum_i \de \abs{z_i}^2 \wedge \de \theta_i.
\end{equation}

Roughly, splitting the coordinates into $\abs{z_i}^2$ and $\theta_i$ gives a factorization
\begin{equation} \label{toric-approximate-splitting}
\C^n \approx \OO^{n+} \times T^n,
\end{equation}
where $\OO^{n+}$ denotes the ``positive orthant'' $\{\abs{z_i}^2 \ge 0\}$,
represented (for $n=3$) in Figure \ref{fig-toric-picture-c3}.
\figscaled{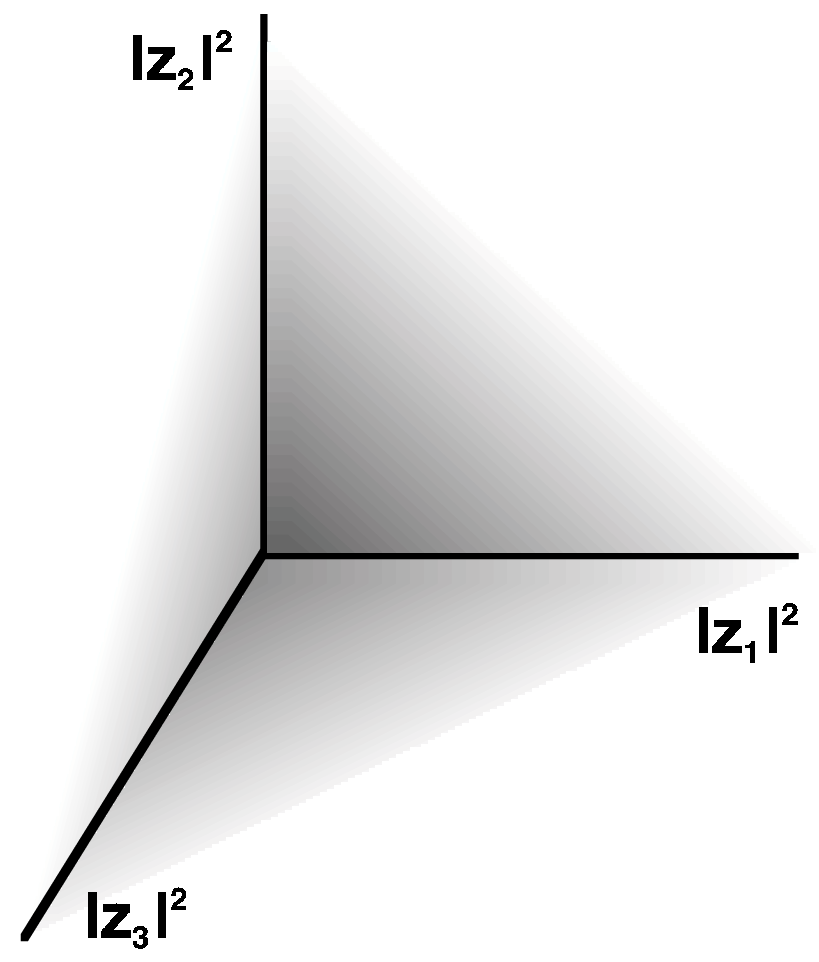}{The positive octant $\OO^{3+}$, which is the toric
base of $\C^3$.}{height=3in}
Namely, at each point of $\OO^{n+}$ we have the product of $n$ circles obtained by fixing
$\abs{z_i}$ and letting $\theta_i$ vary.
However, when $\abs{z_i}^2 = 0$
the circle $\abs{z_i} e^{\I \theta_i}$ degenerates to a single point.  Therefore 
\eqref{toric-approximate-splitting} is not quite precise, because the ``fiber'' $T^n$
degenerates at each
boundary of the ``base'' $\OO^{n+}$; which circle of $T^n$ degenerates is determined by which $\abs{z_i}^2$
vanishes, or more geometrically, by the direction of the unit normal to the boundary.
When $m > 1$
of the $\abs{z_i}^2$ vanish, which occurs at the intersection locus of $m$ faces of the orthant, the corresponding $m$ circles of $T^n$ degenerate.  At the origin
all $n$ cycles have degenerated and $T^n$ shrinks to a single point.

In this sense all the information about the symplectic manifold 
$\C^3$ is contained in Figure \ref{fig-toric-picture-c3}, which is called the ``toric diagram'' for $\C^3$.
When looking at this diagram one always has to remember that there is a $T^3$ over 
the generic point, and that this
$T^3$ degenerates at the boundaries, in a way determined by the unit normal.
Despite the fact that the $T^3$ becomes singular at the boundaries, the full geometry of $\C^3$ 
is of course smooth. (Of course, all this holds for general $n$ as well as $n=3$, but the
analogue of Figure \ref{fig-toric-picture-c3} would be hard to draw in the general case.)

\figscaled{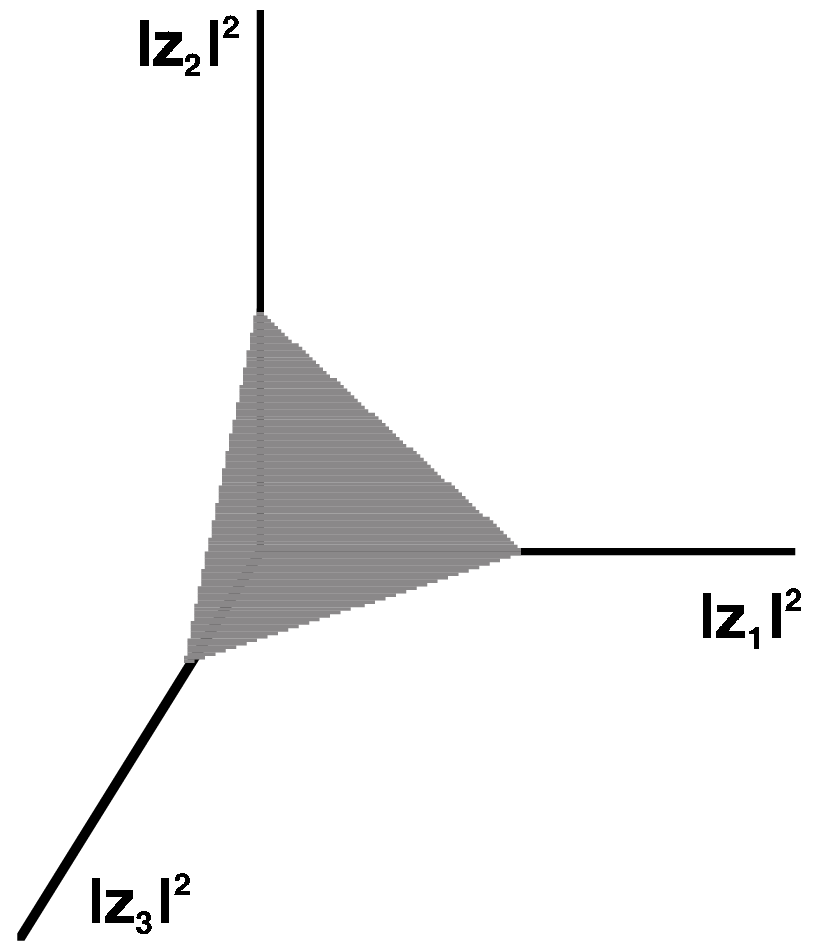}{The toric base of $\C\PP^2$; geometrically it is just the two-dimensional 
interior of a triangle, but here we show it naturally embedded in $\R^3$
and cut out by the condition \eqref{moment-map-constraint}.}{height=3in}
\end{example}

\begin{example}[Complex projective space] \label{example-cpn-toric}
Next we want to give a toric representation for $\C\PP^n$.
We first give a slightly different quotient presentation of this space than the one
we used in \eqref{projective-identification}:  namely, for any $r > 0$, we start
with the $2n+1$-sphere
\begin{equation} \label{moment-map-constraint}
\abs{z_1}^2 + \cdots + \abs{z_{n+1}}^2 = r,
\end{equation}
and then make the identification
\begin{equation} \label{phase-identification}
(z_1, \dots, z_{n+1}) \sim (e^{\I \theta} z_1, \dots, e^{\I \theta} z_{n+1})
\end{equation}
for all real $\theta$.  This is equivalent to our original ``holomorphic quotient'' 
definition, where we did
not impose \eqref{moment-map-constraint} but worked modulo arbitrary rescalings of the $z_i$
instead of just phase rescalings; 
indeed, starting from that definition
one can make a rescaling to impose \eqref{moment-map-constraint}, and afterward one still
has the freedom to rescale by a phase as in \eqref{phase-identification}.  The presentation
we are using now is more closely rooted in symplectic geometry.  

This toric presentation is also natural from the physical point 
of view, as we now briefly discuss.  The physical theory which describes the worldsheet of the superstring
propagating on $\C\PP^n$ is a two-dimensional quantum field theory known as the ``supersymmetric
nonlinear sigma model into $\C\PP^n$.''  We will not discuss this sigma model in detail, but the crucial point is that in this case
it can be obtained as the IR limit of an $\N=(2,2)$ supersymmetric \ti{linear} sigma model with $U(1)$ gauge 
symmetry \cite{Witten:1993yc}.  Specifically, the coordinates $z_i$ appear as the scalar components of
$4$ chiral superfields, all with $U(1)$ charge $1$.  Then the physics of the vacua of the 
linear sigma model exactly mirrors our toric construction of $\C\PP^n$; 
namely, the constraint \eqref{moment-map-constraint} is imposed by the D-terms, and the quotient \eqref{phase-identification}
is the identification of gauge equivalent field configurations.  This construction, which we will generalize
below when we discuss other toric varieties, turns out to be extremely useful for the study of the
topological string on such spaces; we will see some examples of its utility in later sections.

Note that in our toric presentation of $\C\PP^n$ we have the parameter $r > 0$, which did not appear
in the holomorphic quotient.  This parameter appears naturally in the gauged linear sigma model 
(as a Fayet-Iliopoulos parameter),
where one sees directly that it corresponds to the size of $\C\PP^n$.

Now we want to use this presentation to draw the toric diagram.  As for $\C^n$, 
the toric base lies in the space coordinatized by the $\abs{z_i}^2$.  In the present case we have to
impose \eqref{moment-map-constraint}, so the base turns out to be an $n$-dimensional simplex; for example,
in the case of $\C\PP^2$ it is just a triangle, as shown in Figure \ref{fig-toric-base-cp2}.
Over each point of the base we have a $T^2$ fiber generated by shifts of $\theta_i$ (naively
this would give a $T^3$ for $\theta_1, \theta_2, \theta_3$, but the identification \eqref{phase-identification} reduces this to $T^2$.)  
A cycle of $T^2$ collapses over each boundary
of the triangle, as indicated in Figure \ref{fig-toric-base-cp2-collapsing-cycles}.
\end{example}

\figscaled{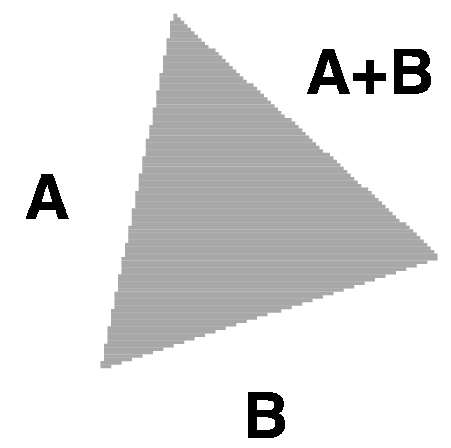}{The toric base of $\C\PP^2$.  Over each boundary
a cycle of the fiber $T^2$ collapses; if we label the basis cycles as $A$ and $B$, then the
collapsing cycle over each boundary is as indicated.}{height=1.4in}

\figscaled{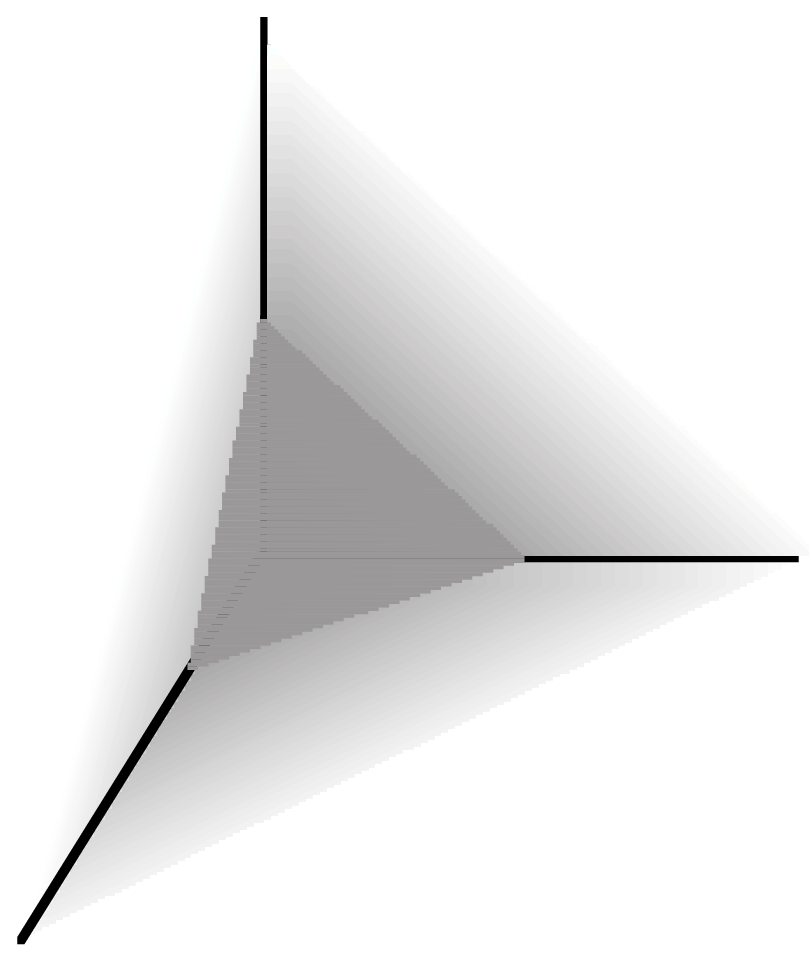}{The toric base of the local $\C\PP^2$ geometry.}{height=3in}

\begin{example}[Local $\C\PP^2$]  \label{example-local-cp2-toric} 
To get a toric presentation of a Calabi-Yau manifold we have to choose
a non-compact example.  The construction is closely analogous to what we did above to
construct $\C\PP^n$;
namely, for $r>0$, we start with
\begin{equation} \label{local-cp2-constraint}
-3 \abs{z_0}^2 + \abs{z_1}^2 + \abs{z_2}^2 + \abs{z_3}^2 = r,
\end{equation}
and then make the additional identification
\begin{equation}
(z_0, z_1, z_2, z_3) \sim (e^{-3 \I \theta} z_0, e^{\I \theta} z_1, e^{\I \theta} z_2, e^{\I \theta} z_3),
\end{equation}
for any real $\theta$.  In the gauged linear sigma model of \cite{Witten:1993yc} this is 
realized by taking four chiral superfields with $U(1)$ charges $(-3, 1, 1, 1)$.  Actually,
the fact that the local $\C\PP^2$ geometry is Calabi-Yau can also be understood naturally in
the gauged linear sigma model:  the condition $c_1 = 0$ turns out to be equivalent to the
statement that the sum of the $U(1)$ charges vanishes, which in turn implies vanishing of the 1-loop
beta function.

We can also draw the toric diagram for this case.  Introducing the notation $p_i = \abs{z_i}^2$,
the base is spanned by the four real coordinates $p_0, p_1, p_2, p_3$, subject to the
condition \eqref{local-cp2-constraint}, which can be solved to eliminate $p_0$,
\begin{equation}
p_0 = \frac{1}{3}(p_1 + p_2 + p_3 - r).
\end{equation}
The condition that all $p_i > 0$ then becomes
\begin{align}
p_1 + p_2 + p_3 &> r,\\
p_1 &> 0,\\
p_2 &> 0,\\
p_3 &> 0.
\end{align}
So the toric base is the positive octant in $\R^3$ with a corner chopped off, as shown 
in Figure \ref{fig-toric-base-local-cp2}.  The triangle at the corner represents
the $\C\PP^2$ at the core of the geometry, just as in the previous example.

\figscaled{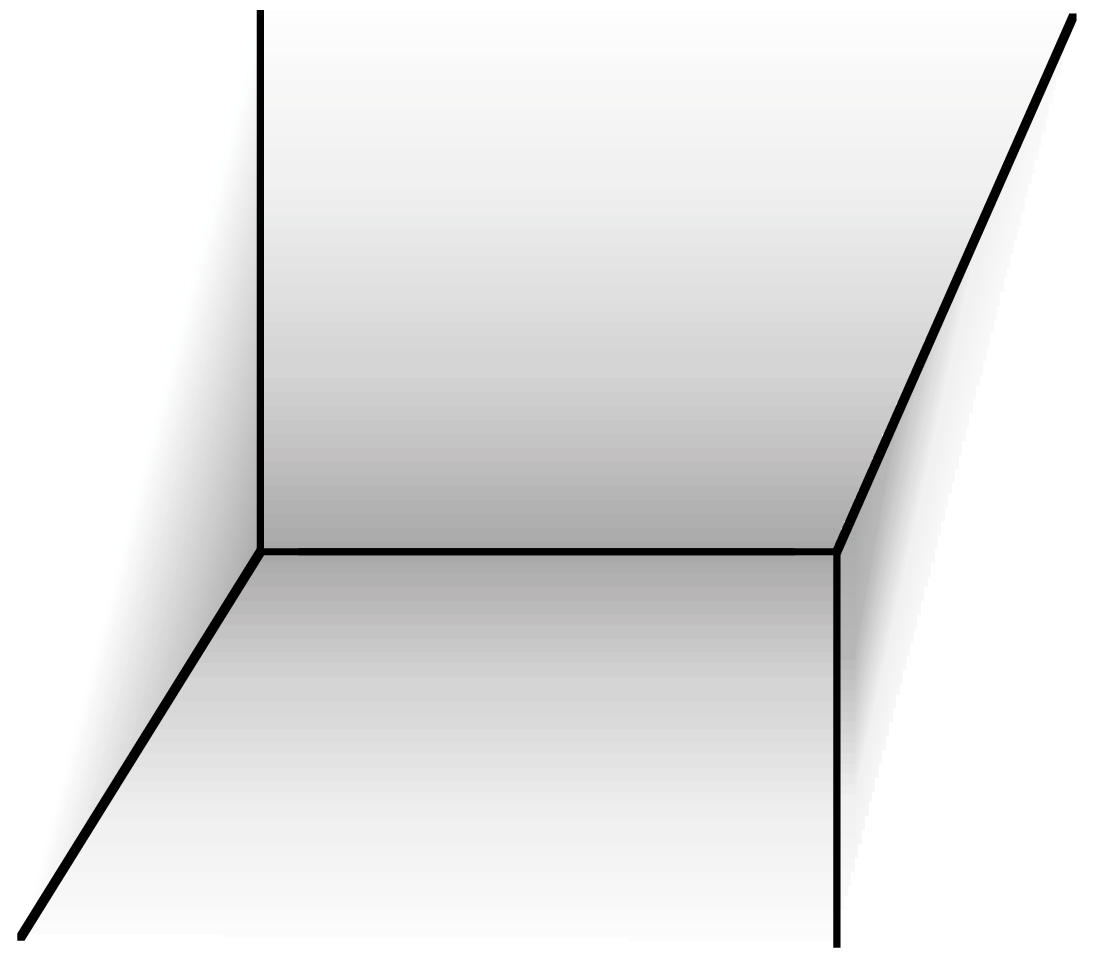}{The toric base of the local $\C\PP^1$ geometry.}{height=3in}
\end{example}

\begin{example}[Local $\C\PP^1$]  A similar construction gives the toric diagram for the local $\C\PP^1$
geometry, $\OO(-1) \oplus \OO(-1) \to \C\PP^1$, from Example \ref{example-local-cp1}.  
One obtains in this case Figure \ref{fig-toric-base-local-cp1}.  One feature
of interest is the $\C\PP^1$ at the core of the geometry, which can be easily seen as the
line segment in the middle.  (To see that the line segment indeed represents the topology
of $\C\PP^1$, recall that along this segment two of the three circles of the fiber $T^3$ 
are degenerate,
so that one just has an $S^1$ in the fiber; moving along the segment, 
this $S^1$ then sweeps out a $\C\PP^1$; indeed, the $S^1$ degenerates at
the two ends of the segment, which are identified with the north and south
poles of $\C\PP^1$.)  Furthermore it is easy to read off the volume of this $\C\PP^1$ from the
toric diagram:  the \kahler form in this geometry is $k = \de p_i \wedge \de \theta_i$,
and integrating it just gives $2 \pi \Delta p$, i.e. the length of the line segment!\footnote{We
are using a fact about \kahler geometry, namely, the volume of a holomorphic cycle is just obtained by integrating $k$ over the cycle.}

This example illustrates a general feature:  finite segments (or more generally finite simplices)
of the toric diagram correspond to compact cycles in the geometry, and the sizes of the simplices
correspond to the volumes of the cycles.
\end{example}

\begin{example}[\bf Local $\C\PP^1 \times \C\PP^1$] \label{example-local-cp1-cp1-toric} 
We can give a toric construction for this case as well,
again parallel to the holomorphic construction we gave above; in gauged linear sigma model terms
it would correspond to having $5$ chiral superfields and two $U(1)$ gauge groups, with the
charges $(-2,1,1,0,0)$ and $(-2,0,0,1,1)$.  (Note that the charges under both $U(1)$ groups
sum to zero as required for one-loop conformality.)  The corresponding toric diagram is
the ``oubliette'' shown in Figure \ref{fig-toric-base-local-cp1-cp1}.

\figscaled{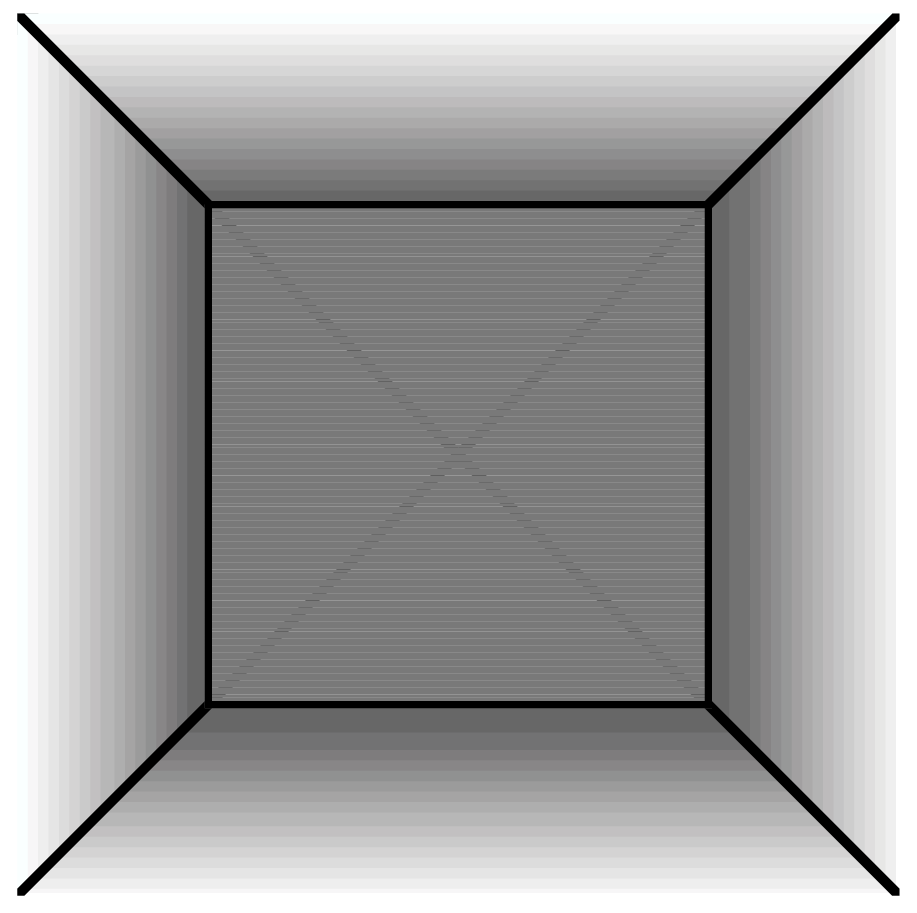}{The toric base of the local $\C\PP^1 \times \C\PP^1$ geometry.}{height=3in}
\end{example}

Our list of toric Calabi-Yaus has included only non-compact examples, 
but we should note that it is also 
possible to construct compact Calabi-Yaus using the techniques of toric geometry.  Indeed, we have
already done so in Examples \ref{example-k3} and \ref{example-quintic}, 
where we started with the toric manifolds $\C\PP^3$ and $\C\PP^4$ respectively and then
imposed some extra algebraic relations on the coordinates 
to obtain a Calabi-Yau.  A similar construction can be performed
starting with a more general toric manifold, and this gives a large class of interesting examples of
compact Calabi-Yau spaces.
This construction is also natural from the physical point of view:
in the gauged linear sigma model, imposing an algebraic relation on the coordinates
corresponds to introducing a superpotential.

\section{The topological string} \label{sec-top-string}

With the geometrical preliminaries behind us, we are now ready to move on to physics.  In this
section we will sketch the definition of the topological string.  First we describe the
two-dimensional field theories which are underlying the physical string theory.  Next we
discuss the ``twisting'' procedure which converts the ordinary field theory into its
topological cousin, and how to extend this field theory to the full-fledged string theory.
After this discussion we will be in a position to appreciate why Calabi-Yau threefolds are
particularly relevant spaces for the topological string.  We then plunge into a discussion
of the two different variations of the topological string (A and B models) and their observables, 
with a brief intermezzo on their holomorphic properties, and finish with
a description of exactly what is computed by the topological string at genus zero.

\subsection{Sigma models and $\N=(2,2)$ supersymmetry}

The string theories in which we will be interested
(both the ordinary physical version and the topological version) have to do
with maps from a surface $\Sigma$ to a target space $X$.  Roughly, in string theory one
integrates over all such maps $\phi: \Sigma \to X$ as well as over metrics on $\Sigma$, 
weighing each map by the Polyakov 
action:\footnote{Actually, this is the Polyakov action for the bosonic string;
we are really interested in the superstring, for which there are extra fermionic degrees
of freedom, but we are suppressing those for simplicity.}
\begin{equation} \label{polyakov-action}
\int \De \phi\ \De g\  e^{- \int_\Sigma \abs{\partial \phi}^2 }.
\end{equation}
The integral over $\phi$ alone defines a two-dimensional quantum field theory which is called 
a ``sigma model into $X$''; its saddle points are 
locally area-minimizing surfaces in $X$.  Because we are integrating both over $\phi$
and over metrics on $\Sigma$, one often describes the string theory as obtained by coupling the sigma
model to two-dimensional quantum gravity.

Classically, the sigma model action depends only on the conformal class of the metric $g$, so that the
integral over metrics can be reduced to an integral over conformal structures --- or equivalently, 
to an integral over complex structures on $\Sigma$.  For the string theory to be well defined we need
this property to persist at the quantum level, but this turns out to be a nontrivial restriction
on the allowed $X$;
namely, requiring that the sigma model should be conformally invariant even after including 
one-loop quantum effects on $\Sigma$, one finds 
the condition that $X$ should be Ricci flat.  

For generic $X$ one might
expect even more conditions to appear when one considers higher-loop quantum effects; this does
happen in the bosonic string, but mercifully not in the superstring provided that $X$
is \kahler.
The reason why the \kahler condition is so effective in suppressing quantum corrections is that
it is related to $(2,2)$ supersymmetry of the 2-dimensional sigma model, and hence implies
bose/fermi cancellations in loops on the worldsheet.\footnote{Note that this 
``worldsheet'' supersymmetry
is different from the spacetime supersymmetry we discussed in the previous section, although the
\kahler condition on $X$ is ultimately responsible for both, and there are arguments which relate one to the other.} This $(2,2)$ supersymmetry is also
crucial for the definition of the topological string, so we now discuss it in more detail.

The statement of $\N=2$ supersymmetry means that there are $4$ worldsheet currents 
\begin{equation} \label{n2susy-operators}
J, G^+, G^-, T,
\end{equation}
with spins $1, \threehalf, \threehalf, 2$ respectively, and with prescribed operator product relations.  
These operators get interpreted as follows:  $T$ is the usual energy-momentum tensor; $G^{\pm}$ are
conserved supercurrents for two worldsheet supersymmetries; $J$ is the conserved current for the $U(1)$
R-symmetry of the $\N=2$ algebra, 
under which $G^\pm$ have charges $\pm 1$.  The modes of these currents act on the Hilbert space
of the worldsheet theory.

In the case of the sigma model on $X$, these currents are analogous 
(in the ``B-model'' case --- see below) to the operators
\begin{equation} \label{n2susy-operators-geometric}
\deg, \bp, \bp^\dagger, \Delta
\end{equation}
acting on $\Omega^*(LX)$, the space of differential forms on the loop space of $X$.
(This analogy arises because the 
loop space is roughly the configuration space of the sigma model on $X$.)
This identification suggests that among the operator
product relations of the $\N=2$ algebra should be
\begin{align}
(G^+)^2 &\sim 0,\\
(G^-)^2 &\sim 0,\\
G^+ G^- &\sim T + J;
\end{align}
these relations indeed hold and they will play a particularly important role in
what follows.

In the case where $X$ is Calabi-Yau, so that the sigma model is conformal, 
we can make a further refinement:  each of the currents \eqref{n2susy-operators}
is a sum of two commuting copies, one ``left-moving'' (holomorphic) and one ``right-moving'' (antiholomorphic).
We thus obtain two copies of the $\N=2$ algebra, which we write $(J, G^\pm, T)$ and
$(\overline{J}, \overline{G}^\pm, \overline{T})$; 
this split structure is referred to as $\N=(2,2)$ supersymmetry.
This structure of $\N=(2,2)$ superconformal field theory --- the operators listed above as well as the Hilbert 
space on which they act --- should be regarded as an invariant associated to the 
Calabi-Yau manifold $X$; from it
one can recover various more well-known invariants such as the Dolbeault cohomology groups of $X$, 
but the full superconformal field theory is a considerably more subtle object, as we will see.

\subsection{Twisting the $\N=(2,2)$ supersymmetry} \label{sec-twisting}

Given an $\N=(2,2)$ 
superconformal field theory as described in the previous section, there is an important construction
which produces a ``topological'' version of the theory.  One can think of this procedure
as analogous to the passage from the de Rham complex $\Omega^*(X)$ to its cohomology $H^*(X)$:  while the
cohomology contains less information than the full de Rham complex, the information it does contain is 
far more easily organized and understood.  So how do we construct this topological version 
of the SCFT?  Guided by the relation $(G^+)^2 \sim 0$ and the above analogy, 
we might try to form the cohomology of the zero mode of $G^+$.
In fact this is not quite possible, because $G^+$ has the wrong spin, namely $3/2$; 
in order to obtain a scalar zero mode we need to begin with an operator of spin $1$.

This problem can be overcome,
as explained in \cite{Witten:1988xj} (see also \cite{Witten:1991zz}), by ``twisting'' the sigma model.  The twist can be understood in various ways, but one way to describe it is as a shift in the operator $T$:
\begin{equation} \label{twist-T}
T_{\mathrm{new}} = T_{\mathrm{old}} - \half \partial J.
\end{equation}
This shift has the effect of changing the spins of all operators by an amount proportional to their $U(1)$ charge $q$,
\begin{equation} \label{twist-S}
S_{\mathrm{new}} = S_{\mathrm{old}} - \half q.
\end{equation}
After this shift the operators $(G^+, J)$ have spin $1$ while $(T, G^-)$ have spin $2$.\footnote{Note
that although $G^\pm$ now have integer spin, they still obey fermionic statistics!}  Now we can 
define $Q = G^+_0$, which makes sense on arbitrary $\Sigma$ and obeys $Q^2 = 0$, and restrict
our attention to only observables which are annihilated by $Q$.

In this context one often calls $Q$ a ``BRST operator,'' since the restriction to observables annihilated
by a nilpotent fermionic $Q$ is precisely how one implements gauge invariance in the BRST formalism for quantization of gauge theories.
Here we have not obtained
$Q$ from the BRST procedure.  Nevertheless, the structure of the twisted
$\N=2$ algebra is isomorphic to one which \ti{is} obtained from the usual BRST procedure,
namely that of the \ti{bosonic} string.  In that case one has currents $(Q, J_{\mathrm{ghost}})$ of
spin $1$ and $(T, b)$ of spin 2, where $(Q,b)$ are the BRST current and antighost corresponding
to the diffeomorphism symmetry on the bosonic string worldsheet\footnote{We are using the notation $Q$ both for the
current in the bosonic string and for its zero mode.}; the isomorphism to the twisted
$\N=(2,2)$ algebra is
\begin{equation}
(G^+, J, T, G^-) \leftrightarrow (Q, J_{\mathrm{ghost}}, T, b).
\end{equation}

\subsection{Constructing the string correlation functions}

In the last subsection we noted that the twisted $\N=2$ algebra is isomorphic to a subalgebra of
the symmetry algebra of the bosonic string.  In particular, this subalgebra includes the 
$b$ antighost, which is the crucial
element needed for the computation of correlation functions in the bosonic string.  Namely,
the $b$ antighost provides the link between CFT correlators, computed on a fixed worldsheet $\Sigma$,
and string correlators, which involve integrating over all metrics on $\Sigma$; one sees this link
by performing the Faddeev-Popov
procedure, which reduces the integral over metrics on $\Sigma$ to an integral over the moduli space
$\M_g$ of genus $g$ Riemann surfaces, with the $b$ ghosts providing the measure.  The genus $g$
free energy of the bosonic string obtained in this way is\footnote{Strictly speaking this is the 
answer for $g > 1$; the expression \eqref{bosonic-string-partition-function}
has to be slightly modified for $g = 0,1$ because the sphere and torus admit nonzero holomorphic vector fields.}
\begin{equation} \label{bosonic-string-partition-function}
\int_{\M_g} \IP{\abs{\prod_{i=1}^{3g-3} b(\mu_i)}^2}.
\end{equation}
Here the symbol $\IP{\cdots}$ denotes a CFT correlation function.  The $3g-3$ $\mu_i$ are ``Beltrami
differentials,'' anti-holomorphic $1$-forms on $\Sigma$ with values in the holomorphic tangent bundle; they span the space
of infinitesimal deformations of the $\bp$ operator on $\Sigma$, which is the tangent space
to $\M_g$.  Then $b(\mu_i)$ is an operator obtained
by integrating the $b$-ghost against $\mu_i$:
\begin{equation}
b(\mu) = \int_\Sigma b_{zz} \mu_{\overline{z}}^z.
\end{equation}
More abstractly, $b$ is an operator-valued $1$-form on $\M_g$, so the expectation value of 
the product of $3g-3$ copies of $b$ gives a holomorphic $(3g-3)$-form; taking both the holomorphic and
antiholomorphic pieces we then get a $(6g-6)$-form, which can be integrated over $\M_g$.

Now comes the important point:  since the twisted $\N=2$ superconformal algebra is isomorphic to the
algebra appearing in the bosonic string, we can now define a ``topological string''
from the correlation functions of the 
$\N=(2,2)$ SCFT on fixed $\Sigma$, by repeating \eqref{bosonic-string-partition-function} 
with $b$ replaced by $G^-$:
\begin{equation} \label{topological-string-partition-function}
F_g = \int_{\M_g} \IP{\abs{\prod_{i=1}^{3g-3} G^-(\mu_i)}^2}.
\end{equation}
The formula \eqref{topological-string-partition-function} should also be understood as coming from
coupling the twisted $\N=(2,2)$ theory to topological gravity --- see \cite{Witten:1988xj} for some discussion.

One then defines the full topological string free energy to be
\begin{equation}
\F = \sum_{g=0}^\infty \lambda^{2-2g} F_g,
\end{equation}
where $\lambda$ is the ``string coupling constant'' weighing the contributions at
different genera.\footnote{This expression is only perturbative; it should be understood 
in the sense of an asymptotic series in $\lambda$.}  Finally, the topological 
string partition function is
defined as
\begin{equation}
Z = \exp \F.
\end{equation}

\subsection{Why Calabi-Yau threefolds?} \label{sec-why-cy3}

From our present point of view, the construction of the topological string would have made
sense starting from any $\N=(2,2)$ SCFT, and in particular, the sigma model on any Calabi-Yau 
space $X$ would suffice.  
On the other hand, for the physical string, there is a good reason to focus on Calabi-Yau 
threefolds.  Namely, if we look for backgrounds which could resemble the real world, we find an obvious
constraint:  to a first approximation, the real world looks like $4$-dimensional Minkowski space
$M$.  On the other hand, conformal invariance of the SCFT coupled to worldsheet supergravity
requires the total dimension of spacetime to be $10$.
To reconcile these two statements one is naturally led to consider
backgrounds $M \times X$, where $X$ is some compact 6-dimensional
space, small enough that it cannot be seen directly, either by the naked eye or by any experiment
we have so far been able to do.  Studying string theory on $M \times X$,  one 
finds that the \ti{internal} properties of $X$ lead to
physical consequences for the observers living in $M$.  Conversely,
the four-dimensional perspective on the string theory computations
sheds a great deal of light on the geometry of $X$, as we will see.

Remarkably, it turns out that the case of Calabi-Yau threefolds is special for the topological
string as well.  Namely, although one can define $F_g$ for any Calabi-Yau $d$-fold, this $F_g$
actually vanishes for all $g \neq 1$ unless $d=3$!  This follows from considerations of
charge conservation:  namely, the topological twisting turns out to introduce a background $U(1)$
charge $d(g-1)$.  In order for the correlator appearing in \eqref{topological-string-partition-function}
to be nonvanishing, the insertions which appear must exactly compensate this background charge; but
the insertions consist of $3g-3$ $G^-$ operators, so they have total charge $-3(g-1)$, hence the
correlator vanishes unless $d=3$.

\subsection{A and B twists}

We are almost ready to discuss the geometric meaning of the topological string, but there is
one subtlety to take care of first.
In Section \ref{sec-twisting} 
we glossed over an important point:  although we chose the operator $G^+$ for our BRST
supercharge $Q$, we could equally well have chosen $G^-$.  The latter possibility corresponds to
an opposite twist where we replace \eqref{twist-T} by
\begin{equation}
T_{\mathrm{new}} = T_{\mathrm{old}} + \half \partial J.
\end{equation}
With this twist it is $G^-$ rather than $G^+$ which will have spin $1$.
We have a similar freedom in the antiholomorphic sector, so altogether there are four possible choices
of twist, corresponding to choosing for the BRST operators
\begin{align}
(G^+, \overline{G}^+):&\      {\mathrm A}\ \text{model}\\
(G^-, \overline{G}^-):&\  \overline{\mathrm A}\ \text{model}\\
(G^+, \overline{G}^-):&\      {\mathrm B}\ \text{model}\\
(G^-, \overline{G}^+):&\  \overline{\mathrm B}\ \text{model}
\end{align}
We have listed each choice together with the name usually given to the corresponding topological
string.  The $\overline{\mathrm A}$ ($\overline{\mathrm B}$) model is related to the A (B) model in a trivial way, namely, all 
correlators are just related by an overall complex conjugation; so essentially we have two
distinct ways to make a topological string theory from a given Calabi-Yau $X$, 
namely the A and B models.

\subsection{Observables and correlation functions}

So far we have described how to start with the Calabi-Yau space $X$ and construct
two topological string theories called the A and B models.
Now let us begin to discuss the observables of these models and the meaning of the
correlation functions.

In the A model case, the combined BRST 
operator $Q + \overline{Q}$ turns out to be the $\de$ operator on $X$, and its cohomology is the de Rham cohomology $H^*_{\mathrm dR}(X)$.  It is natural to impose an additional ``physical state'' constraint which leads
to considering only the degree $(1,1)$ part of this cohomology.  
A $(1,1)$ form corresponds to a
deformation of the \kahler form on $X$, so finally, the observables of the A model which we are
considering are deformations
of the \kahler moduli of $X$.  Furthermore, one can show directly that correlation functions computed in 
the A model are \ti{independent} of the chosen complex structure on $X$;
namely, one shows that the operators which deform the complex structure
are $Q$-exact, so that they decouple from the computation of the string amplitudes.

In the B model case the space of physical states in 
the BRST cohomology again consists of objects of bidegree $(1,1)$, but
this time the complex in question is the $\bp$ cohomology with values in $\wedge^* TX$, so the observables
are $(0,1)$-forms with values in $TX$, i.e. Beltrami differentials on $X$.  As we discussed before,
these Beltrami differentials correspond to deformations of the complex structure of $X$; so 
the observables of the B model are deformations of complex structure.  Similarly to the A model case,
one shows that the B model correlation functions are independent of the \kahler structure.

In sum,
\begin{align}
\text{A model on}\ X &\leftrightarrow \text{\kahler moduli of}\ X, \\
\text{B model on}\ X &\leftrightarrow \text{complex moduli of}\ X.
\end{align}

Now, what do the correlation functions in the A and B models actually mean mathematically?
Usually the correlation functions in a quantum field theory are hard to define because of the
complexity inherent in the path integral over an infinite-dimensional field space.
In the present case we are indeed computing a path integral $\int e^{-S}$, 
but this path integral is significantly
simplified by the fermionic $Q$ symmetry \cite{Witten:1991zz}:  it reduces to a sum of local contributions 
from the fixed points of $Q$!  The rest of the field space 
contributes zero, because one can introduce coordinates in which $Q$ acts by an infinitesimal shift of
a Grassmann coordinate $\theta$, and then note that the integral over that one coordinate gives
\begin{equation}
\int \de \theta\,e^{-S} = 0.
\end{equation}
This follows from the standard rules for Grassmann integration, and the fact that $Q$ is a symmetry of the
path integral, so that $S$ is independent of $\theta$.

So the path integral is localized on $Q$-invariant configurations.  In the B model these turn out
to be simply the constant maps $\phi: \Sigma \to X$, obeying $\de \phi = 0$.  
In this sense the string worldsheet reduces to a point on $X$, so the B model is ``local,'' and its 
correlation functions are those of a field theory on $X$.  It turns out that these correlation functions
compute quantities determined by the \ti{periods} of the holomorphic 3-form $\Omega$, which
are sensitive to changes in the complex structure.  

In the A model, on the other hand, one
finds the condition $\bp \phi = 0$, which requires only that the map $\phi: \Sigma \to X$ be holomorphic;
such a map is called a \ti{worldsheet instanton}.
In nontrivial instanton sectors the 
string worldsheet does not reduce to a point.
From this point of view the fact that the A model depends on \kahler moduli is easy to understand; 
it arises simply because each worldsheet instanton is weighted by the factor
\begin{equation}
e^{- \int_C k}
\end{equation}
i.e. the area of the curve $C \subset X$ which is the image of the string worldsheet in $X$.  
The sum over instanton sectors 
is a complicated structure, non-local from the point of view of $X$, and therefore the A model does not reduce straightforwardly to a field theory on $X$.

Note that the B model moduli (the periods) are naturally
complex numbers themselves, while the A model moduli (volumes of 2-cycles) are real numbers, 
so we seem to have a serious asymmetry between the two moduli spaces and hence between
the A and B models.  As we 
mentioned earlier, the symmetry between the two moduli spaces is restored by including an extra
class $B \in H^2(X,\R)$.  When $B$ is included, the weighting factor for a worldsheet instanton 
becomes
\begin{equation}
e^{- \int_C k + iB}.
\end{equation}
We will combine $k$ and $B$ into a single modulus $t = k + iB \in H^2(X,\C)$.

\subsection{Holomorphic anomaly} \label{sec-holomorphic-anomaly}

As we have discussed above, 
the A and B models each depend on only ``half'' the moduli of $X$, namely the
\kahler and complex moduli respectively.  In fact even more is true:
in each case the partition function \ti{formally} depends only holomorphically on its moduli.  One sees this
by computing the antiholomorphic derivative of a correlator, which amounts to 
inserting the operator corresponding to the antiholomorphic deformation 
into the correlation function.  This operator is BRST-exact, so one might expect
that it is decoupled
from correlation functions of BRST-invariant operators.  However, the $G^-$ insertions in 
the definition \eqref{topological-string-partition-function} of the correlation function are not 
BRST-exact; taking this into account one finds that the antiholomorphic derivative of the correlator
is the integral of a total derivative over the moduli space $\M_g$.  Such an integral would vanish if the moduli
space were compact, but since it is not compact one has to worry about contributions from the boundary; indeed
\figscaled{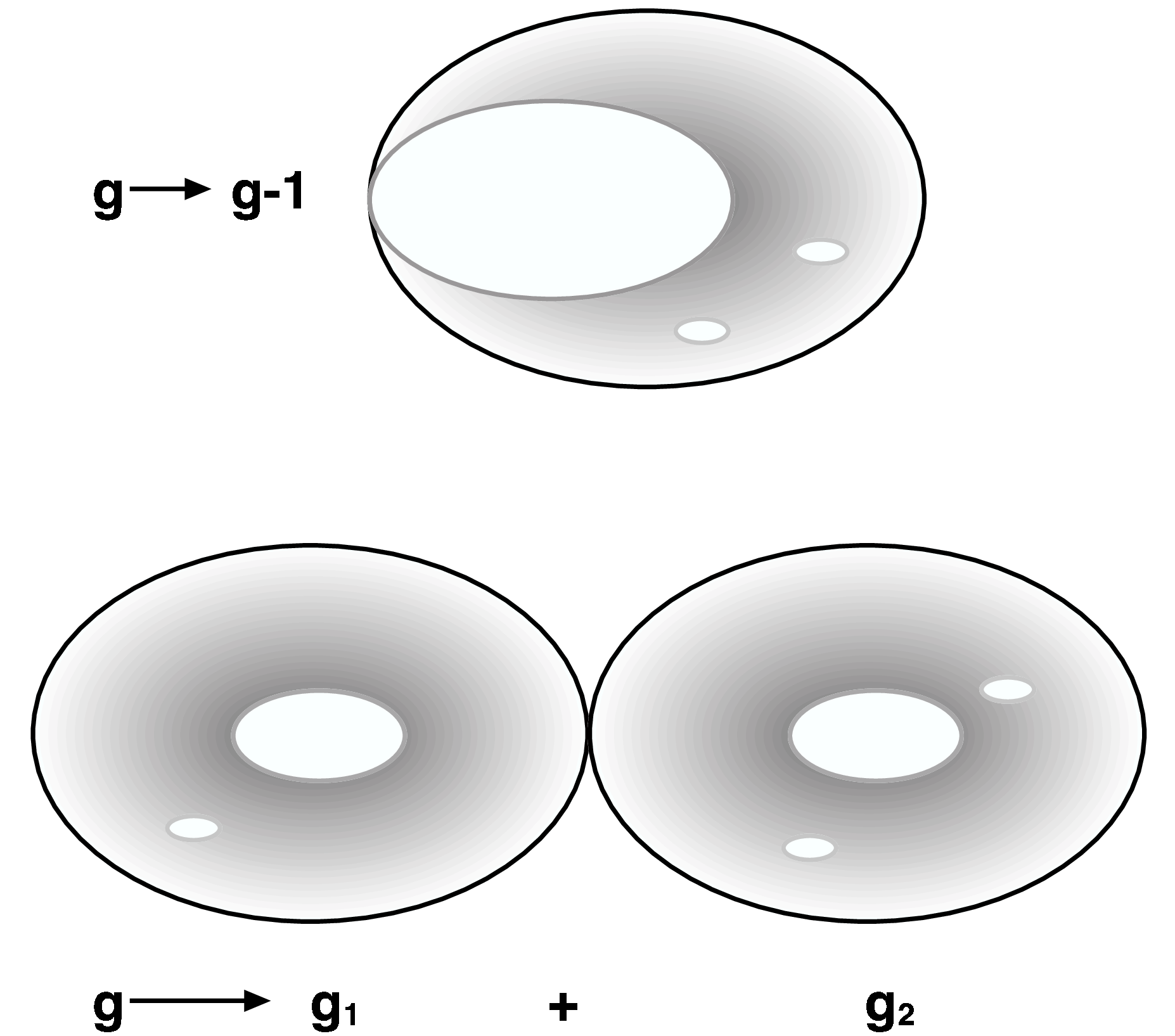}{Degenerations of a Riemann surface of genus $g$, corresponding
to boundary components of the moduli space $\M_g$.}{height=3in}
there are such contributions, so the partition function is not quite holomorphic as a function
of the moduli.  Nevertheless its antiholomorphic
dependence can be determined precisely; it is expressed in terms of a ``holomorphic anomaly equation'' derived in \cite{Bershadsky:1993ta,Bershadsky:1994cx}.
Through the anomaly equation $\bp F_g$ gets related to the $F_{g'}$ with $g' < g$, corresponding to 
boundaries of moduli space where some cycle of the genus $g$ surface shrinks --- see Figure
\ref{fig-boundaries-of-moduli}.

In the case of the B model in genus 1, the holomorphic anomaly
is familiar to mathematicians; it is related to the curvature of the determinant line bundle
which obstructs the construction of a holomorphic $\det \bp$ \cite{MR86g:32035}.
The full holomorphic anomaly in the B model, including all genera, can be interpreted 
as the statement that the partition function transforms as a \ti{wavefunction} obtained by
quantizing the symplectic space $H^3(X,\R)$ \cite{Witten:1993ed,Dijkgraaf:2002ac}.

\subsection{Genus zero} \label{sec-genus-zero}

After all these preliminaries, we can begin to discuss the geometric content of the topological string.
It is natural to begin with the simplest case, namely
genus zero; it turns out that this case already contains a lot of interesting
information about $X$.

\subsubsection{A model}

In the A model one finds for the genus zero free energy
\begin{equation} \label{f0-a-model}
F_0 = \int_X k \wedge k \wedge k + \sum_{n \in H_2(X,\Z)} \sum_{m=1}^\infty d_n \frac{e^{-\IP{n,t}m}}{m^3}.
\end{equation}
The first term is the classical contribution in the sense of worldsheet perturbation theory; it 
corresponds to the zero-instanton sector, where the string reduces to a point, and just
gives the volume of $X$.  The second term is more interesting since it contains information about
worldsheet instantons.  Its form is intuitive, at least if we focus on the $m=1$ term:  we sum over all $n \in H_2(X,\Z)$, the homology classes of the image of the worldsheet, and weigh each instanton by the
factor $e^{-\IP{n,t}}$ giving the complexified area.  The interesting information is then contained in
the number $d_n$ which counts the number of holomorphic maps in the homology class $n$.\footnote{Sometimes
this number needs some extra interpreting from the mathematical point of view:  it could be that the
holomorphic maps are not isolated, so that there is a whole moduli space of such maps.  Nevertheless,
the virtual or ``expected'' dimension of this moduli space is always zero (when $X$ is a Calabi-Yau threefold); roughly this means that one can define a sensible ``number of maps'' 
even when the actual dimension 
happens to be nonzero.  The index computation showing that the virtual dimension vanishes when $d=3$ 
is in fact isomorphic to the charge-conservation computation which showed that general $F_g$ are 
only nontrivial when $d=3$.}

The sum over $m$ reflects the subtlety that there are contributions from ``multi-wrappings,'' 
maps $\Sigma \to X$ which are $m$-to-one; these lead to a universal correction, 
determined by the geometry of maps $S^2 \to S^2$, captured by the factor $1 / m^3$.

\subsubsection{B model}

To write the B model partition function we introduce a convenient coordinate system for the 
complex moduli space.  To describe it we first discuss the space $H^3(X,\C)$, which has the
Hodge decomposition
\begin{equation}
\begin{tabular}{ccccccccc}
$H^3$ & = & $H^{3,0}$ & $\oplus$ & $H^{2,1}$ & $\oplus$ & $H^{1,2}$ & $\oplus$ & $H^{0,3}$, \\
$h^3$ & = &   $1$     &    +     & $h^{2,1}$ &    +     & $h^{2,1}$ &    +     &    $1$.          
\end{tabular}
\end{equation}
%FIXME: make this table look nicer
(The fact that $h^{3,0} = h^{0,3} = 1$ reflects the fact that a Calabi-Yau space has a unique
nonvanishing holomorphic 3-form up to scalar multiple.)
Therefore $H_3(X,\R)$ has real dimension $2 h^{2,1} + 2$.
Now we choose a symplectic basis of $H_3(X,\Z)$; this amounts to choosing
3-cycles $A^i$, $B_j$, for $i=1,\dots,h^{2,1}+1$ and $j=1,\dots,h^{2,1}+1$, with intersection numbers
\begin{equation} \label{symplectic-marking}
A^i \cap A^j = 0, \quad B_i \cap B_j = 0, \quad A^i \cap B_j = \delta^i_j.
\end{equation}
Note that $h^{2,1}(X)$ is the complex dimension of the moduli space of complex structures
(this identification is obtained by using the holomorphic 3-form to convert Beltrami differentials
to $(2,1)$-forms.)  This suggests that we could try to get coordinates on the moduli space by defining
\begin{equation} \label{A-cycle-periods}
X^i = \int_{A^i} \Omega.
\end{equation}
Actually this gives $h^{2,1}+1$ complex coordinates corresponding to the $h^{2,1}+1$ A cycles, one more 
than the $h^{2,1}$ needed to cover the moduli space.  The reason for this overcounting is 
that $\Omega$ is not quite unique for a given complex 
structure --- it is unique only up to an overall complex rescaling,
so from \eqref{A-cycle-periods}, the $X^i$ are also ambiguous up to an overall rescaling.  Thus
we have the right number of coordinates after accounting for this rescaling; and indeed the
periods over the A cycles do determine the complex structure.  Thus we say that the $X^i$ give
``homogeneous coordinates'' on the moduli space.

What about the periods over the B cycles?  Writing\footnote{There is an unfortunate clash of notation here; the $F_i$ we define here are \ti{not} the genus $i$ free energy, although below
we will consider the genus $0$ free energy, which we will write simply as $F$!}
\begin{equation} \label{B-cycle-periods}
F_i = \int_{B_i} \Omega
\end{equation}
it follows from the above that they must be expressible in terms of the A periods,
\begin{equation}
F_i = F_i(X^j).
\end{equation}
(Of course, since our choice of symplectic basis was arbitrary, and in particular we could have
interchanged the A and B cycles, one could equally well write $X^i = X^i(F_j)$.)

We are almost ready to write the B model genus zero free energy, but we 
need one more fact, namely the statement of ``Griffiths transversality.''
Recall that $\Omega \in H^{3,0}(X,\C)$.  Now work in a local complex coordinate
system in which $\Omega = f(z) \de z_1 \wedge \de z_2 \wedge \de z_3$, and consider a variation
of complex structure given by a Beltrami differential $\mu$, 
which changes the local complex coordinates by 
$\de z_i \mapsto \de z_i + \mu_i^{\overline{j}} \de \overline{z}_j$.
Then expanding in $\de z$ and $\de \overline{z}$ one sees that 
to first order in $\mu$, the variation of $\Omega$ satisfies $\delta \Omega \in H^{3,0} \oplus H^{2,1}$, and the second-order variations similarly have
$\delta \delta \Omega \in H^{3,0} \oplus H^{2,1} \oplus H^{1,2}$.
This implies
\begin{align}
\int_X \delta \Omega \wedge \Omega &= 0, \\
\int_X \delta \delta \Omega \wedge \Omega &= 0.
\end{align}
Using this fact and the ``Riemann bilinear identity,'' which states that for closed 3-forms
$\alpha$, $\beta$ one has
\begin{equation} \label{riemann-bilinear-identity}
\int_X \alpha \wedge \beta = \int_{A^i} \alpha \int_{B_i} \beta - \int_{A^i} \beta \int_{B_i} \alpha,
\end{equation}
one can prove that
\begin{equation}
\dwrt{X^i} F_j = \dwrt{X^j} F_i.
\end{equation}
This is the integrability condition which allows one to define a new function $F$:
\begin{equation}
F_i = \dwrt{X^i} F.
\end{equation}
The $F$ so defined is the genus zero free energy of the B model.  Strictly speaking, $F$ is not
quite a function on the complex moduli space, because it depends on the choice of the overall
scaling of $\Omega$; under $\Omega \mapsto \xi \Omega$ one has $F \mapsto \xi^2 F$.  So
$F$ is homogeneous of degree $2$ in the homogeneous coordinates $X^i$ on the moduli space;
geometrically speaking, it is a section of a line bundle over the moduli space rather than an honest function.\footnote{Even this more refined description
is still a little misleading, because $F$ also depends on the choice of A and B cycles, i.e. the choice
of a special coordinate system.  For a fixed such choice one obtains a homogeneous section $F$ as we 
described; if one makes a symplectic transformation of the basis, $F$ transforms
by an appropriate Legendre transform.}  It is given by a simple formula
\begin{equation} \label{f0-b-model}
F = \half X_i F^i.
\end{equation}

\subsubsection{Comparing the A and B models}

We have just described the content of the A and B models at genus zero.
Note that in contrast to the A model,
which involved an infinite sum over worldsheet instantons, weighed by the integral
coefficients $d_n$,
the B model free energy was determined purely by ``classical'' geometry (the periods) 
and has no obvious underlying integral structure.  These properties also persist to
higher genera.
In this sense one could say that the B model is easy 
to compute, and contains relatively boring information, while the A model is hard to compute
but contains more interesting information.  
On the other hand, it is the A model free energy which is
easier to define; at least formally it just counts holomorphic maps, whereas even to 
define the B model we had to introduce the notion of special coordinates!

\section{Computing the topological amplitudes} \label{sec-computing}

Having defined the topological string theory and seen that it is related to some quantities of
geometric interest, the next step is to learn how to compute the topological amplitudes at all genera.  In
principle they could be computed using their definition \eqref{topological-string-partition-function},
i.e. by direct integration over the moduli space of Riemann surfaces.  But this is too hard for all
but the very simplest amplitudes; if this were the only method at our disposal, 
topological string theory would be just a mathematical curiosity.  Instead it is a powerful
tool, because a variety of techniques have been discovered which allow one to compute topological string amplitudes
not only at tree level but to all genera!
In this section we will summarize the various major techniques for computation of topological string amplitudes.

First we describe mirror symmetry, a technique which allows one to exploit the simplicity of the B model
for computations in the A model.  It was first applied in genus zero, since that is where the B model
amplitudes are easiest to compute.  The B model computation was subsequently extended to
higher genera using the holomorphy of the amplitudes, thus effectively
solving the mirror A model at higher genera; we briefly indicate how this extension goes.  Next we
discuss an alternative approach to computation of topological amplitudes which exploits a duality
between the topological open and closed string; this approach yields results at all genera for a particular class of
non-compact geometries.  Along the way we sketch the meaning of branes in the topological string and their
target space field theories.  The results obtained from the open/closed duality suggest the existence of
a more powerful method for computations of A model amplitudes in arbitrary toric geometries; 
the last three subsections are devoted to this method, known as the ``topological vertex.''  First we 
describe what the vertex is; next we sketch a method of computing it using mirror symmetry and the
symmetries of the B model; and finally we describe an interpretation directly in the A model, where
the vertex is understood as a sum over fluctuations of \kahler geometry at the Planck scale, i.e., the
quantum foam.

\subsection{Mirror symmetry} \label{sec-mirror-symmetry}

In the last section we concluded that while the A model on a Calabi-Yau threefold $M$ contains
some very interesting geometric information about holomorphic curves in $M$,
it is the B model which is easier to compute.  Remarkably, it is possible to exploit the simplicity of
the B model to make computations in the A model!  Namely, the A model on a Calabi-Yau space $M$ is
often equivalent to a B model on a ``mirror'' Calabi-Yau space $W$.  Therefore computations of 
the periods of $W$ can be exploited to count holomorphic curves in $M$.

A good general reference for mirror symmetry is \cite{MR2003030}.

\subsubsection{T-duality}

To understand how such a surprising duality could be true, we consider an example which is in some
sense underlying the whole phenomenon:  bosonic string theory on a circle $S^1$ of radius $R$.  
The spectrum of physical states of
this theory has one obvious quantum number, namely the number $w$ of times the string is wound
around $S^1$.  It also has a second quantum number $n$ corresponding to the momentum of
the center of mass of the string going around the circle; this momentum is quantized in 
units of $1/R$, as is familiar from point particle quantum mechanics in 
compact spaces.  The contribution to the worldsheet energy of a state from these two quantum numbers
is (in units with $\alpha' = 1$)
\begin{equation}
E_{n,w} = (wR)^2 + \left(\frac{n}{R}\right)^2.
\end{equation}
Note that the set of possible $E_{n,w}$ is invariant under the interchange $R \leftrightarrow 1/R$ ---
namely $E_{n,w}$ at radius $R$ is the same as $E_{w,n}$ at radius $1/R$!
This is the first clue that this interchange might be a symmetry of the full string theory; indeed,
there is such a symmetry, called ``T-duality,'' which can be rigorously understood from the worldsheet
point of view, and has deep consequences for the target space physics.
Indeed, all of the different approaches to understanding mirror symmetry involve
T-duality in some essential way \cite{Strominger:1996it,Hori:2000kt}.

\begin{example}[Mirror symmetry for $T^2$]
The simplest example is one we already mentioned in Section
\ref{sec-examples-cy-d1}.  Namely, given a rectangular torus $T^2$ with radii $R_1$, $R_2$
and defining 
\begin{align} \label{torus-moduli-reprise}
A &= \I R_1 R_2, \\
\tau &= \I R_2 / R_1,
\end{align}
exchanging $R_1 \leftrightarrow 1 / R_1$ is equivalent to exchanging $A \leftrightarrow \tau$.  This
is an example of mirror symmetry for which 
$M$ and its mirror $W$ are both $T^2$, but with different metrics, i.e. different values of the
moduli.  Anyway, given that the physical string has this T-duality symmetry, one could ask
how it gets implemented in the topological theory.  Since T-duality exchanges complex and \kahler moduli
it would be natural to conjecture that it exchanges the A and B models, and this is indeed the
case; the A model on $T^2$ with \kahler modulus $A$ computes exactly the same quantity as the
B model on $T^2$ with complex modulus $\tau = A$.

Since $T^2$ has complex dimension $1 \neq 3$, most of the topological string
is trivial as we explained in Section \ref{sec-why-cy3}.  
However, one can still look at the one-loop free energy $F_1$, and
mirror symmetry turns out to be an interesting statement already here.  Namely, 
the B model at one loop
computes the inverse of the determinant of the $\bp$ operator acting on $T^2$, in keeping with the general principle that 
the B model has to do with \ti{local} expressions on the target space.  This determinant is
the Dedekind $\eta$ function,
\begin{equation} \label{eta-function}
\eta(q) = q^{1/24} \prod_{n=1}^\infty (1 - q^n),
\end{equation}
where $q = e^{2 \pi i \tau}$.  On the other hand, the A model at one loop counts maps $T^2 \to T^2$,
but according to mirror symmetry, it should also give the $\eta$ function.
This gives a natural interpretation of the integrality of the coefficients in the 
$q$-expansion of $1/\eta(q)$.  Namely, $q$ gets related to
$e^{-A}$ by the mirror map, and from the A model point of view 
the coefficient of $e^{-nA}$ counts maps which wrap $T^2$ over itself $n$ times.
It can be checked directly that this counting is indeed correct.
\end{example}

\subsubsection{Mirror symmetry for threefolds}

Now what about the case of maximal interest, namely Calabi-Yau threefolds?  Here also one might expect
a mirror duality.  Indeed, this duality was conjectured before a single non-trivial example was known, on 
the basis of lower-dimensional examples like the one discussed above, and also because 
from the point of view of the $\N=(2,2)$ algebra the difference between A and B models
is purely a matter of convention --- considered abstractly, the SCFT has no way of knowing whether it is an A model or a B model.  This conjecture turned out to be spectacularly true, and by now many examples of mirror pairs are known, both compact and non-compact.

Here we sketch a physical proof given in \cite{Hori:2000kt}
which encapsulates all known examples of mirror symmetry.
Like the T-duality example we gave above, 
the proof is most naturally stated directly in the physical superstring rather than the topological string; but after twisting it reduces to an equivalence between a topological A model and a
topological B model.

So we begin with a toric Calabi-Yau threefold $M$ and realize 
it concretely via the gauged linear sigma model of \cite{Witten:1993yc}, as we described
in Example \ref{example-cpn-toric}.  Recall that this model is 
constructed from a set of chiral superfields $Z_i$ representing 
the homogeneous coordinates of $M$, and that its space of
vacua is $M$ itself.  Then to get the mirror theory to the sigma model on 
$M$ one splits each $Z_i$ into its modulus and phase as we did 
before when discussing the toric diagram,
\begin{equation}
Z_i \to (\abs{Z_i}^2, \theta_i),
\end{equation}
and then performs T-duality on the circle coordinatized by $\theta_i$.  The T-duality gives a new dual periodic
coordinate $\phi_i$, and we organize this coordinate together with $\abs{Z_i}^2$ into a new ``twisted chiral'' superfield
\begin{equation}
Y_i = \abs{Z_i}^2 + \I \phi_i.
\end{equation}

Crucially, the dual description in terms of the $Y_i$ has a superpotential:
\begin{equation} \label{dualized-superpotential-with-sigma}
W(Y) = (\sum_i Q_i Y_i - t)\Sigma + \sum_i e^{- Y_i}.
\end{equation}
Here $\Sigma$ is the twisted chiral superfield in the $U(1)$ vector multiplet, and $Q_i$ are
the $U(1)$ charges of the $Z_i$.  The first term in \eqref{dualized-superpotential-with-sigma} follows from
a classical T-duality computation; the really interesting part is the second term.
This term was derived in \cite{Hori:2000kt} from an instanton computation in the gauged linear 
sigma model.  It
can also be determined more indirectly (and more easily), as 
follows \cite{Aganagic:2004yh}.\footnote{For simplicity we just discuss one chiral superfield $Z$.}
One compares masses of BPS particle states in the original theory and in the mirror.
In the original theory the field $Z$ has momentum modes with BPS mass $\abs{Q \Sigma}$.
After T-duality these momentum modes become winding modes along the T-dual circle,
i.e. they should correspond to classical BPS solitons where $\phi$ increases by $2 \pi \I$.
For such a soliton interpolating between vacua to exist, $W(Y)$ must have critical points
which are spaced by $2 \pi \I$.  Moreover, since the BPS mass of a soliton interpolating
from $y_1$ to $y_2$ is $\abs{W(y_1) - W(y_2)}$, we see that this difference must be
equal to $\abs{Q \Sigma}$.  Finally,
because of the periodicity of $\phi_i$, the instanton-generated superpotential
can only involve $Y_i$ through its exponential $e^{- Y_i}$.  These conditions are
enough to fix $W(Y)$ as in \eqref{dualized-superpotential-with-sigma}.

Integrating out $\Sigma$ then yields the
holomorphic constraints in the dual model,
\begin{equation} \label{holomorphic-constraint}
\sum_i Q_i Y_i = t,
\end{equation}
and the reduced superpotential,
\begin{equation} \label{dualized-superpotential}
W(Y) = \sum_i e^{- Y_i}.
\end{equation}
The two equations \eqref{holomorphic-constraint}, \eqref{dualized-superpotential}
contain all the information about the dual theory, as we now see in an example.
  
\begin{example}[Mirror symmetry for local $\C\PP^2$.]  
Consider the local $\C\PP^2$ geometry
$\OO(-3) \to \C\PP^2$, for which we discussed the toric realization
in Example \ref{example-local-cp2-toric}.
The gauged linear sigma model for this geometry involves four chiral superfields $Z_i$ with charges $Q_i$,
\begin{eqnarray}
\begin{tabular}{ccccc}
$Z =$ &($Z_0$,&$Z_1$,&$Z_2$,&$Z_3$), \\
$Q =$ &($ -3$,&$1  $,&$1  $,&$1$).
\end{tabular}
\end{eqnarray}
% FIXME: make that look nicer
The holomorphic constraint in the dual model is
\begin{equation} \label{holomorphic-constraint-local-cp2}
-3 Y_0 + \sum_{i=1}^3 Y_i = t,
\end{equation}
and the superpotential is
\begin{equation}
W = \sum_{i=0}^3 e^{- Y_i}.
\end{equation}
It is convenient to make the change of variables
\begin{equation} \label{dual-change-of-variables}
y_i = e^{-Y_i / 3}.
\end{equation}
Then, after eliminating $Y_0$ using \eqref{holomorphic-constraint-local-cp2}, we are left with
the superpotential
\begin{equation} \label{dual-superpotential-local-cp2}
W = y_1^3 + y_2^3 + y_3^3 + e^{t/3} y_1 y_2 y_3.
\end{equation}
So the mirror of the gauged linear sigma model is a gauged Landau-Ginzburg model, describing 
uncharged twisted chiral superfields $Y_i$ which interact via the superpotential \eqref{dualized-superpotential}.
More precisely, the mirror is an \ti{orbifold} of the Landau-Ginzburg model, 
because the change of variables \eqref{dual-change-of-variables} is not quite one-to-one; the $y_i$ are
ambiguous by cube roots of unity, and therefore we have to divide out by the group $\Z_3^2$ which multiplies the $y_i$ by cube roots of unity while leaving $W$ invariant.
This is the generic situation:  the mirror to an $\N=(2,2)$ gauged linear sigma model
is an orbifolded Landau-Ginzburg model.  Note that the complexified 
\kahler modulus $t$ of the original theory
appears in the Landau-Ginzburg model as a modulus of the holomorphic superpotential.

From this Landau-Ginzburg realization one can directly compute the 
desired genus zero partition function.  Nevertheless, one might ask:  how 
is the Landau-Ginzburg theory related to our original claim that
the sigma model on the Calabi-Yau geometry should have a mirror which is also
a sigma model on a Calabi-Yau?  The point is that the
Landau-Ginzburg model with superpotential \eqref{dual-superpotential-local-cp2} is
actually equivalent to a sigma model with Calabi-Yau target space:  more precisely, 
one can interpolate from one to the other just by varying \kahler parameters, which
are decoupled from the B model correlation functions.  
After so doing we obtain the mirror to the local $\C\PP^2$ geometry; it is simply
given by the equation $W=0$, modulo the orbifold action.

Let us look at this geometry a bit more closely.
If the $y_i$ are considered as homogeneous coordinates
in projective space, then $W=0$ describes an elliptic curve (torus)
since it is a cubic equation in $\C\PP^2$.  
Passing to inhomogeneous coordinates we could rewrite it as
an equation in two variables,
\begin{equation} \label{mirror-riemann-surface}
x^3 + z^3 + 1 + e^{t/3}xz = 0.
\end{equation}
Indeed, the mirror geometry in this case 
is effectively an elliptic curve rather than a Calabi-Yau threefold, in the sense that the
B model partition function can be computed solely 
from the geometry of the elliptic curve.  This is a
common phenomenon when computing mirrors of noncompact Calabi-Yaus.
Nevertheless, the usual statement of mirror symmetry requires a threefold
mirror to a threefold; to make contact with that formulation we should add two extra variables $u,v$ which enter the geometry in a rather trivial way, replacing \eqref{mirror-riemann-surface} by
\begin{equation}
x^3 + z^3 + 1 + e^{t/3}xz = uv.
\end{equation}
These two variables $u,v$ just contribute a quadratic term to the superpotential $W$ in the Landau-Ginzburg
realization, so they do not couple to the rest of the physics.
\end{example}

One can similarly derive mirror symmetry for compact Calabi-Yaus with linear sigma model realizations.
\begin{example}[Mirror symmetry for the quintic threefold.]
Recall the quintic threefold from Example \ref{example-quintic}.
This space can be obtained by
starting with the gauged linear sigma model for $\OO(-5) \to \C\PP^4$ and then introducing a 
superpotential
which reduces the space of vacua to the quintic hypersurface in $\C\PP^4$.  Temporarily
ignoring this superpotential and repeating the steps above, we get a
Landau-Ginzburg model with
\begin{equation}
W = y_1^5 + y_2^5 + y_3^5 + y_4^5 + y_5^5 + e^{t/5}y_1y_2y_3y_4y_5,
\end{equation}
modulo a $\Z_5^4$ symmetry multiplying the $y_i$ by fifth roots of unity.  Now what changes in 
the mirror if we include the superpotential in the original theory?  
Remarkably, it turns out that the only effect is to change the
fundamental variables of the theory to the $y_i$ instead of $Y_i$.  (One might think that what
is the ``fundamental variable'' is a matter of terminology, but concretely, it affects the 
measures of integration one uses when computing the B model periods.)
\end{example}

\subsubsection{Super mirror symmetry}

There is another point of view on mirror symmetry for compact Calabi-Yaus realized torically,
which is in a sense more direct.  Namely, it was observed in \cite{Schwarz:1995ak}
that the A model on the quintic threefold is in fact equivalent to the A model on a weighted
\ti{super} projective space $\C\PP^{1,1,1,1,1|5}$, with five bosonic directions and one 
fermionic one.  This space is compact 
but nevertheless can be constructed in a gauged linear sigma model without the need for a superpotential.  
Since it has $U(1)$ isometries, unlike the quintic threefold,
one can T-dualize on phases directly to obtain the mirror.  
This requires a generalization of the mirror techniques of 
\cite{Hori:2000kt} to the case of a chiral superfield $\Theta$ whose lowest component is fermionic, which
was worked out in \cite{Aganagic:2004yh}.  

The main difference from the bosonic case is that the
number of fields is not conserved:
namely, since the phase of $\Theta$ is bosonic,
dualizing it gives a new bosonic chiral superfield $X$.  But since $\Theta$ contributes central charge
$-1$ instead of $+1$ in the sigma model, 
one also has to get two more fermionic fields $\eta,\chi$ on the mirror side,
since the central charges on the two sides must be equal.
As in the bosonic case the superpotential can be determined by comparing BPS masses, and it 
turns out to be
\begin{equation}
W(\Sigma,X,\eta,\chi) = -Q \Sigma(X - \eta \chi) + e^{-X}.
\end{equation}
This superpotential defines the mirror Landau-Ginzburg model.

In addition to providing a streamlined derivation of the mirror periods for hypersurfaces in toric
varieties, super mirror symmetry is important in its own right, particularly in light of a recent
application of topological strings on supermanifolds to a twistorial reformulation of
$\N=4$ super Yang-Mills theory \cite{Witten:2003nn}.
In that case the supermanifold in question is the super twistor space 
$\C\PP^{3|4}$, and computations in \cite{Aganagic:2004yh}
showed that its mirror is (at least in the limit where $\C\PP^{3|4}$ has large volume) a quadric hypersurface
in $\C\PP^{3|3} \times \C\PP^{3|3}$.  This result may be relevant for gauge theory, since $\C\PP^{3|3} \times \C\PP^{3|3}$ can also be viewed as a twistor space, which is related at least classically to $\N=4$
super Yang-Mills \cite{Witten:1978xx}; one might expect that a topological string
on $\C\PP^{3|3} \times \C\PP^{3|3}$ could give an alternative twistorial version of $\N=4$ super Yang-Mills
\cite{Aganagic:2004yh,Witten:2003nn,Neitzke:2004pf}.

\subsection{Holomorphy and higher genera}

So far we have discussed the topological amplitudes only at genus zero.  
More generally, one can compute all the $F_g$
using the fact that they depend only holomorphically on moduli.\footnote{Actually, 
as we mentioned in Section \ref{sec-holomorphic-anomaly}, 
the $F_g$ are not quite holomorphic; but the antiholomorphic 
dependence is completely determined by the anomaly equation of \cite{Bershadsky:1994cx} and
does not qualitatively affect the discussion to follow.} We think of $F_g$ as a holomorphic
section of a line bundle over the moduli space.  Such objects are highly constrained --- recall 
that a holomorphic line bundle over a compact space has only a finite-dimensional space of
holomorphic sections.  The Calabi-Yau moduli spaces under consideration are compact,
or can be compactified by adding some points at infinity, where the singular behavior of the $F_g$ can
be constrained by geometrical considerations; hence the $F_g$ are basically
determined by holomorphy, up to a finite-dimensional ambiguity at each $g$ \cite{Bershadsky:1994cx}.  
Using some integrality properties of the $F_g$ which we discuss in Section
\ref{sec-black-hole-5d}, this ambiguity can
also be fixed; this leads to a practical method for computing the $F_g$, which has been applied
to high degrees and genera \cite{Katz:1999xq}.

\subsection{Branes and large $N$ dualities}

Another approach to computing the $F_g$ depends on the notion of ``large $N$ duality.''  Such 
dualities have played a starring role in the physical string theory over the last few years
\cite{Aharony:1999ti,Maldacena:1998re}; as it turns out, they are equally important in the topological string \cite{Gopakumar:1998ki,Ooguri:2002gx}.  We now turn to an overview of how they are realized
in this context.

\subsubsection{D-branes in the topological string} \label{sec-branes}

Large $N$ dualities relate open string theory in the presence
of $N$ D-branes to closed string theory in the gravitational background those D-branes produce; so in 
order to discuss their topological realization, we have to begin by explaining the notion of D-brane
in the topological string.

From the worldsheet perspective, a D-brane simply corresponds to a boundary condition which can be
consistently imposed on worldsheets with boundaries.  In the topological case what we mean by
``consistency'' is that the
boundary condition preserves the BRST symmetry.  In the A model this condition implies that the boundary
should be mapped into a Lagrangian submanifold $L$ of the target Calabi-Yau $X$ \cite{Witten:1992fb} (``Lagrangian'' means that the dimension of $L$ is half that of $X$ and the \kahler form $\omega$ vanishes when restricted to $L$).  Such an $L$ should be thought of as a real section of $X$ --- 
a typical $1$-dimensional model is the upper half-plane, which ends on the real axis $L$.
If we allow open strings with boundaries on $L$, we say that we have a D-brane which is ``wrapped'' on $L$.
We can also include a weighting factor $N$ for each boundary, in which case we say we have $N$ D-branes
instead of one.

\figscaled{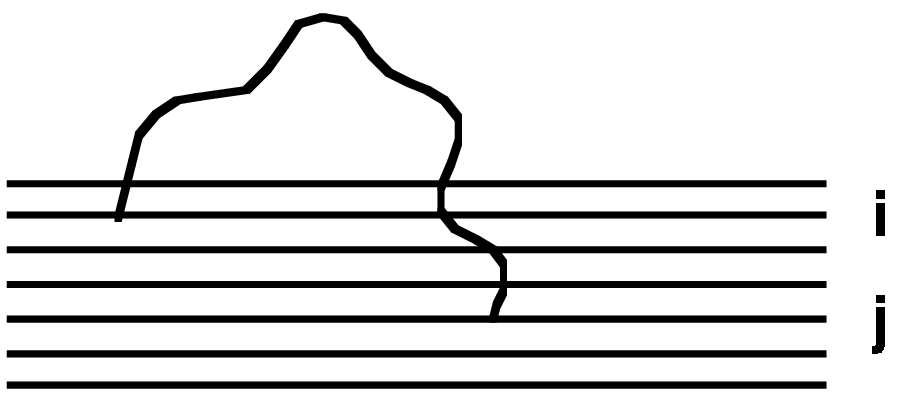}{A stack of $N$ branes carries a $U(N)$ gauge symmetry; the fundamental
and antifundamental gauge indices arise from strings which can end on any of the $N$ branes.}{width=4in}

We will be interested in computing the partition function of the topological 
open string theory with branes.  For this purpose
it turns out that taking a target space viewpoint is very convenient:  the dynamics of the open strings
ending on branes 
can be completely described in terms of a string field theory on the branes.  What field theory is it?
Both in the physical and the topological string theory, the open strings produce a gauge theory on the branes in the low energy limit; for example, in the case of a stack of $N$ coincident branes in oriented
string theory in flat space, the fact that strings can end on any of the $N$ branes
leads to a $U(N)$ gauge theory.  See Figure \ref{fig-stack-of-branes}.

In the physical string the gauge theory of the open strings is rather complicated, 
although at low energies it reduces to Yang-Mills theory.
But in the topological A model the situation is much simpler and one can work out 
the exact open string field theory describing a stack of $N$ branes; it is again a gauge
theory, but this time a topological gauge theory, namely $U(N)$ Chern-Simons theory.  
To see this we first note
that our construction of the topological string (and specifically its coupling to worldsheet 
gravity) was modeled on the bosonic string, and therefore the open string field theory should
also be the obtained by the same procedure one uses for the open bosonic string.  
In the open bosonic string it was shown in \cite{Witten:1986cc} that the string 
field theory is an abstract version of Chern-Simons, written
\begin{equation}
S = \int A * Q A + \frac{2}{3} A * A * A.
\end{equation}
Specializing to the case of the topological A model, using the dictionary $Q \leftrightarrow \de$, 
one can show \cite{Witten:1992fb} that this abstract Chern-Simons in this case boils down to 
the standard Chern-Simons action for a $U(N)$ connection $A$,
\begin{equation}
S = \int_L \Tr(A \wedge \de A + \frac{2}{3} A \wedge A \wedge A),
\end{equation}
possibly corrected by terms involving holomorphic instantons ending on $L$.\footnote{In 
fact, one might ask how the appearance of the Chern-Simons action is consistent with the localization
of the open string path integral on holomorphic configurations --- one might have expected to get
\ti{only} the terms from holomorphic instantons.  The resolution is that the localization
has to be interpreted carefully because of the non-compactness of the field space; one has to include
contributions from ``degenerate instantons'' in which the Riemann surface has collapsed to a Feynman
diagram (with lines replaced by infinitesimal ribbons), and these diagrams precisely account for the Chern-Simons action.}  In some interesting cases
there are no holomorphic instantons and we just get pure Chern-Simons; this happens in particular
in the case where $L$ is the $S^3$ in the deformed conifold $T^* S^3$.

One can similarly consider the open string field theory on $N$ B model branes.  In the case
where the branes wrap the full Calabi-Yau threefold $X$, one gets a holomorphic version of Chern-Simons,
with action \cite{Witten:1986cc}
\begin{equation} \label{holomorphic-chern-simons}
\int_X \Omega \wedge \Tr (A \bp A + \frac{2}{3} A^3).
\end{equation}
Here $A$ is a $u(N)$-valued $(0,1)$ form on $X$, which we are combining with the $(3,0)$ form
$\Omega$ so that the full action is a $(3,3)$ form as required.  Starting from
\eqref{holomorphic-chern-simons}, one can also obtain the action 
for B model branes which wrap holomorphic 0,2,4-cycles inside $X$, by realizing such lower-dimensional 
branes as defects in the gauge field on a brane that fills $X$:  the brane charges correspond respectively to the Chern 
classes $c_3$, $c_2$, $c_1$ of the gauge field.

As an aside, it is interesting that the branes which appear in the A model are wrapping Lagrangian
cycles, which are 3-cycles for which the volume is naturally measured by the holomorphic 3-form
$\Omega$ --- the natural object in the B model!  Similarly, in the B model the branes turn out to
wrap holomorphic cycles, whose volume is measured by the A model field $k$.  This crossover between
the A and B models may be a hint of a deeper relation, possibly an S-duality, which is currently under investigation \cite{Neitzke:2004pf,Nekrasov:2004js}.

\subsubsection{The geometric transition} \label{sec-geometric-transition}

After these preliminaries on branes in the topological string, we are ready to use them to compute
closed string amplitudes.  The crucial point which makes such a computation possible 
is that the topological D-branes affect the closed string background; so we first explain how 
this works.  

In the physical superstring D-branes
are sources of Ramond-Ramond flux.  In the A or B model topological string we expect something
similar, but now the flux in question should be 
the \kahler 2-form or holomorphic 3-form respectively.  More precisely, consider a Lagrangian subspace
$L$, on which an A model brane could be wrapped.  
Since the total dimension of $X$ is $6$, we can consider a $2$-cycle
$C$ which links the $3$-dimensional $L$, similar 
to the way two curves can link one another inside a space of total dimension $3$.
The precise meaning of ``link'' is that 
$C = \p S$ for some 3-cycle $S$ which intersects $L$ once; so $C$ is homologically
trivial as a cycle in $X$, although it becomes nontrivial if considered as a cycle in $X \setminus L$.
Because $C$ is homologically trivial we must have $\int_C k = 0$ in $X$, since $\de k = 0$.
Now the effect of wrapping $N$ branes on $L$ is to create a flux of the \kahler form through $C$, namely
\begin{equation}
\int_C k = Ng_s.
\end{equation}
This can be understood by saying that the branes act as a $\delta$-function
source for $k$, i.e., the usual $\de k = 0$ is replaced by 
\begin{equation}
\de k = N g_s \delta(L).
\end{equation}
Similarly, a B model brane on a 2-cycle $Y$ induces a flux of $\Omega$ over a 3-cycle linking $Y$.
Note that this phenomenon actually suggests a privileged role for 2-cycles; we could also have B 
model branes on 0, 4, or 6-cycles, but these branes do not induce gravitational backreaction since
there is no candidate field for them to source.

\figscaled{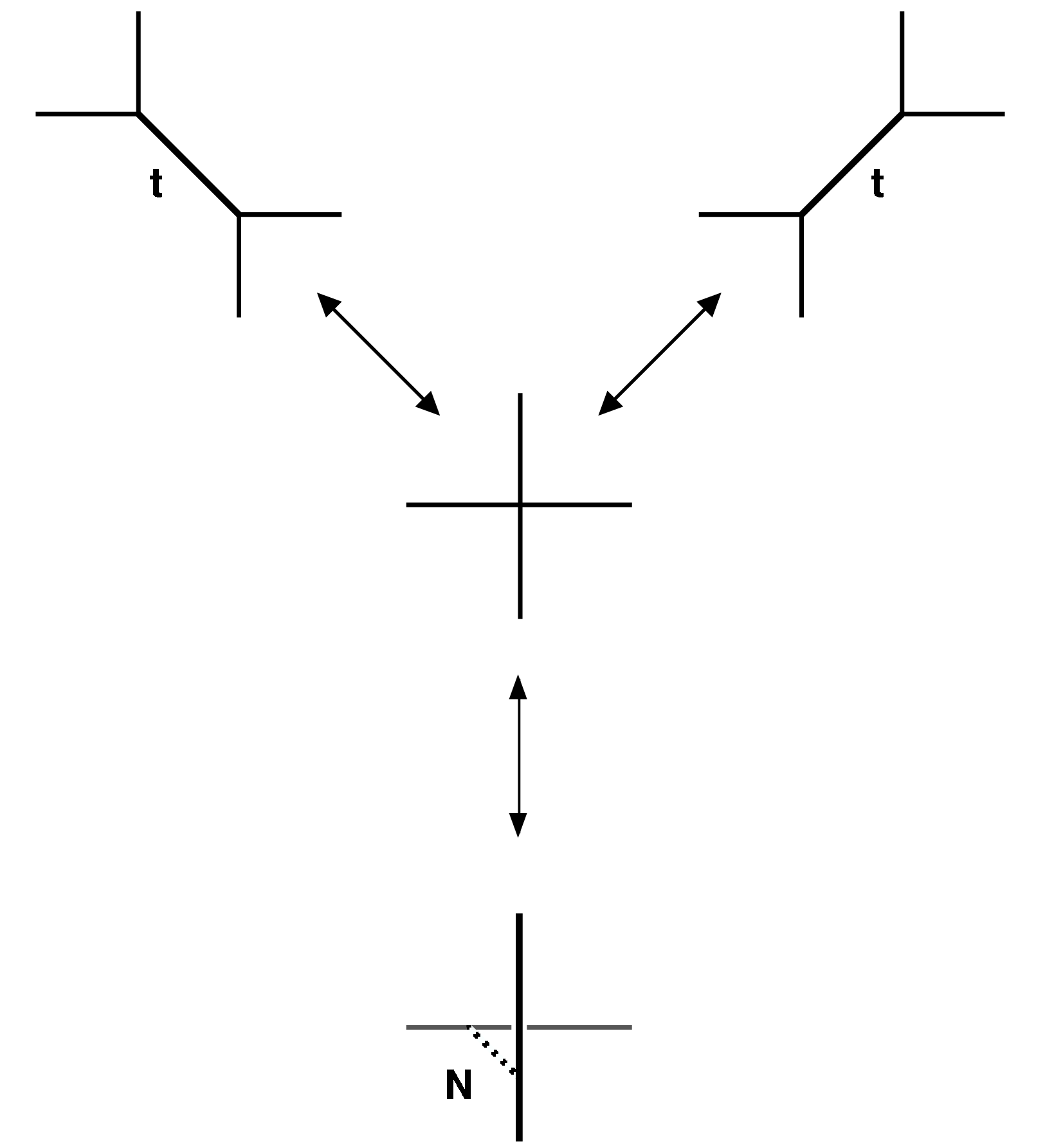}{The geometric transition between the resolved conifold with \kahler parameter $t = N g_s$ (above)
and deformed conifold with $N$ branes (below).}{height=4in}

Now let us describe an example in which the 
closed string backreaction from branes can be used to compute A model closed string amplitudes.
\begin{example}[Large $N$ duality for the conifold.]
Consider the A model on the deformed conifold $T^*S^3$.  This geometry is uninteresting
from the point of view of the closed A model, since it has no 2-cycles and hence no \kahler moduli; but 
it contains the Lagrangian 3-cycle $S^3$ on which we can wrap A model branes.  The effect of these 
branes on the closed string geometry is to create a flux $N g_s$ of the \kahler form $k$ on the $S^2$
which links $S^3$.  Now, \'{a} la AdS/CFT, let us try to describe the string theory on this 
geometry in terms of a background
without branes.  There is an obvious guess for the answer:  as we discussed earlier, in addition to
the deformed conifold which has a nontrivial $S^3$ at its core, there is also the resolved
conifold which has a nontrivial $S^2$, and both geometries look the same at long distances.  So
it is natural to conjecture \cite{Gopakumar:1998ki} that the dual geometry is the resolved conifold, where the nontrivial $S^2$ has volume $t = N g_s$.
In the resolved conifold there are no branes anymore, and indeed there is not even a nontrivial cycle where the branes could have been wrapped!  The passage from one geometry to the other is referred to as a ``geometric transition,'' and the key is that
the A model partition function is the same both before and after the geometric transition.  

The geometric transition is summarized in 
Figure \ref{fig-conifold-geometric-transition}.  The pictures appearing
in that figure require a bit of explanation, though:
they are similar, but not identical, to the toric pictures which appeared in 
Section \ref{sec-toric-geometry}.  The full geometry in this case is a $T^2 \times \R$ fibration 
over the whole of $\R^3$, rather than a $T^3$ fibration over some bounded region inside $\R^3$.
The solid lines represent loci where one of the circles of the $T^2$ fiber degenerates.  
At the top of the figure we have the resolved conifold, with its \kahler modulus $t$
(actually there are two different versions of the resolved
conifold, related by a relatively mild topology changing transition called a ``flop.'')  
Here all the degeneration loci line in a common plane.  As $t \to 0$ the resolved conifold
approaches the singular conifold, shown in the middle of the figure; the degeneration locus 
then consists simply of two intersecting lines.  Finally separating these two lines in space
gives the deformed conifold $T^* S^3$.  
The Lagrangian submanifold $S^3$ can be seen in the resulting picture:  namely, it is the
$T^2$ fibration over the dotted line connecting the two degeneration loci.

On the deformed conifold the closed A model is trivial, because there are no \kahler moduli; so
the partition function is just the partition function of $U(N)$ Chern-Simons theory on $S^3$, with level $k$
determined by $g_s = 2 \pi i / (k+N)$.   On the resolved conifold there are no open strings, so one gets
a prediction for the partition function of the closed A model there:  namely, from 
the Chern-Simons side one expects
\begin{equation}
Z(g_s, t) = \prod_{n=1}^\infty (1 - q^n Q)^n,
\end{equation}
where $q = e^{- g_s}$ and $Q = e^{-t}$.
Note that this expansion has integral coefficients!  This seems remarkable from the point of 
view of the closed string, and might make us wonder whether the closed string partition function
has an interpretation as the answer to some counting problem.  The answer
is ``yes,'' as we will see in Section \ref{sec-black-hole-5d}
when we discuss the application of the topological string to counting BPS states
in five dimensions.
\end{example}

One can also use open/closed duality to compute the 
closed string partition function in more complicated geometries \cite{Aganagic:2002qg}, as we
now discuss. 
\figscaled{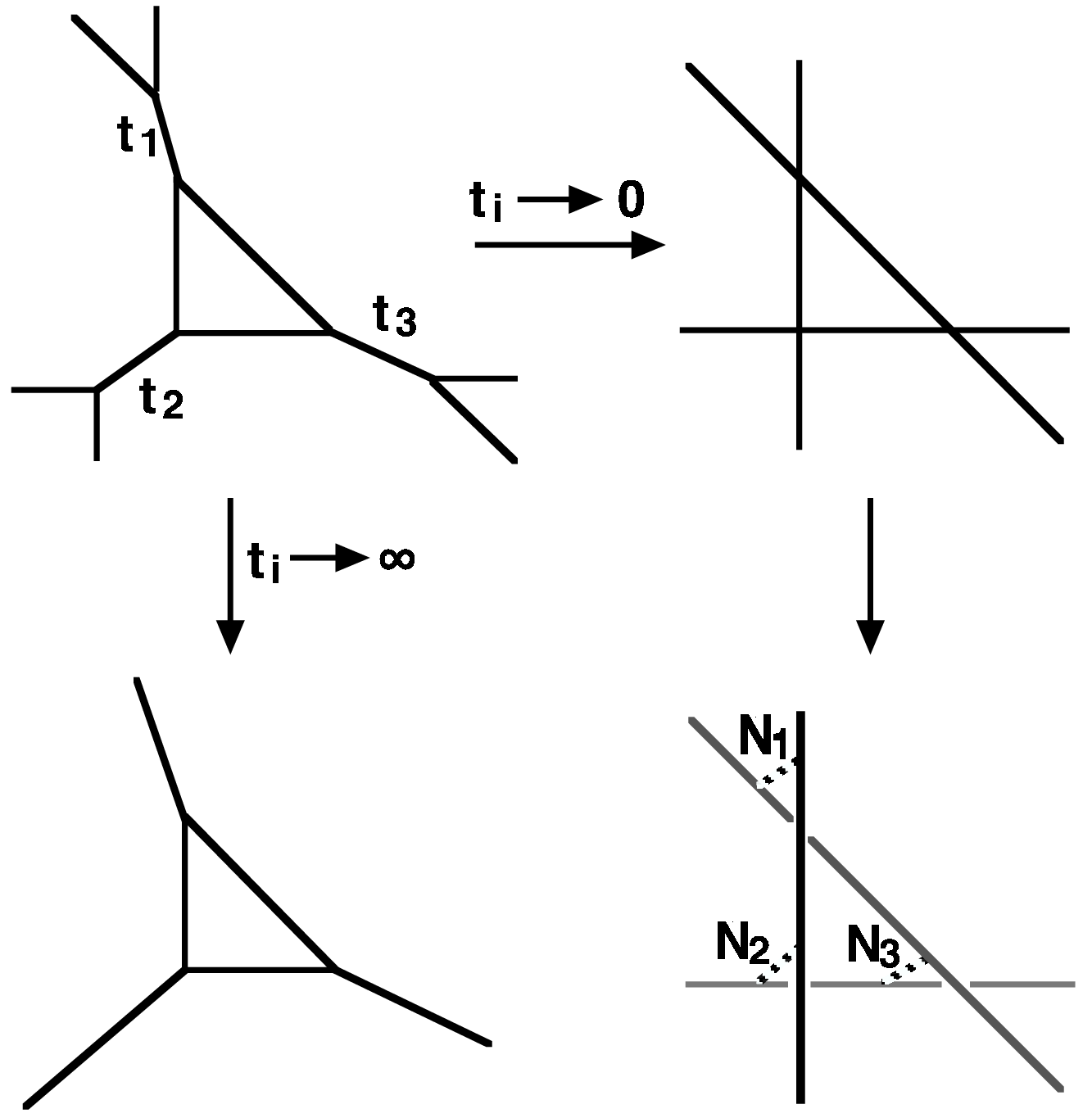}{The geometric transition 
relating local $\C\PP^2$ (lower left) to a rigid geometry with $N_i \to \infty$ branes (lower right).}{height=4in}
\begin{example}[Large $N$ duality for local $\C\PP^2$.]
For example, consider the local
$\C\PP^2$ geometry.  As shown in Figure \ref{fig-local-cp2-geometric-transition}, we can
obtain this geometry as the $t_i=N_ig_s \to \infty$ limit of a geometry with three compact
$\C\PP^1$'s.  Namely, in the 
lower left corner we have local $\C\PP^2$, which we consider as the $t_i \to \infty$
limit of the more complicated geometry at upper left.  This geometry is in turn related
by three geometric transitions to the geometry at lower right, which has three Lagrangian $S^3$'s
represented by the dotted lines, each supporting $N_i$ A model branes.
In this way the closed A model partition function on local $\C\PP^2$ is identified with the open 
string partition function on these three stacks of branes;   
no \kahler moduli remain after the transitions, 
so the closed string does not contribute anything.  Naively, this open string 
partition function would be just the product of three copies of 
the Chern-Simons partition function, coming from the three $S^3$'s.  
However, we have to remember that the open string field theory of the A model 
is not pure Chern-Simons theory; it includes corrections due to worldsheet instantons.  
In this toric case one can show that the only instantons which contribute
are ones in which the worldsheets form tubes connecting two of the Lagrangian
$S^3$'s, as shown in Figure
\ref{fig-tube-instanton}.  Each such tube ends on an unknotted circle in $S^3$; so in a 
generic instanton sector each $S^3$ has
two such circles on it, and a careful analysis shows that these circles are in fact linked, forming 
the ``Hopf link.''  One therefore has to compute the Chern-Simons partition function including 
an operator associated to the link.  This operator was determined in \cite{Ooguri:1999bv} and
turns out to be given by a sum of Chern-Simons link invariants.
Putting everything together \cite{Aganagic:2002qg},
the full partition function at all genera is a sum over irreducible representations of $U(N)$:
\begin{equation}
Z = \sum_{R_1, R_2, R_3} e^{-t \abs{R_1}} S_{R_1 R_2} e^{-t \abs{R_2}} S_{R_2 R_3} e^{-t \abs{R_3}} S_{R_3 R_1},
\end{equation}
where $S_{R R'}$ is the Chern-Simons knot invariant of the Hopf link with representations $R$ and $R'$ on the two circles, as defined in \cite{Witten:1989hf}, and $\abs{R}$ is the number of boxes in a Young
diagram representing $R$.
\figscaled{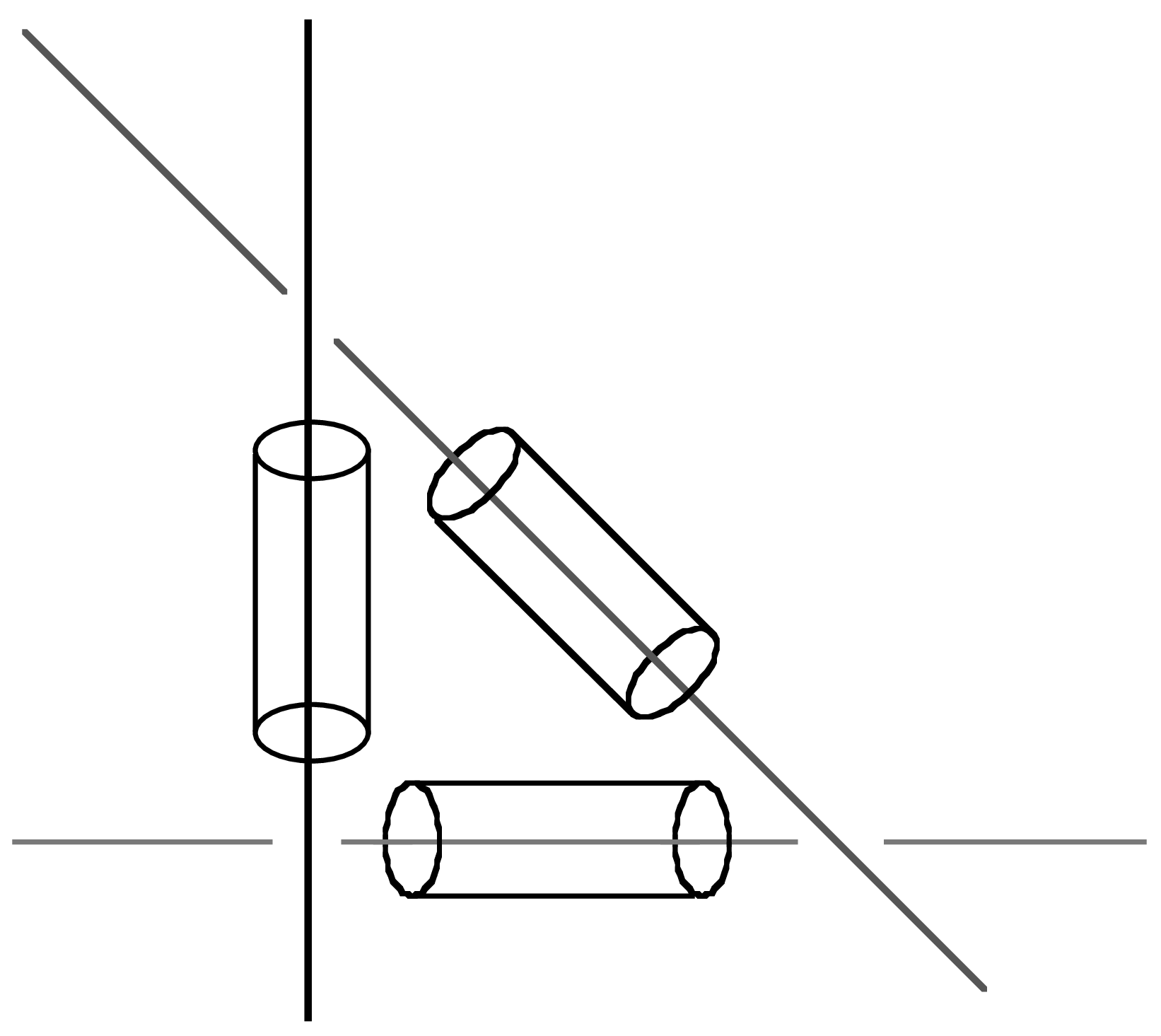}{Worldsheet instantons, with each boundary on an $S^3$, which
contribute to the A model amplitudes after the transition, or dually, to the
A model amplitudes on local $\C\PP^2$.}{height=3in}
\end{example}

\subsection{The topological vertex}

Although the geometric transitions we described above lead to an all-genus formula for the A model
partition function in the local $\C\PP^2$ geometry, the method of computation is somewhat unsatisfactory:  
one obtains local $\C\PP^2$ only after taking the $t_i \to \infty$ limit of a more complicated geometry.  One
might have hoped for a more intrinsic method of computation.  Indeed there is such a method, and
it generalizes to arbitrary toric diagrams, whether or not they come from geometric transitions! 
The method essentially involves treating the toric diagram (with fixed \kahler parameters) 
as if it were a Feynman diagram, with trivalent vertices and fixed Schwinger parameters.  Namely, one can define
a ``topological vertex,'' $C_{R_1 R_2 R_3}(g_s)$, depending on three Young diagrams $R_1, R_2, R_3$ and on the string coupling $g_s$ \cite{Aganagic:2003db}.  See Figure \ref{fig-topological-vertex}.
\figscaled{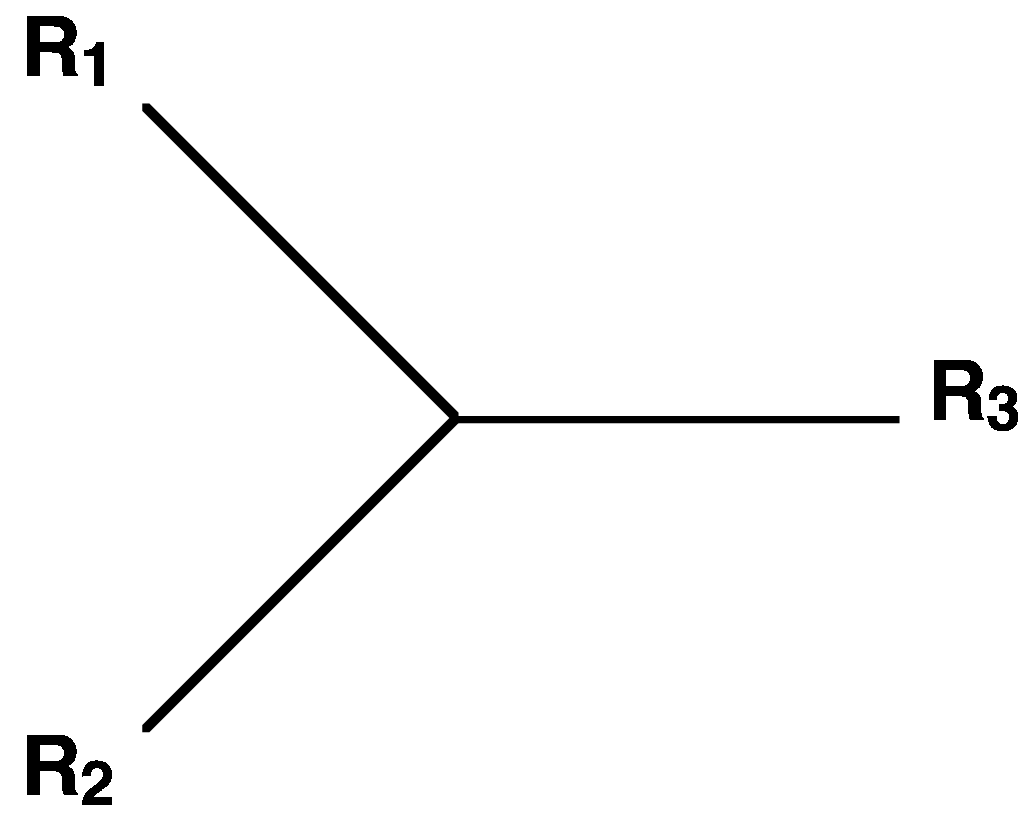}{The topological vertex, which assigns a function of $g_s$
to any three Young diagrams $R_1$, $R_2$, $R_3$.}{width=2in}
Then one assigns a Young diagram $R$
to each edge of the toric diagram, with a propagator $e^{-t \abs{R} + m C_2(R)}$ for each
internal edge, and a factor $C_{R_1 R_2 R_3}(g_s)$ for each vertex.\footnote{The integer $m$ appearing in the propagator is related to the relative orientation of the 2-surfaces on which the propagator ends.}  The assignment of
representations to edges of the toric diagram is as follows:  external edges always carry the trivial representation,
while for internal edges one sums over all $R$.

Of course, the actual vertex $C_{R_1 R_2 R_3}(g_s)$ is rather complicated!  It was originally determined
in \cite{Aganagic:2003db} using Chern-Simons theory along the lines discussed in Section
\ref{sec-geometric-transition}.  Since then two other
methods of computing the vertex have appeared, which we will describe in the next two subsections.

\subsection{Computing the vertex from $W_\infty$ symmetries}

First we briefly describe a target space approach to computing the topological string partition function
\cite{top-integ}.
Namely, suppose we study the A model on a non-compact threefold which has a toric realization
as we discussed in Section \ref{sec-toric-geometry}.  
By mirror symmetry this is equivalent to the B model on a Calabi-Yau of the form
\begin{equation}
F(x,z) = uv,
\end{equation}
with the corresponding holomorphic 3-form
\begin{equation}
\Omega = \frac{\de u \wedge \de x \wedge \de z}{u}.
\end{equation}
We view this geometry as a fibration over the $(x,z)$ plane, with 1-complex-dimensional
fibers.  At points $(x,z)$ with $F(x,z) = 0$
the fiber degenerates to $uv = 0$, which has two components
$u=0$ and $v=0$; so $F(x,z) = 0$ characterizes the degeneration locus of the fibration.  
Contour integration around $u=0$ on the fiber
reduces $\Omega$ to 
\begin{equation}
\omega = \de x \wedge \de z.
\end{equation}

So the geometry of the Calabi-Yau
threefold is captured by an algebraic curve $F(x,z) = 0$,
embedded in the $(x,z)$ space; this ambient space is furthermore 
equipped with the two-form $\omega$.  What are the symmetries
of this structure?  If $F$ were identically zero, then we would just have the group of $\omega$-preserving
diffeomorphisms, which form the so-called ``$W_\infty$'' symmetry.  This infinite-dimensional symmetry
is extremely powerful.  Indeed, even when $F \neq 0$ and the $W_\infty$ symmetry is spontaneously broken, it nevertheless gives constraints on the dynamics of the Goldstone modes which 
describe deformations of $F$.  But these
deformations exactly correspond to complex structure deformations of the Calabi-Yau geometry, which
are the objects of study in the B model!  Hence this $W_\infty$ symmetry generates Ward identities which
act on the closed string field theory of the B model (the ``Kodaira-Spencer theory of gravity,'' described
in \cite{Bershadsky:1994cx}.)
In fact, these Ward identities are sufficient to
completely determine the B model partition function at all genera (and hence the A model partition function
on the original toric threefold) --- see \cite{top-integ}.

\subsection{Quantum foam} \label{sec-quantum-foam}

In the last subsection we sketched a derivation of the topological vertex by applying mirror symmetry
and then using the B model closed string target space theory.
However, one can also obtain the vertex by a direct A model
closed string target space computation \cite{Okounkov:2003sp,Iqbal:2003ds,gw-dt}.  
The string field theory in question is
a theory of ``\kahler gravity'' \cite{Bershadsky:1996sr}, which roughly sums over \kahler geometries with
the weight $e^{- \int k^3 / g_s^2}$.  One can think of this summing over geometries as a kind of
``quantum foam'' --- the spacetime itself is wildly fluctuating and ``foamy'' at small scales.  This
feature has long been expected for theories of quantum gravity, but in the case of the topological A model
it turns out that one can describe this quantum foam very precisely; namely,
there is a simple description of exactly which \kahler geometries should be 
summed over, and this description enables us to compute the topological vertex.

\figscaled{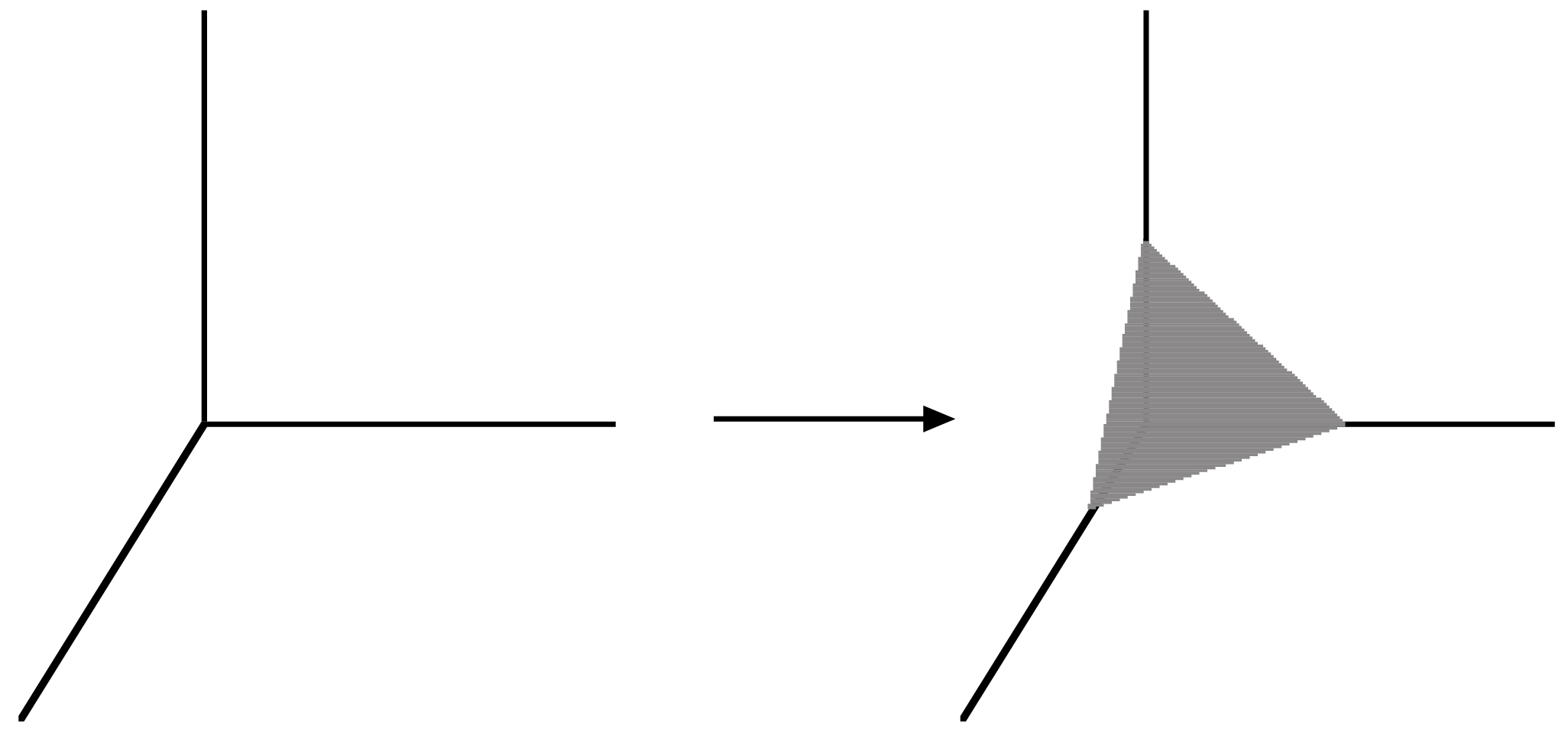}{Blowing up the origin in $\C^3$ gives a new geometry which is not Calabi-Yau
but still contributes to the target space sum in the A model.}{width=6in}
So let us begin with the problem of computing the A model partition function 
on the non-compact Calabi-Yau $\C^3$.  The simplest geometry which contributes to
the quantum foam in this case is simply $\C^3$ itself.  The rest of the geometries
that contribute may be obtained by making various blow-ups involving the origin $(0,0,0) \in \C^3$.\footnote{One 
might wonder what is special about the origin, since $\C^3$ has
a translation symmetry.  Actually, there is nothing special about the origin.  We are using a 
toric realization of $\C^3$ to get a $U(1)^3$ action on the space of possible blow-ups, and the claim
is that by standard localization techniques, the partition function can be computed considering
only blow-ups which are torically invariant.  Since the origin is the only point of $\C^3$
that is invariant in the toric representation we chose, 
this implies that we only consider blow-ups of the origin.}  The simplest possibility
is to just blow up the origin once; this leads to the toric diagram shown in
Figure \ref{fig-cp2-blow-up}, where the origin has been replaced by a single $\C\PP^2$.
This new geometry is not Calabi-Yau; the only Calabi-Yau 
geometry which is asymptotically $\C^3$ is $\C^3$ itself.  Nevertheless, it should be included in the
target space A model sum; this is not unexpected, since a theory of quantum gravity should sum over
off-shell configurations as well as on-shell ones.  
After blowing up the origin there
is a new \kahler modulus $t$ for the size of $\C\PP^2$; in the A model partition sum this modulus turns
out to be quantized, $t = n g_s$, and we sum over all $n$.  In the toric diagram the modulus $t$ is
reflected in the size of the triangle representing $\C\PP^2$, as we discussed in Section \ref{sec-toric-geometry}.

\figscaled{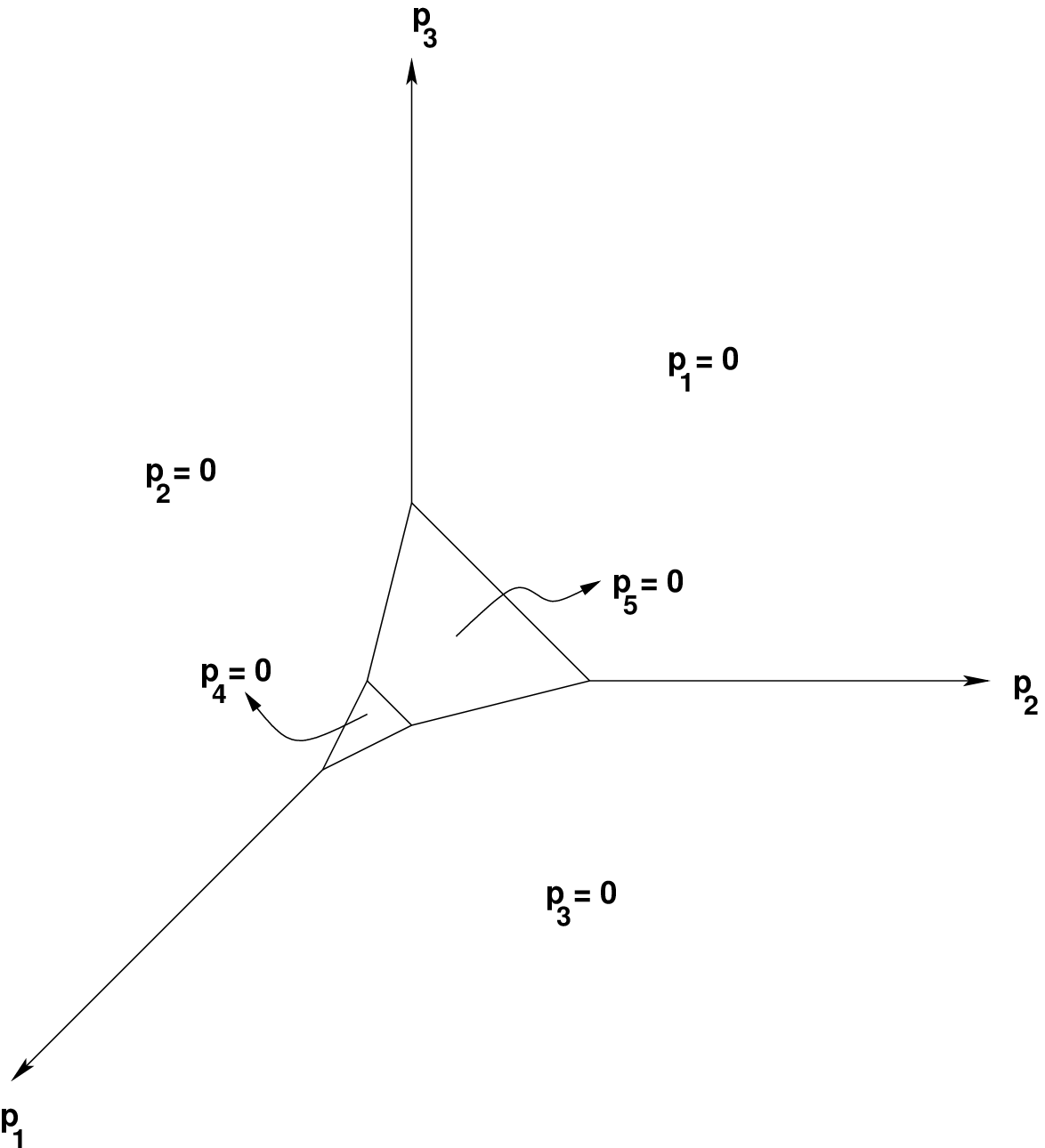}{This toric diagram is the result of blowing up the origin of $\C^3$
and then blowing up a torically invariant point on the exceptional divisor $\C\PP^2$.}{height=3in}
One can also do more complicated blow-ups.  For example, after blowing up the origin of $\C^3$, one
could then blow up a fixed point on the exceptional divisor $\C\PP^2$, 
as shown in Figure \ref{fig-cp2-blow-up-exc}.  We could then blow up another point on the resulting surface,
then another, and so on.  Continuing in this way one obtains a large class of toric manifolds which are
asymptotically $\C^3$; a typical example is shown in Figure \ref{fig-cp2-many-blowups}.
\figscaled{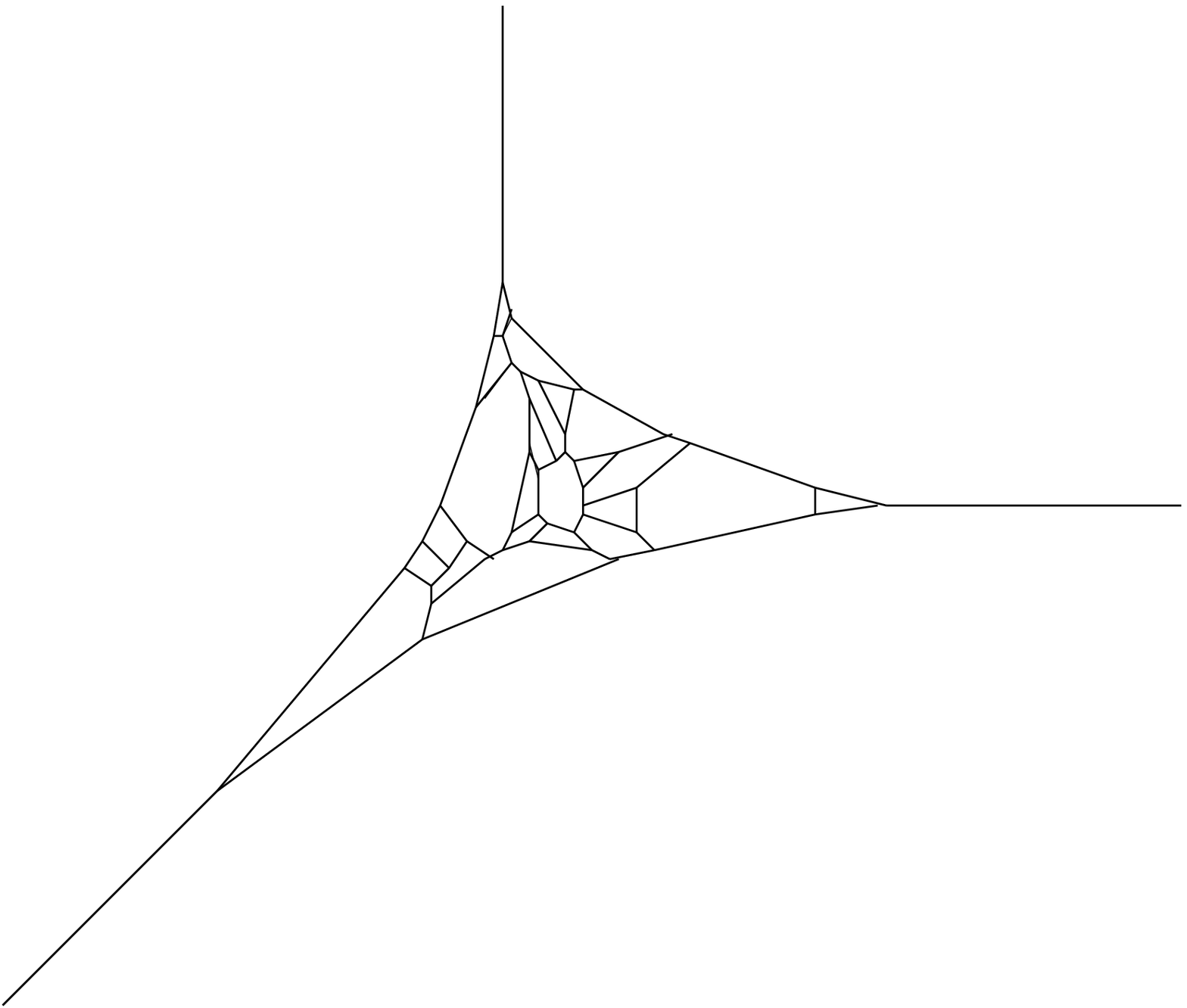}{This toric diagram represents a typical result of blowing up the origin
in $\C^3$, then blowing up a point on the exceptional divisor, then blowing up another torically
invariant point on the exceptional set, and repeating many times.}{height=3in}

However, it turns out that these blow-ups are not the only configurations that contribute to the 
A model partition sum.  
Namely, for any toric manifold obtained by successive blow-ups of points, 
the interior of the toric diagram is always a convex set;
but to reproduce the A model partition function
one also has to include generalizations of Figure \ref{fig-cp2-many-blowups} in which 
the interior of the diagram is not required to be convex.  These generalizations still have 
an algebro-geometric meaning, which can be roughly explained
as follows \cite{Iqbal:2003ds}.
Consider the ring $R = \C[X,Y,Z]$ of algebraic functions on $\C^3$; these 
are just polynomials in the three complex coordinates.  Given any ideal $I$ in 
$R$, there is an construction known as ``blowing up along $I$'' \cite{MR57:3116}, which yields a new algebraic variety, equipped with a line bundle ${\mathcal L}$.  Holomorphic sections
of this line bundle correspond precisely to elements of $I$.  Note in particular that there are many
ideals $I$ which give the same algebraic variety but different bundles ${\mathcal L}$.  We identify
the first Chern class of ${\mathcal L}$ with the \kahler class $k$ (so $k$ is naturally quantized!)

In the partition sum we want to blow up not along arbitrary ideals but only over torically invariant ones;
the coordinate ring $R$ has a natural action of $U(1)^3$ which just multiplies $X$,$Y$ and $Z$ by phases,
and we restrict to ideals $I$ which are invariant under that action.
These ideals are in 1-1 correspondence with 3-dimensional Young diagrams $D$ 
(or equivalently to configurations of a ``melting
crystal,'' as described in \cite{Okounkov:2003sp}.)   
The weight $e^{- \int k^3 / g_s^2}$
for such a geometry obtained by blowing up an ideal is simply $q^{\abs{D}}$, where $q = e^{- g_s}$ and $\abs{D}$ is
the number of boxes of the 3-dimensional Young diagram $D$, or equivalently the codimension of the corresponding ideal, or equivalently the relative number of sections of the line bundle ${\mathcal L}$.

Amazingly, the sum over all 3-dimensional Young diagrams with this simple weight 
gives the exact A model partition function on $\C^3$,
\begin{equation}
Z_A(\C^3) = \sum_{D} q^{\abs{D}} = \prod_{i=1}^n (1 - q^n)^{-n}.
\end{equation}
This is the special case $C_{\cdot \cdot \cdot}$
of the topological vertex where the representations $R_1, R_2, R_3$ on the legs are
trivial.  More generally, one could consider infinite 3-d Young diagrams, which asymptote to fixed 2-d diagrams $R_1, R_2, R_3$ along the $x, y, z$ directions; in this case the sum over diagrams gives the
full topological vertex $C_{R_1 R_2 R_3}$!

\section{Physical applications} \label{sec-physical-applications}

So far we have mostly discussed the topological string in its own right.
Now we turn to its physical applications.  At first it might be a surprise that there are
any physical applications at all.  Remarkably, they do exist, and they are quite spectacular! 
 
How is such a link possible?
The topological string can be considered as a \ti{localized} version of the physical string, i.e. it 
receives contributions only from special path-integral configurations, which can be identified with 
special configurations of the physical string.  At the same time, there
are some ``BPS'' observables of the physical string for which the physical string computation localizes
on these same special configurations.  In these cases the computations in the topological string
and the physical string simply become isomorphic!

The main examples which have
been explored so far are summarized in the table below:

\medskip

\begin{tabular}{|c|c|c|} \hline
physical theory & physical observable & topological theory \\ \hline \hline
$\N=2$, $d=5,4$ gauge theory & prepotential & A model \\ \hline
$\N=1$, $d=4$ gauge theory & superpotential & B model with branes/fluxes \\ \hline
spinning black holes in $d=5$ & BPS states & perturbative A model \\ \hline
charged black holes in $d=4$ & BPS states & nonperturbative A/B model? \\ \hline
\end{tabular}

\medskip

Now we will discuss these applications in turn.

\subsection{$\N=2$ gauge theories} \label{sec-geom-eng}

We begin with the application to $\N=2$ gauge theories.  First we describe the physical amplitudes
of $\N=2$ theories which are captured by the topological string; then we explain the particular
geometries which give rise to interesting gauge theories; and finally we show how to use mirror
symmetry to recover the Seiberg-Witten solution of $\N=2$ theories.

\subsubsection{What the topological string computes}

To understand the connection between the topological string
and $\N=2$ gauge theories in $d=4$, we begin by discussing the
physical theory obtained by compactifying the Type II (A or B) 
superstring on a Calabi-Yau $X$.
The holonomy of $X$ breaks $3/4$ of the supersymmetry, leaving $8$
supercharges which make up the $\N=2$ algebra in $d=4$;
the massless field content in $d=4$ can then be organized into multiplets of $\N=2$ 
supergravity as follows:
\medskip
\begin{center}
\begin{tabular}{|c|c|c|c|} \hline
            & vector       & hyper          & gravity \\ \hline \hline
 IIA on $X$ & $h^{1,1}(X)$ & $h^{2,1}(X)+1$ &    $1$  \\ \hline
 IIB on $X$ & $h^{2,1}(X)$ & $h^{1,1}(X)+1$ &    $1$  \\ \hline
\end{tabular}
\end{center}
\medskip
We will focus on the vector multiplets, for which the effective action is 
better understood.  Each vector multiplet contains a single complex scalar, 
and these scalars correspond to the \kahler moduli of $X$ in the Type IIA case,
or the complex moduli in the Type IIB case.

The topological string computes particular F-terms in the effective action 
which involve the vector multiplets \cite{Antoniadis:1994ze,Bershadsky:1994cx}.  
These terms can
be written conveniently in terms of the $\N=2$ Weyl multiplet, which is a chiral superfield
$\W_{\alpha \beta}$ with lowest component $F_{\alpha \beta}$.\footnote{Here the ``graviphoton'' $F$ is 
the field strength for the $U(1)$ vector in the supergravity multiplet, and $\alpha$, $\beta$
are spinor indices labeling the self-dual part of the full field strength $F_{\mu \nu}$, i.e. 
$F_{\mu \nu} = F_{\alpha \beta} (\gamma_\mu)^{\alpha \dot{\sigma}} (\gamma_\nu)_{\dot{\sigma}}^\beta + F_{\dot{\alpha} \dot{\beta}} (\gamma_\mu)^{\dot{\alpha}}_\sigma (\gamma_\nu)^{\dot{\beta} \sigma}$.}  Namely, forming the
combination
\begin{equation}
\W^2 = \W_{\alpha \beta} \W_{\alpha' \beta'} \eps^{\alpha \alpha'} \eps^{\beta \beta'},
\end{equation}
the terms in question can be written as
\begin{equation} \label{gravity-f-terms}
\int \de^4 x \int \de^4 \theta\, F_g(X^I) (\W^2)^g.
\end{equation}
Now we can state the crucial link between physical and topological strings:  the $F_g(X^I)$ which
appears in \eqref{gravity-f-terms} is precisely the genus $g$ topological string free energy, 
written as a function of the vector multiplets $X^I$ 
(so if we study Type IIB then the $F_g$ appearing is the B model free energy, since
the vector multiplets in that case parameterize the complex deformations, while
for Type IIA $F_g$ is the A model free energy.)

Note that each $F_g$ contributes to a different term in the effective action and hence
to a different physical process.  To see this more clearly we can expand \eqref{gravity-f-terms} in components;
one term which appears is (for $g>1$)
\begin{equation} \label{gravity-f-terms-expanded}
\int \de^4x\, F_g(X^I) (R_+^2 F_+^{2g-2}),
\end{equation}
so $F_g(X^I)$ is the coefficient of
a gravitational correction to the amplitude for scattering of $2g-2$ graviphotons.
In the application to $\N=2$ gauge theory we will mostly be interested in $F_0$, which gets identified
with the prepotential of the gauge theory, as one sees from \eqref{gravity-f-terms}.

\subsubsection{Compactifying on ALE fibrations}

Now let us focus on the specific geometries which will lead to interesting $\N=2$ gauge theories.  
In order
to decouple gravity we should consider a non-compact Calabi-Yau space.  The simplest example is an ALE
singularity $\C^2 / G$, as we discussed in Example \ref{example-ale}.
Recall from that example that one can think of the singularity 
of $\C^2 / G$ as containing a number of zero size $\C\PP^1$'s, which naturally correspond
to the simple roots of a Lie algebra $\gl$.  Then considering Type IIA string 
theory on $\C^2 / G$, one obtains massless states from
D2-branes which wrap around these zero size $\C\PP^1$'s.  These massless states get identified
with gauge bosons in six dimensions, and it turns out that one gets a gauge theory
with gauge symmetry $\gl$ (note
in particular that the number of these gauge bosons agrees with the rank of $\gl$ as expected.)

But $\C^2 / G$ is not quite the example we want; we want to get down to $d=4$ rather than $d=6$, and we
also want to get down to $8$ supercharges rather than $16$.  These goals can be simultaneously accomplished by
fibering $\C^2 / G$ over a genus $g$ Riemann surface $\Sigma_g$; this can be done in a way so that the
resulting six-dimensional space is a Calabi-Yau threefold $X$.  Compactifying the Type IIA string on $X$ gives
an $\N=2$ theory with gauge group determined by $G$ and with $g$ adjoint hypermultiplets \cite{Katz:1997xe}.  
(The origin of these hypermultiplets can be roughly 
understood by starting with the gauge theory in $d=6$ and compactifying 
it on $\Sigma_g$; the electric and magnetic Wilson lines of the gauge theory give rise to the $4g$ 
scalar components of the $g$ hypermultiplets.)  

We first consider the special case $g=1$.

\begin{example}[$\C^2 / G \times T^2$]
In this case the fibration
of $\C^2 / G$ over the Riemann surface $T^2$ is trivial, so the $\N=2$ supersymmetry should be enhanced to $\N=4$;
this agrees with the fact that we get a single adjoint hypermultiplet, which is the required matter content for
the $\N=4$ theory.  Furthermore, there is a relation
\begin{equation}
\vol(T^2) = 1 / g^2_{YM}.
\end{equation}
T-dualizing on the two circles of $T^2$ then implies that the theory with coupling $g_{YM}$ is equivalent
to the theory with coupling $1 / g_{YM}$ --- so the existence of a string theory realization already implies 
the highly nontrivial Montonen-Olive duality of $\N=4$ super Yang-Mills!
\end{example}

One could also consider the case $g>1$, but in this case the gauge theory is not asymptotically free.
We therefore focus on $g=0$, and for simplicity we consider the case 
$G = \Z_2$.

\begin{example}[$\C^2 / \Z_2$ fibered over $\C\PP^1$]
This geometry turns out to be just the local $\C\PP^1 \times \C\PP^1$ geometry we discussed in Example \ref{example-local-cp1-cp1}; one of the $\C\PP^1$'s is the base of the fibration, while the other is sitting 
in the fiber (obtained by resolving the singularity $\C^2 / \Z_2$.)  We call their sizes $t_b$ and $t_f$ respectively.
Type II string theory on this geometry gives pure $\N=2$ Yang-Mills in four dimensions, with gauge group $SU(2)$.

To ``solve'' this gauge theory \'{a} la Seiberg and Witten \cite{Seiberg:1994rs}, one wants to compute its prepotential $F_0$, as a function of the Coulomb branch modulus.  This modulus determines the mass of the $W$ bosons, so in our geometric 
setup it gets identified with the
\kahler parameter $t_f$ (recall that the $W$ bosons are obtained by wrapping branes over the fiber $\C\PP^1$.)  
The other \kahler parameter $t_b$ is identified with the Yang-Mills coupling, through the relation
\begin{equation} \label{geom-eng-gauge-couplings}
t_b = 1 / g^2_{YM}.
\end{equation}

Now, as we remarked above, the prepotential $F_0$ of the gauge theory
should coincide with the $F_0$ computed by the genus zero A model topological string.
We can obtain the exact solution for $F_0$ using mirror symmetry; 
namely, recalling that we have a toric realization for this geometry as discussed in Example
\ref{example-local-cp1-cp1-toric}, the techniques we illustrated in Section \ref{sec-mirror-symmetry}
can be straightforwardly applied.  The mirror geometry is of the form
$F(x,z) = uv$, where the Riemann surface $F(x,z) = 0$
turns out to be precisely the Seiberg-Witten curve encoding the solution of the model \cite{Katz:1997fh}!
From this Seiberg-Witten curve one can read off all the desired information.

One frequently describes the Seiberg-Witten solution as counting gauge theory instantons in four dimensions,
whereas in Section \ref{sec-genus-zero} we described the A model $F_0$ 
as counting genus zero worldsheet instantons in $X$.  
The connection between these two languages is clear:  indeed, from \eqref{geom-eng-gauge-couplings}
one sees that the worldsheet instantons which wrap $n$ times around the base $\C\PP^1$ 
contribute with a factor $e^{-n / g^2_{YM}}$ to $F_0$, and hence they 
correspond precisely to $n$-instanton effects in four dimensions.
\end{example}

One can similarly obtain any ADE gauge group just by making an appropriate choice of 
the finite group $G$.
Conversely, anytime we have a toric geometry where the \kahler parameters arise by resolving some
singularity, we expect that that toric geometry can be interpreted in terms of gauge theory.  The
zoo of $\N=2$ theories one can ``geometrically engineer'' in this way includes cases with arbitrarily complicated product gauge groups and bifundamental matter content, as well as some
exotic conformal fixed points in higher dimensions; see e.g.
\cite{Katz:1997xe,Katz:1997fh,Kachru:1996fv,Katz:1998eq,Morrison:1997xf,Shapere:1999xr}.
To obtain the prepotentials for the geometrically-engineered theories 
is in principle straightforward via mirror symmetry,
and it has been worked out in many cases, but it is not always easy --- e.g. for the $E_k$ singularities
one would have a more difficult job, because to realize these geometries torically one has to include a superpotential, which makes the mirror procedure and computation of the mirror periods less straightforward.

Finally we should mention an important subtlety which we have so far glossed over:  
at generic values of $g^2_{YM}$ and the fiber moduli $t_i$, the string theory
actually contains more information than just the four-dimensional gauge theory.  This is to be expected
since the $F_0$ of the gauge theory depends just on the Coulomb branch 
moduli $t_i$, while the $F_0$ of the A model has one more parameter:  it 
also depends on the size of the base, which we identified with $g^2_{YM}$ at the string 
scale.  To isolate the four-dimensional theory we have to take a decoupling limit in 
which $g^2_{YM}$ and $t_i$ approach zero, which sends the string scale to infinity while keeping the
masses of the $W$ bosons on the Coulomb branch fixed \cite{Kachru:1996fv}.  If we do not take this
decoupling limit, we get a theory which includes information about compactification 
on $S^1$ from five to four dimensions;
from that point of view the four-dimensional 
instantons can be interpreted as particles of the five-dimensional theory which
are running in loops, as was explained in \cite{Lawrence:1997jr}.

\subsection{$\N=1$ gauge theories}

So far we have seen that the IR dynamics in a large class of $\N=2$ gauge theories can be completely solved
using mirror symmetry.  Now we want to move on to the $\N=1$ case, where we will see that the topological
string is similarly powerful.

How can we geometrically engineer an $\N=1$ theory?
Starting with compactification of Type II string theory
on a Calabi-Yau space, we need to introduce an extra ingredient
which reduces the supersymmetry by half.  There are two natural possibilities:  we can add either
D-branes or fluxes.  In both cases we want to preserve the four-dimensional Poincar\`{e} invariance; so
if we use D-branes we have to choose them to fill the four uncompactified dimensions, and if we 
use fluxes we have to choose them entirely in the Calabi-Yau directions (i.e. the $0$, $1$, $2$, $3$  components of the flux should vanish.)  In fact, the two possibilities are sometimes equivalent
via a geometric transition in which branes are replaced by flux,
as we discussed in Section \ref{sec-geometric-transition}.  

In the next two subsections we describe the superpotentials which arise from these two ways of breaking from $\N=2$ to $\N=1$; these superpotentials can be computed by the topological string, and they 
are the basic objects we want to understand in the $\N=1$ context, since they
determine much of the IR physics.
The form of the superpotentials obtained in the two cases is quite similar, and as we explain in the
following section, this is not an accident; it follows from the equality of topological
string partition functions before and after the geometric transition.  This geometric transition
is a practical tool for computation of the superpotentials, and we discuss some basic examples.  Finally
we discuss an alternative method of computing the superpotentials via holomorphic matrix models, which also gives
an interesting new perspective on the geometric transition:  the dual geometry emerges as a kind of effective
theory of a density of eigenvalues in the large $N$ limit!

\subsubsection{Breaking to $\N=1$ with branes}

To engineer $\N=1$ gauge theories, we begin with Type II string theory on a Calabi-Yau space $X$.  This would
give $\N=2$ supersymmetry, but let us reduce it to $\N=1$ by introducing $N$ D-branes, which are wrapped on
some cycle in the Calabi-Yau and also fill the four dimensions of spacetime.  Then we obtain an $\N=1$
theory in four dimensions, with $U(N)$ gauge symmetry, as we discussed in 
Section \ref{sec-branes}.  (Note the difference from the geometric
engineering we did in the $\N=2$ case;
there we obtained the gauge symmetry from a geometric singularity, but in the $\N=1$ case it just
comes from the $N$ branes.  As we will see, in this case the geometry 
is responsible for details of the gauge theory, specifically the form of the bare superpotential.)

We now want to expose a connection between this gauge theory and the topological string on $X$.
In the $\N=2$ case we saw that the genus zero topological string free energy $F_0$ computed the prepotential.  
After introducing D-branes in the topological string, we need
not consider only closed worldsheets anymore; we can also consider open strings, i.e. Riemann surfaces 
with boundaries.  Therefore we can
define a free energy $F_{g,h}$, obtained by integrating over worldsheets with genus $g$
and $h$ holes, with each hole mapped to one of the D-branes; and we can ask whether this $F_{g,h}$
computes something relevant for the $\N=1$ theory.
The answer is of course ``yes.''  (More precisely, as in the $\N=2$ case, it turns out that $g=0$ is the case
relevant to the pure gauge theory; higher genera are related to gravitational corrections, which
we will not discuss here.)

To write the terms which the topological string computes in the $\N=1$ theory with branes, we need
the ``glueball'' superfield $S$; this is a chiral superfield 
with lowest component $\Tr \psi_\alpha \psi^\alpha$, where $\psi_\alpha$ is the gluino.
Organize the $F_{0,h}$ into a generating function:
\begin{equation} \label{open-free-energy}
F(S) = \sum_{h=0}^\infty F_{0,h} S^h.
\end{equation}
The F-term the genus zero topological string computes in the $\N=1$ theory can then be written \cite{Bershadsky:1994cx}
\begin{equation} \label{n1-fterm}
\int \de^4 x \int \de^2 \theta\, N \frac{\partial F}{\partial S}.
\end{equation}
This term gives a superpotential for the glueball $S$, and it turns out that this superpotential
captures a lot of the infrared dynamics of the gauge theory.  More precisely, in addition
to \eqref{n1-fterm}, one also has to include the term
\begin{equation}
\int \de^4 x \int \de^2 \theta\, \tau S,
\end{equation}
which is simply the classical super Yang-Mills action in superfield notation, with
\begin{equation}
\tau = \frac{4 \pi \I}{g^2_{YM}} + \frac{\theta}{2 \pi}.
\end{equation}
After including this extra term, one then has the glueball superpotential
\begin{equation} \label{glueball-superpotential}
W(S) = N \frac{\partial F}{\partial S} + \tau S.
\end{equation}
In the IR one expects that the glueball field will condense to some value
with $W'(S) = 0$, so one
can determine the vacuum structure of the theory just by extremizing this $W(S)$,
as we will see below in some examples.

\subsubsection{Breaking to $\N=1$ with fluxes}

Now what about the case where we introduce fluxes instead of branes?  Consider the Type IIB
superstring on a Calabi-Yau $X$.  Recall from the last section that this theory has a prepotential term
\begin{equation} \label{sugra-prepotential}
\int \de^4 x \int \de^4 \theta\, F_0(X^I),
\end{equation}
where $F_0$ is the B model topological string free energy at genus zero, and the $X^I$ are the vector superfields, whose lowest components parameterize the complex structure moduli of $X$.  How 
does this term change if we introduce $N^I$ units of Ramond-Ramond three-form flux on the $I$-th A cycle?\footnote{Recall
that in writing the $\N=2$ supergravity Lagrangian we have chosen a splitting of $H_3(X)$ 
into A and B cycles, with the $X^I$ representing the A cycle periods.}
In the $\N=2$ supergravity language, it turns out that this flux corresponds to the $\theta^2$
component of the superfield $X^I$; turning on a vacuum expectation value for this component
absorbs two $\theta$ integrals from \eqref{sugra-prepotential}, leaving behind an F-term 
in the $\N=1$ language \cite{Vafa:2000wi},
\begin{equation} \label{moduli-superpotential}
\int \de^4 x \int \de^2 \theta\, N^I \frac{\partial F_0}{\partial X^I}.
\end{equation}
As above, this term can be interpreted as a superpotential, this time for the moduli $X^I$.
There is a natural extension to include a flux $\tau_I$ on the $I$-th B cycle:
\begin{equation} \label{moduli-superpotential-extended}
W(X^I) = N^I \frac{\partial F_0}{\partial X^I} + \tau_I X^I.
\end{equation}
This form of the superpotential was derived in \cite{Gukov:1999ya,Taylor:1999ii}.

\subsubsection{The geometric transition, redux}

There is an obvious analogy between \eqref{glueball-superpotential} and \eqref{moduli-superpotential-extended}.
Note though that the lowest component of the $X^I$ which appears in \eqref{moduli-superpotential-extended} is
a scalar field parameterizing a complex structure modulus, while the $S$ which appears
in \eqref{glueball-superpotential} is a fermion bilinear, which naively cannot have a classical
vacuum expectation value.  Nevertheless, the analogy between the two sides seems to be suggesting that
we should treat $S$ also as an honest scalar, and we will do so in what follows.

So what do \eqref{glueball-superpotential} and \eqref{moduli-superpotential-extended} have to do with one 
another?  The crucial link is provided by the notion of ``geometric transition,'' which we discussed
in Section \ref{sec-geometric-transition}, 
but now in the context of the Type IIB superstring rather than the topological string:\footnote{
See \cite{Vafa:2000wi} for a detailed discussion of the superstring version of the large $N$ duality
in the Type IIA case.}
start with a Calabi-Yau $X$ which has a nontrivial 2-cycle.  Then wrap $N$ D5-branes on this 2-cycle,
obtaining a $U(N)$ gauge theory.  There is a dual geometry where the D5-branes disappear and are replaced by a 3-cycle $A$; in this dual geometry there are $N$ units of Ramond-Ramond flux on the dual cycle $B$.  
The claim is that the physical string theories on these two geometries
are equivalent in the IR, after we identify the glueball superfield $S$ with the period of $\Omega$ over
the A cycle in the dual geometry.\footnote{On the face of it this claim might sound
bizarre since the theory with branes should have $U(N)$ 
gauge symmetry in four dimensions; but since we are now talking about the effective theory in $d=4$, 
what we should really compare is the IR dynamics, and we know
that $\N=1$ gauge theories confine, which reduces the $U(N)$ to $U(1)$ in the IR.}  With this
identification \eqref{glueball-superpotential} and \eqref{moduli-superpotential-extended} are identical.

One can therefore use either the brane picture or the flux picture to compute the glueball superpotential.
In this section we will discuss some examples of the use of the flux picture.

\begin{example}[D5-branes on the resolved conifold]
The simplest example of a geometric transition from branes to flux is provided by the resolved conifold, 
which just has a single 2-cycle $\C\PP^1$.  So suppose we wrap $M$ D5-branes on 
the $\C\PP^1$ of the resolved conifold.  As one might expect, this simplest possible geometry leads 
to the simplest possible gauge theory in $d=4$, namely $\N=1$ super Yang-Mills.
This theory has a well-known glueball superpotential, which we now derive
from the flux picture and \eqref{moduli-superpotential}.  
The dual geometry after the transition is the deformed conifold, 
which has a compact $S^3$ and its dual B cycle, with corresponding periods
\begin{align}
X = \int_A \Omega &= \mu, \\
F = \int_B \Omega &= \mu \log \mu.
\end{align}
(A simple way to check the formula for $F$ is to note 
that it has the correct monodromy;
as $\mu \to e^{2 \pi i} \mu$ the B cycle gets transformed into a linear combination of the 
B cycle and the A cycle, corresponding to the fact that $F$ gets shifted by the A period $\mu$.)
From the periods we immediately obtain the closed string $F_0$, via \eqref{f0-b-model},
% oops, subtleties in non-compact case.... you really get it by integrating up to a cutoff
\begin{equation}
F_0 = \half XF = \half \mu^2 \log \mu.
\end{equation}

Now to compare with the gauge theory we have to identify $\mu = S$ as we stated above.  This
leads to the superpotential
\begin{equation}
W(S) = N \frac{\partial F_0}{\partial S} - 2 \pi \I \tau S = NS \log S - 2 \pi \I \tau S.
\end{equation}
This is the standard Veneziano-Yankielowicz glueball superpotential for $\N=1$ super Yang-Mills
\cite{Veneziano:1982ah}.  

By extremizing $W(S)$ one finds the expected 
$N$ vacua of $\N=1$ super Yang-Mills,\footnote{We have not
been careful to keep track of the cutoff $\Lambda_0$; if one does keep track of it,
one finds that it combines with the bare coupling $\tau$ to give the QCD scale
$\Lambda$ which appears in \eqref{sym-vacua}.}
\begin{equation} \label{sym-vacua}
S = \Lambda_0^3 \exp(2 \pi \I j \tau / N) = \Lambda^3 \exp \left( 2 \pi \I j / N \right),
\end{equation}
where $j = 1, \dots, N$.
\end{example}

So far we have not used much of our topological-string machinery.  But now we can consider
a more elaborate example.

\begin{example}[D5-branes on the multi-conifold]  \label{example-multi-conifold}
Instead of the singular conifold geometry
\begin{equation}
u^2 + v^2 + y^2 + x^2 = 0,
\end{equation}
which just has a single zero size $\C\PP^1$, consider
\begin{equation} \label{multi-conifold}
u^2 + v^2 + y^2 + W'(x)^2 = 0,
\end{equation}
for some polynomial $W(x)$ of degree $n+1$.  Writing
\begin{equation}
W'(x) = \prod_{i=1}^n (x - x_n),
\end{equation}
the geometry has $n$ conifold singularities located at
the critical points $x_1, \dots, x_n$ of $W$.  
The singularities can be resolved by blowing up to obtain $n$
$\C\PP^1$'s at these $n$ points (all these $\C\PP^1$'s are homologous, however, so in particular
there is only one \kahler modulus describing the resolution.)  

We want to use this geometry to engineer an interesting $\N=1$ gauge theory.
To construct this gauge theory we consider $M$ D5-branes.  What are the possible supersymmetric
configurations?
We should expect that we can get a supersymmetric configuration by
wrapping $M_1$ branes on the first $\C\PP^1$, $M_2$ on the second, and so on, and in 
this configuration we expect to realize a gauge symmetry
$U(M_1) \times \cdots \times U(M_n)$.
All these configurations can be naturally understood as
different sectors of a single UV theory, which describes the dynamics 
of the $M$ branes and includes a $U(M)$ adjoint chiral multiplet $\Phi$, 
whose lowest component represents the $x$-coordinate of the branes.\footnote{The 
adjoint scalar $\Phi$ is present even in the conifold case which we
considered above, but there (as we will shortly see) it is accompanied by a quadratic 
superpotential $W(\Phi) = \Phi^2$, so $\Phi$ can be harmlessly integrated out to leave pure $\N=1$
super Yang-Mills.}
The supersymmetric vacua described above then 
arise from configurations in which $M_1$ of the eigenvalues of $\Phi$ are
equal to $x_1$, $M_2$ are equal to $x_2$ and so on.

A very natural way for this vacuum structure to arise is if the $U(M)$ gauge theory describing the branes 
has a bare superpotential $\Tr W(\Phi)$.  
This is indeed the case; one can derive this result from the holomorphic Chern-Simons action which,
as we discussed earlier, is the topological open string field theory of the brane \cite{Aganagic:2000gs}.
Namely, one shows from the holomorphic Chern-Simons action 
that, as one moves the 2-brane along a path, sweeping out a 3-cycle $C$, the classical action is shifted by $\int_C \Omega$; combined with the explicit form of $\Omega$ in the geometry
\eqref{multi-conifold} this gives the classical action for the brane at $x$ as $W(x)$.  This
classical action in the topological string turns out to be the superpotential of the physical superstring.  This superpotential computation 
can also be interpreted directly in the worldsheet language as coming from disc diagrams with
boundary on the brane; to see this from the topological string one notes that $F_{0,1}$ contributes
an S-independent term to \eqref{glueball-superpotential}, which gets interpreted
as the desired bare superpotential.

Thus we have geometrically engineered an 
$\N=1$ gauge theory, with $U(M_1) \times \cdots \times U(M_n)$
gauge group, one adjoint chiral multiplet $\Phi$, and a superpotential $\Tr W(\Phi)$.  To answer questions
about the vacuum structure
of this theory we now want to find the appropriate glueball superpotential, which is now a function of
$n$ different glueball fields $S_i$ for the $n$ gauge factors.
As in the case of the conifold, one way to compute the superpotential
is to consider the dual geometry in which the
branes have disappeared and each of the $n$ $\C\PP^1$'s has been replaced by an $S^3$.  This
geometry is written
\begin{equation}
u^2 + v^2 + y^2 + W'(x)^2 = f(x),
\end{equation}
where $f(x)$ is a polynomial of degree $n-1$ characterizing the deformation. This $f(x)$ depends
on the $M_i$, and is completely fixed 
by the requirement that the period of $\Omega$ over the $i$-th $S^3$ is $M_i g_s$ (in keeping
with the principle that the B model branes produce precisely this flux of $\Omega$ --- this is 
precisely analogous to the fact that A model branes produce a flux of $k$, which we
used in Section \ref{sec-geometric-transition} to compute A model amplitudes.)  This
approach was followed in \cite{Cachazo:2001jy}, and leads to a complete computation
of the glueball superpotential.
\end{example}

\subsubsection{Holomorphic matrix models}

So far we have shown how to compute the glueball superpotential from a transition
to a geometry where D5-branes are replaced by fluxes.
Alternatively, one can avoid the geometric transitions altogether and compute directly
in the gauge theory on the D5-branes.  The idea is that since the glueball superpotential is computed by
the topological string, one can avoid all the complexities of Yang-Mills theory, and use
instead the topological open string field theory; as we explained above, in the case of the 
B model this is (the dimensional reduction of) holomorphic Chern-Simons.  One 
finds that the whole computation of the topological string free energy is reduced to 
a computation in a holomorphic matrix model \cite{Dijkgraaf:2002fc,Dijkgraaf:2002vw,Dijkgraaf:2002dh}.
For example, in the case of the multi-conifold geometry of Example \ref{example-multi-conifold},
one just has to integrate over a single $N \times N$ matrix $\Phi$, with action $W(\Phi) / g_s$:
\begin{equation}
Z = \int \de^{N^2} \Phi\, e^{- W(\Phi) / g_s}.
\end{equation}

The matrix model contains all the information that can be obtained from the open topological string in
this background.  For example, to compute the genus zero open topological string partition function --- 
which determines the glueball superpotential --- one just has to study the large $N$ (planar) limit
of the matrix model!

These models have turned out to be a quite powerful tool, which is applicable to geometries
more general than the case we described here.  They are also related in a beautiful way to
the geometric transitions we described above:  namely, the planar limit of the matrix model
can be described as a saddle-point expansion around a particular distribution of the infinitely
many eigenvalues,
and this distribution turns out to capture the dual geometry in a precise way.  In this sense
the smooth geometry seems to be an emergent property, which only makes sense in the planar (classical)
limit of the string theory.

\subsection{BPS black holes in $d=5$} \label{sec-black-hole-5d}

So far we have discussed applications of the topological string to gauge theory, which
involved only the genus zero 
free energy $F_0$.  Now we want to discuss an application to black hole entropy, which is
more sophisticated in the sense that it naturally involves all of the $F_g$.
We ask the following question:  given a compactification of M theory to five dimensions
on a Calabi-Yau threefold $X$, how many BPS states are there with a particular
spin and charge?

First, what do we mean by ``charge''?  M-theory compactified on $X$ has a $U(1)$ gauge field for
each 2-cycle of $X$, obtained by dimensional reduction of the M-theory 3-form $C$ on the 2-cycle,
i.e. via the ansatz $C_{\mu \alpha \beta} = A_\mu \omega_{\alpha \beta}$, where $\omega_{\alpha \beta}$
is the harmonic 2-form dual to the cycle in question.  So we get $U(1)^n$ gauge symmetry, where
$n = b_2(X)$ is the number of independent 2-cycles.  We also naturally get states which are charged
under this $U(1)^n$; namely, an M2-brane wrapped on a 2-cycle gives a particle charged under
the corresponding $U(1)$.  Hence the charges in the theory are classified by the
second homology of $X$, $Q \in H_2(X,\Z)$.

So we could ask for the number of BPS states with given $Q$.  But actually there
is a finer question we can ask:  namely, it turns out that in five dimensions it is possible for
a state to have spin and still be BPS.  The little group for a massive particle in this dimension
is $SO(4) = SU(2)_L \times SU(2)_R$, giving rise to spins $(j_L,j_R)$, and one can get BPS states so
long as one requires either $j_L = 0$ or $j_R = 0$.  So fixing, say, $j_R = 0$, we can ask for the
number of BPS states with charge $Q$ and spin $j_L$.

A convenient way of packaging this information is suggested by the notion of \ti{elliptic genus},
which we now quickly recall in a related context.

\begin{example}[The $\N=(1,1)$ elliptic genus]

Consider a theory with $\N=(1,1)$ supersymmetry in two dimensions.
The partition function on a torus with modular parameter $\tau$, with
the natural boundary conditions, is
\begin{equation}
\Tr (-1)^F q^{L_0} \overline{q}^{\overline{L_0}}.
\end{equation}
This partition function is relatively ``boring'' in the sense that it just computes the Witten index \cite{Witten:1982df}, which is independent of $q$ and $\overline{q}$.  But in an $\N=(1,1)$ theory
one can define separate left and right-moving fermion number
operators $F_L$, $F_R$, and we can use these to define a more interesting object, the
\ti{elliptic genus} \cite{Witten:1994jg},
\begin{equation} \label{elliptic-genus-2d}
\Tr (-1)^{F_R} q^{L_0} \overline{q}^{\overline{L_0}}.
\end{equation}
The usual argument shows that \eqref{elliptic-genus-2d}
gets contributions only from states which have $\overline{L_0} = 0$,
so it is independent of $\overline{q}$, but it is a nontrivial function of $q$, which has modular properties.
Like the usual Witten index it has some rigidity properties, namely, it does not depend on small deformations of the theory; this follows from the fact that the coefficients in the $q$
expansion are integral.

\end{example} 

Now we turn to the case of interest for us.

\begin{example}[The $d=5$, $\N=2$ elliptic genus]
Returning to the $d=5$ BPS state counting, note that we have a splitting into left and right similar to
the one for $\N=(1,1)$ theories, so instead of computing the ordinary index
\begin{equation}
\Tr (-1)^{J} e^{- \beta H} 
\end{equation}
we can consider an elliptic genus analogous to \eqref{elliptic-genus-2d},
\begin{equation} \label{elliptic-genus-5d}
\Tr (-1)^{J_R} q^{J_L} e^{- \beta H}.
\end{equation}
Like \eqref{elliptic-genus-2d}, this elliptic genus has a rigidity property:  it is independent 
of the complex structure moduli of $X$, although it can and does depend continuously on the \kahler
moduli $t_i$.  This property is reminiscent of the A model topological string, and indeed it turns 
out that the A model partition function $Z_A(g_s, t_i)$ is precisely the elliptic genus \eqref{elliptic-genus-5d}, with the identification
\begin{equation} \label{spin-vs-genus}
q = e^{- g_s},
\end{equation}
as we will see below.\footnote{Strictly speaking, this is true once we rescale $t_i$ by a factor $\beta$,
which completely absorbs the $\beta$ dependence.}
In this sense the spin-dependence of the BPS state counting gets related to the genus-dependence
of the topological string.
\end{example}

Now, \ti{why} is the A model partition function counting BPS states?  Such a connection
seems reasonable; after all, the A model counts holomorphic maps, and the
image of a holomorphic map is a supersymmetric cycle on which a brane could be wrapped to give a BPS state. 
There is a more precise argument which explains the relation; it was worked out in \cite{Gopakumar:1998ii,Gopakumar:1998jq} and goes roughly as follows.

Consider the Type IIA string on $X$.  As we mentioned earlier,
there are certain F-terms in the effective four-dimensional action of this theory which are computed 
by the A model topological string, namely
\begin{equation} \label{gravity-f-terms-reprise-2}
\int \de^4 x \int \de^4 \theta\, F_g(t) (\W^2)^g + \mathrm{c.c.},
\end{equation}
which when expanded in components give
\begin{equation} \label{gravity-f-terms-expanded-reprise}
\int \de^4 x\, F_g(t) (R_+^2 F_+^{2g-2} + R_-^2 F_-^{2g-2}).
\end{equation}
If we consider the Euclidean version of the theory, then in four dimensions we can turn on a self-dual
graviphoton background $F_+ \neq 0$, $F_- = 0$, 
i.e. $\W \neq 0$, $\overline{\W} = 0$.  Substituting this background into \eqref{gravity-f-terms-expanded-reprise}
we get a correction to the $R_+^2$ term,
\begin{equation} \label{curvature-correction}
\left( \sum_{g=0}^\infty F_g(t) F_+^{2g-2} \right) R_+^2.
\end{equation}
Note that in \eqref{curvature-correction} we have a sum over all genus topological string amplitudes,
with the role of the topological string coupling played by the graviphoton field strength $F_+$.

To establish the relation between the topological string and the elliptic genus, we now want to show
that one can compute the same $R_+^2$ correction in a graviphoton background
in a different way which gives the elliptic genus.
This second computation is based on Schwinger's computation of the correction to the vacuum energy
from pair-production of charged particles in a background electric field.  In the present context
the relevant charged particles are the quanta of charged $\N=2$ hypermultiplet fields obtained by 
quantization of the wrapped D2- and D0-branes; 
for a D2-brane wrapped on 
the cycle $Q$, bound to $k$ D0-branes, the central charge is
\begin{equation}
Z = \IP{Q,t} + ik,
\end{equation}
and the mass of the corresponding BPS state is $m = \abs{Z}$.  
We compute the corrections to the effective action
due to pair production of such states in the self-dual graviphoton background $F$.  Since these states
come in hypermultiplets, their contribution to the vacuum energy cancels,
but it turns out that they make a nonzero contribution to the $R_+^2$ term:  for example, a multiplet whose
lowest component is scalar contributes to $R_+^2$ precisely as a scalar would have contributed to the vacuum energy.

Let us focus on the contribution to the $R_+^2$ correction
from a BPS hypermultiplet with lowest component scalar, arising from a quantization of a 
D2-brane in homology class $Q$.  Actually, since the D2-brane can 
be bound to D0-branes, these hypermultiplets will come in families: 
we will get one for each value of the D0 brane charge $k$.
The Schwinger computation expresses the contribution from each of these hypermultiplets as 
a one-loop determinant; summing over $k$ to treat the whole family at once gives
\begin{equation} \label{curvature-correction-scalar}
\sum_{k} \log \det (\Delta + m_k^2) = \sum_{k = -\infty}^\infty \int_\eps^\infty \frac{\de s}{s} \frac{e^{-s (\IP{Q,t} + ik)}}{(2 \sinh \frac{s F_+}{2})^2}.
\end{equation}
(Here $F_+$ enters the determinant through the non-commutation of the covariant derivatives which
appear in $\Delta$.)  The integral appearing in \eqref{curvature-correction-scalar} looks formidable, but luckily
we do not have to do it:  the sum over $k$ gives a $\delta$-function which cancels the
integral and also removes the awkward dependence on the cutoff $\eps$.  We get a simple result,
\begin{equation} \label{r2-contribution-scalar-multiplet}
\sum_{n=1}^\infty \frac{1}{n} \frac{e^{-n \IP{Q,t}}}{(2 \sinh \frac{n F_+}{2})^2}.
\end{equation}
This is the contribution to the $R_+^2$ correction coming from a single family of BPS multiplets
with lowest component scalar; 
alternatively, setting the topological string coupling 
$g_s = F_+$, we could interpret it as the contribution to the topological string free energy ${\mathcal F}(g_s, t_i)$ from this family,
\begin{equation} \label{top-free-energy-contribution-scalar-multiplet}
\sum_{n=1}^\infty \frac{1}{n} \frac{e^{-n \IP{Q,t}}}{(2 \sinh \frac{n g_s}{2})^2}.
\end{equation}

Now what does all this have to do with the elliptic genus \eqref{elliptic-genus-5d} 
in M-theory on $X$?
We will argue that the exponential of \eqref{top-free-energy-contribution-scalar-multiplet} 
in fact agrees with the contribution to the elliptic genus from a single BPS
hypermultiplet in five dimensions, obtained from quantization of an M2 brane
in homology class $Q$.
The first promising sign is that the exponential of \eqref{top-free-energy-contribution-scalar-multiplet} 
has a nice integer expansion:  namely, it is
\begin{equation} \label{one-m2-brane-contribution}
\prod_{n=1}^\infty (1 - q^n e^{-\IP{Q,t}})^n.
\end{equation}
To reproduce this, write the hypermultiplet field as $\phi$ in five dimensions.  This
$\phi$ can have excitations which are not Poincar\`{e} invariant but are still BPS.  
Namely, choosing complex coordinates $z_1, z_2$ for the Euclidean time-slice 
$\R^4$, we can write
\begin{equation}
\phi = \phi(z_1, z_2, \overline{z_1}, \overline{z_2}),
\end{equation}
and the BPS excitations are the ones independent of $\overline{z_i}$.
Expanding
\begin{equation}
\phi = \sum_{l,m \ge 0} \phi_{lm} z_1^l z_2^m,
\end{equation}
we get a collection of creation operators $\phi_{lm}$.  The operator
$\phi_{lm}$ creates $SU(2)_L$ spin $l+m+1$, so there are $n$ of them that create spin $n$
(and BPS mass $\IP{Q,t}$.)
The second quantization of these operators then accounts for the factor
\eqref{one-m2-brane-contribution}.

This almost completes the identification between the topological string partition function
and the elliptic genus, except that the hypermultiplets obtained from
quantization of the wrapped M2-brane need not in general have 
lowest component scalar.
From the discussion of the last paragraph, one easily sees how to modify
the contribution to the elliptic genus if the lowest component has $SU(2)_L$ spin $j$:
one just has to replace $q^n$ by $q^{n+j}$ in \eqref{one-m2-brane-contribution}.
We should also note that the creation operators $\phi_{lm}$ may be fermionic or bosonic
depending on the net spin.  Hence the most general form of the contribution
to the elliptic genus is
\begin{equation} \label{one-m2-brane-contribution-general-spin}
\left[\prod_{n=1}^\infty (1 - q^{n+j} e^{-\IP{Q,t}})^n\right]^{\pm 1}.
\end{equation}
One can check that this also agrees with the result of the Schwinger computation of
the $R_+^2$ correction in this more general case, and hence with the topological string.

What have we learned about the topological string?
We can already obtain an interesting result by taking the $g_s \to 0$ limit in
the contribution \eqref{top-free-energy-contribution-scalar-multiplet} 
to ${\mathcal F}$ from a single five-dimensional hypermultiplet:  namely, we recover
\begin{equation}
\frac{1}{g_s^2} \sum_{n=1}^\infty \frac{e^{-n \IP{Q,t}}}{n^3},
\end{equation}
which is precisely the formula \eqref{f0-a-model} for the contribution of an isolated
genus zero curve to the A model $F_0$!  So the counting of BPS states automatically reproduces the
tricky $\sum_n 1/n^3$, which arose from multi-covering maps $S^2 \to S^2$ in the A model.  

Indeed, from counting BPS states one obtains formulae for the multi-covering
contributions at all genera, as well as ``bubbling'' terms which occur when
part of the worldsheet degenerates to a surface of lower genus.
All these terms are encapsulated in the general form of the topological A model
free energy in terms of the five-dimensional BPS content, which we now write.
It is convenient to choose a 
slightly exotic basis for the representation content:  namely, we introduce the
symbol $[j]$ for the $SU(2)_L$ representation $[2(0) \oplus (\half)]^{\otimes j}$.
Any representation of $SU(2)_L$ can be written as a sum of the representations $[j]$
with integer coefficients (not necessarily positive).
Then write $\N_{j,Q}$ for the number of times $[j]$ appears in the $SU(2)_L$ content
of the BPS spectrum obtained by wrapping M2 branes on $Q$.  Combining the results we 
catalogued above, one obtains
\begin{equation} \label{gopakumar-vafa-formula}
F(t,g_s) = \sum_{j \ge 0} \sum_{Q \in H_2(X,\Z)} \N_{j,Q} \left( \sum_{n \ge 0} \left(2 \sinh \frac{n g_s}{2}\right)^{2j-2} e^{-n \IP{Q,t}} \right).
\end{equation}
The formula \eqref{gopakumar-vafa-formula} expresses all the complexity of the A model topological string
at all genera in terms of the integer invariants $\N_{j,Q}$.  Conversely, it gives an algorithm for computing
the numbers $\N_{j,Q}$, which capture the degeneracy of BPS states, using the topological string.

The topological string thus completely captures the counting of BPS black hole states in compactifications 
of M-theory on Calabi-Yau threefolds.  Nevertheless,
despite the formidable computational techniques which are known for the topological string, it has not yet
been possible to use it to verify one of the simplest predictions from black hole physics:  namely, the
asymptotic growth of the $\N_{j,Q}$ with $Q$ should agree with the scaling of the BPS black hole entropy
with the charge in $d=5$, 
\begin{equation}
S\sim \sqrt{Q^3-j^2}.
\end{equation}

\subsection{BPS black holes in $d=4$} \label{sec-black-hole-4d}

In Section \ref{sec-black-hole-5d} we showed that the topological string counts BPS black
hole states in $d=5$.
Remarkably, it turns out that the topological string is also relevant to black hole
entropy in $d=4$!  This application is somewhat subtler than the $d=5$ case, however.
In the $d=5$ case, using \eqref{gopakumar-vafa-formula} 
one could recover the exact number of BPS states with fixed charge and
spin $j$ from the A model amplitudes up to genus $j$.  In $d=4$, the perturbative
topological string will only give us coefficients of the asymptotic \ti{growth} of the number of states
as a function of the charge; to get the actual number of states with a given fixed charge would require
some sort of nonperturbative completion of the topological string.

We are interested in computing the number of BPS states as a function of
the charge --- or more precisely an index, which counts the
BPS states possibly weighed by $\pm$ signs.
The charges in $d=4$ are a little more subtle than in $d=5$; namely, in $d=4$, each $U(1)$
in the gauge group leads to both an electric and a magnetic charge.  In Type IIA on $X$,
there is a natural splitting of the charges into electric and magnetic; namely, D0- and D2-branes
on $X$ can be considered as electrically charged states, while D4- and D6-branes are magnetically charged
states.  In Type IIB, on the other hand, all of the charges are realized by D3-branes
wrapping 3-cycles, so a general combination of electric and magnetic charges can be realized by a D3-brane wrapping a general 3-cycle, i.e. a choice of $C \in H_3(X,\Z)$.  
In this case a splitting into electric and magnetic charges is obtained only
after making a choice of symplectic basis (A and B cycles), as in \eqref{symplectic-marking}.
In this section we will use the IIB language.

How can the Calabi-Yau space $X$
give us the number of BPS states, as a function of the charge $C$?  The answer is very pretty.  We
first describe it to leading order in the limit of large $C$.  It is convenient
to express the answer in terms of $S$, the entropy:  this turns out to be given by the ``holomorphic volume''
of the Calabi-Yau,
\begin{equation} \label{black-hole-classical-entropy}
S(\Omega) = \frac{\I \pi}{4} \int_X \Omega \wedge \overline{\Omega}.
\end{equation}
Here $\Omega$ is the holomorphic 3-form on the Calabi-Yau.  But we know that this $\Omega$ is not
unique:  there is a whole moduli space of possible choices for $\Omega$, with complex dimension
$h^{2,1}+1$.  So \eqref{black-hole-classical-entropy} is not complete until we explain how 
to choose the appropriate $\Omega$.

The crucial ingredient here is the ``attractor mechanism'' of $\N=2$ supergravity \cite{Ferrara:1995ih,Strominger:1996kf}, which we now describe.
Suppose we consider the supergravity theory obtained by compactifying
Type IIB on $X$ and look for classical solutions describing a spherically symmetric BPS black hole
with charge $C$.  The supergravity theory includes scalar fields corresponding to the moduli of $X$, and we
can choose the expectation values of those scalar fields at infinity arbitrarily.  Studying
the evolution of the scalar fields as we move in from infinity toward the black hole horizon, one
finds a remarkable phenomenon:  the vector multiplet scalars and the graviphoton field strength
approach fixed values at the horizon, independent of the boundary condition at infinity, 
depending only on the charge $C$ of the black hole.\footnote{This statement needs to be
slightly qualified:  the moduli at the horizon are \ti{locally} independent of the moduli at infinity,
but there can be multiple basins of attraction \cite{Moore:1998pn}.}  

Since we are in Type IIB, the vector multiplet scalars determine the holomorphic 3-form $\Omega$ 
on $X$ up to an overall rescaling; this overall rescaling is determined by the graviphoton field
strength.  So the attractor mechanism can be viewed as the statement that the charge $C$
determines $\Omega$ at the horizon.  It is not easy to describe the
map from $C$ to $\Omega$, but the inverse map is straightforward: choosing a basis of 3-cycles and corresponding electric-magnetic splitting $C = (P^I, Q_J)$, the relation is
\begin{equation} \label{attractor-relation}
P^I = \int_{A^I} \re \Omega, \qquad Q_J = \int_{B_J} \re \Omega,
\end{equation}
or more invariantly, $\re \Omega \in H^3(X,\R)$ is the Poincar\`{e} dual of $C \in H_3(X,\Z)$.
Note that the counting of parameters works out correctly:  the complex structure moduli, when augmented 
to include the choice of overall scaling of $\Omega$, make up $2 b_3(X)$ real parameters, and this is also
the number of possible black hole charges.

So given this prescription for $\Omega$, \eqref{black-hole-classical-entropy} is a sensible formula for
the black hole entropy.
Note that it has the expected scaling with the size of the black hole:  namely, from \eqref{attractor-relation} we see that a rescaling $C \mapsto \lambda C$ (which also rescales
the size of the black hole by $\lambda$ thanks to the BPS relation between mass and charge) 
rescales the attractor
moduli by $\Omega \mapsto \lambda \Omega$, and hence $S \mapsto \lambda^2 S$.  This is the expected
behavior for the entropy of a black hole in four dimensions.

Now we want to highlight a connection between \eqref{black-hole-classical-entropy} and the
topological string.  To do so, we begin by noting that if we choose an electric-magnetic splitting, we can 
use the Riemann bilinear identity \eqref{riemann-bilinear-identity} to rewrite \eqref{black-hole-classical-entropy} as
\begin{equation} \label{black-hole-classical-entropy-coordinates}
S(P,Q) = \frac{\I \pi}{4} (X^I \overline{F}_I - \overline{X}^I F_I).
\end{equation}
This expression is quadratic in the periods of $\Omega$, which is reminiscent of the tree level B model free
energy $F_0$.  Indeed, it is very close to being the imaginary part of $F_0$,
\begin{equation} \label{classical-free-energy-coordinates}
\frac{\I \pi}{2} \im F_0(X) = \frac{\I \pi}{4} (X^I F_I - \overline{X}^I \overline{F}_I). 
\end{equation}
Now \eqref{black-hole-classical-entropy-coordinates} and \eqref{classical-free-energy-coordinates}
are not quite equal, but they are related, as explained in \cite{Ooguri:2004zv}:  
namely, beginning with \eqref{classical-free-energy-coordinates}, one can 
introduce the notation $\Phi^I = X^I - \overline{X}^I$,
and then make a Legendre transform from $\Phi^I$ to a dual variable $Q_I$ which we identify
as the black hole electric charge.  According to \eqref{attractor-relation}
this charge is given by $Q_I = F_I + \overline{F}_I$, and substituting this for $Q_I$ in the Legendre transform of \eqref{classical-free-energy-coordinates}, one recovers \eqref{black-hole-classical-entropy-coordinates}!

So the black hole entropy is the Legendre transform of $\im F_0$, to leading order in the
charge.  There is a natural extension of this formula beyond the leading order, 
obtained by noting that the Legendre transform is the leading approximation to a Fourier transform:
\begin{equation} \label{black-hole-top-string}
\sum_Q  \rho(P,Q) e^{- Q_I \Phi^I } = \abs{Z_B(P + \I \Phi)}^2.
\end{equation}
On the left side of \eqref{black-hole-top-string}, $\rho(P,Q)$ is the number of BPS black holes with electric
and magnetic charges $(P,Q)$, while on the right side $Z_B(P + \I \Phi)$ is the B model
partition function, evaluated at the $\Omega$ determined by the 
A cycle periods $X^I = P^I + \I \Phi^I$.  (Note that this formula for $X^I$ determines even the 
overall scaling of $\Omega$.  This corresponds to fixing the coupling in the topological string;
the expansion around large black hole charge corresponds to the genus expansion.)
In other words, the partition function of the mixed ensemble of black holes where we fix
the magnetic charges $P$ and the electric potential $\Phi$, then sum over all electric charges,
is given by $\abs{Z_B(P + \I \Phi)}^2$!  This is a beautiful relation and it is very natural to 
conjecture that it indeed holds to all orders in the charge \cite{Ooguri:2004zv}.

What is the evidence for this conjecture beyond leading order?  The major source of evidence comes
from a reconsideration of the corrections to $\N=2$ supergravity 
computed by the topological string, which we wrote in \eqref{gravity-f-terms}:
\begin{equation} \label{gravity-f-terms-reprise}
\int \de^4 x \int \de^4 \theta\, F_g(X^I) (\W^2)^g + {\mathrm{c.c.}}
\end{equation}
In the background of the charged black hole, the graviphoton field $\W$ and $\overline{\W}$ both have
nonzero expectation values (and create a nontrivial gravitational backreaction.)  The terms
\eqref{gravity-f-terms-reprise} therefore lead to a contribution to the free energy proportional to $F_g(X^I)$.
These corrections were studied 
in \cite{LopesCardoso:1998wt,LopesCardoso:1999cv,LopesCardoso:1999ur}, and found to give
a correction to the black hole entropy which is consistent with the conjecture.  
There are also formulas for the one-loop correction to the black hole entropy  \cite{Maldacena:1997de} 
which agree with the conjecture.
Finally, there is one example in which the conjecture can be checked exactly, 
studied in \cite{Vafa:2004qa}, which we now describe.

\begin{example}[Black hole counting and 2-dimensional Yang-Mills.]
For this example it is convenient to switch to the Type IIA language:  namely,
one considers the Type IIA string on the Calabi-Yau threefold geometry
$X = \LL \oplus \LL^{-1} \to T^2$, where $\LL$ is a particular complex line bundle
over $T^2$.

First consider
the counting of black hole BPS states in this geometry.  The relevant black holes
are obtained by wrapping D4-branes on $\LL^{-1} \to T^2$ as well as wrapping
D2-D0 bound states on $T^2$.  One can then argue that the theory on the $N$ D4-branes
is a topological $U(N)$ gauge theory, and furthermore that it localizes to
a bosonic $U(N)$ Yang-Mills theory on $T^2$.  Bound states of D2 and D0 branes with
the D4-branes can be realized as configurations of the gauge field on the D4-brane.
The counting of BPS black hole states,
summing over all D2 and D0 brane charges but fixing D4-brane charge $N$, is
thus determined by the $U(N)$ Yang-Mills partition function on $T^2$.
This theory was studied in detail
in \cite{Gross:1993hu}, where it was shown that the exact partition function can be
obtained as a sum over representations of $U(N)$:
\begin{equation}
Z_{YM} = \sum_R e^{-\lambda C_2(R) + \I \theta \abs{R}},
\end{equation}
where $\abs{R}$ is the number of boxes in the Young diagram representing $R$.  

Expanding around the large $N$ limit, one finds that this $Z_{YM}$ is the square of a holomorphic
function to all orders in $1/N$, $Z_{YM} = \abs{Z}^2$.  (This splitting into ``chiral'' and ``anti-chiral''
parts is obtained by splitting up the Young diagrams $R$ into short diagrams, with a finite number of
boxes, and large diagrams, for which the size of each column differs from $N$ by a finite number;
from this description it is manifest that the splitting only makes sense in the large $N$ limit.)
So the partition function on the D4-branes is indeed of the form $\abs{Z}^2$.
Furthermore, $X$ is a simple enough geometry that one can explicitly 
compute the A model partition function, and one finds that $Z = Z_A(X)$!

The example of Yang-Mills on $T^2$ thus provides a striking
confirmation of the conjecture that $\abs{Z_A}^2$ counts BPS black hole states in the mixed ensemble.  
It also gives us some insight into the difficulties one should expect to face in trying to define a
nonperturbative topological string (i.e. to define $Z_A$ as an honest function rather than as a formal
power series in $g_s$.)  Namely, as we noted above, the
factorization of $Z_{YM}$ into $\abs{Z_A}^2$ is only valid to all orders in $1/N$, which in the
topological string language means the expansion around $g_s = 0$.  But whatever the
nonperturbative topological string is, we want it to count BPS states and hence to agree with $Z_{YM}$.
Therefore we might expect that $Z_A$ itself probably only makes sense as a power series in $g_s$ in general --- the object
that has a chance to have a nonperturbative completion is $\abs{Z_A}^2$, but the nonperturbative completion
probably is \ti{not} generally factorized into chiral and anti-chiral parts.
\end{example}

\section{Topological M-theory} \label{sec-top-mtheory}

In the last subsection, we described a conjectural relation between the square of the topological
string partition function and the counting of black hole microstates.  This relation, once fully 
understood, could be expected to lead to a proper nonperturbative understanding of the topological 
string.  What kind of theory should we expect to find?

In the context of the physical string the answer to this question is rather surprising:  
it turns out that the proper nonperturbative description of the theory involves the dynamical emergence of an extra dimension, in other words, at strong coupling the theory is not 10-dimensional but 11-dimensional.  While the fundamental degrees of freedom of this 11-dimensional ``M-theory'' are not known, we do know its low-energy description:  it is 11-dimensional supergravity.  It is natural
to ask whether something similar might be true in the topological context; could the 6-dimensional theory be a shadow of some 7-dimensional lift?  There are some tantalizing clues
that this may be the case, which have been explored in \cite{Dijkgraaf:2004te,Nekrasov:2004vv} (see
also \cite{Gerasimov:2004yx}); here we briefly
summarize a few aspects of this story emphasized in \cite{Dijkgraaf:2004te}.

The topological string theory is naturally related to backgrounds which preserve some supersymmetry when they
appear in string compactifications --- namely Calabi-Yau threefolds.  From the target space point of view, these
backgrounds should be understood as the solutions to a six-dimensional gravity theory, or more precisely two such gravity
theories:  the B model should be a theory whose classical solutions are complex threefolds equipped with a holomorphic
3-form, while the A model should have classical solutions which are symplectic manifolds.  Giving both structures
together is equivalent to giving the Ricci-flat
metric on the Calabi-Yau threefold.\footnote{Strictly speaking, one has to require a compatibility condition: 
the symplectic form has to be of type $(1,1)$ in the complex structure so that it can be a \kahler form.}
What about topological M-theory?  There is a natural class of 7-dimensional Riemannian manifolds $Y$ which
preserve supersymmetry, namely manifolds of $G_2$ holonomy.  Furthermore, given a Calabi-Yau manifold $X$,
$Y = X \times S^1$ has holonomy contained in $G_2$.  Hence it is natural to conjecture that the classical solutions of 
topological M-theory should be manifolds of $G_2$ holonomy.

With this guess for the classical solutions, one now has to ask:  is there an action for which these are
the extrema?  Indeed there is a very natural candidate, recently discussed by Hitchin in \cite{Hitchin:2000jd,Hitchin:2001rw}.
In Hitchin's theory the fundamental field is a locally defined 2-form $\beta$, which plays the role of an abelian gauge 
potential, from which one constructs the field strength $\Phi$ with some fixed flux (in other words, the field space
consists of all closed 3-forms $\Phi$ in a fixed cohomology class.)  If this $\Phi$ is suitably generic, then it defines
a ``$G_2$ structure'' on $Y$, for which $\Phi$ is the ``associative 3-form''; this 
just means that $\Phi$ picks out a privileged set of coordinate transformations, namely
those which leave $\Phi$ invariant, and the group of such transformations at each point of $Y$ is isomorphic to $G_2$.
Indeed, these transformations also leave invariant a natural metric on $Y$, which we write as $g_\Phi$;
it can be given concretely in terms of $\Phi$ as
\begin{equation}
g_{jk} = B_{jk} \det(B)^{-1/9}
\end{equation}
where
\begin{equation}
B_{jk} = -\frac{1}{144} \Phi_{j i_1 i_2} \Phi_{k i_3 i_4} \Phi_{i_5 i_6 i_7} \epsilon^{i_1 \ldots i_7}.
\end{equation}
Of course, for general $\Phi$ this metric need not have $G_2$ holonomy.  
To get this condition Hitchin now writes a rather remarkable action:  it is simply the volume of $Y$ in the 
metric $g_\Phi$!
\begin{equation}
V_7(\Phi) = \int_Y \sqrt{g_\Phi}.
\end{equation}
The extrema of this $V_7$, when $\Phi$ varies over a fixed cohomology class, turn out to 
give metrics of $G_2$ holonomy.  In this sense $V_7$ is a candidate action for the proposed topological M-theory.

Some support for the conjecture that $V_7$ is related to topological strings comes from studying the theory
on $Y = X \times S^1$.  In this case it is natural to write $\Phi$ as
\begin{equation}
\Phi = \rho + k \wedge dt,
\end{equation}
where $t$ is a coordinate along $S^1$, and $\rho$ and $k$ are a 3-form and 2-form respectively on $X$.  
Then similarly expanding
\begin{equation}
*_\Phi \Phi = \sigma + \hat \rho dt,
\end{equation}
one finds\footnote{Again here we are assuming some compatibility conditions between $\rho$ and $k$.}
\begin{equation}
V_7(\Phi) = V_H(\rho) + V_S(\sigma),
\end{equation}
where $V_H$ and $V_S$ are two volume functionals in six dimensions analogous to $V_7$.  Upon extremization 
these two functionals lead to complex and symplectic manifolds respectively; so they are candidate actions for
the target space description of the B and A model topological string theories.  To investigate this a little further
one can try to compare the \ti{partition function} of the theory with action $V_H$ (call it $Z_H$) to the B model partition function $Z_B$.  At the classical level (comparing the classical value of $V_H$ to the genus zero part of $Z_B$) one finds that the proper conjecture is not $Z_H = Z_B$ but rather 
\begin{equation} \label{zb-conjecture}
Z_H = \abs{Z_B}^2.
\end{equation}  This is an
encouraging result, since we have already seen in Section \ref{sec-black-hole-4d} some evidence that it 
is $\abs{Z_B}^2$ rather than $Z_B$ which is a natural candidate to be nonperturbatively completed; now we are
finding that it is also the object which is naturally related to the 7-dimensional theory.  One can also study
$Z_H$ at one loop; this was done in \cite{Pestun:2005rp}, which found that \eqref{zb-conjecture} is violated,
but one can restore the agreement by replacing $V_H$ with an action for which the extrema are 
\ti{generalized} complex structures
instead of ordinary ones, as given in \cite{Hitchin:2004ut}.  In fact, the topological string makes sense
on generalized Calabi-Yau manifolds, so it is very natural to use this modified $V_H$ as its target
space description; presumably this means that
in 7 dimensions we should also replace $V_7$ by an action describing generalized $G_2$ manifolds as 
defined in \cite{Witt:2004vr}.

So $V_H$ passes a few basic checks as a candidate description of the B model.
Similarly one can argue that the classical value of $V_S$ agrees with the expectation from the A model; for
this comparison one uses the ``quantum foam'' reformulation of that theory which we reviewed briefly in
Section \ref{sec-quantum-foam}.  These checks give some support to the conjecture that $V_7$ is indeed an appropriate
description of a topological M-theory in 7 dimensions, since it is related to these reformulations
of the topological string in 6 dimensions.  Intriguingly, the reformulations of the topological string
given by $V_H$ and $V_S$ seem to be naturally adapted to the problem of counting black hole
microstates.

Even without understanding all of the details of the 7-dimensional theory, its existence can 
already shed some light on some properties of the 6-dimensional topological string.
For example, it provides a natural interpretation of the fact that the topological string partition function
behaves like a wavefunction, as we mentioned in Section \ref{sec-holomorphic-anomaly}; this is what one generally
expects for the partition function of a 7-dimensional theory considered on a 7-manifold with a 6-dimensional boundary.
In addition, the 7-dimensional description seems to unify the A and B model degrees of freedom in a natural way ---
the holomorphic and symplectic structures in 6 dimensions are naturally combined into the associative 3-form in 7 dimensions.
Interestingly, looking at the canonical quantization on a 6-manifold one seems to find that the A model and B model
degrees of freedom appear as canonical conjugates --- this seems to suggest that in some sense these degrees of freedom
cannot be specified simultaneously in the quantum theory, and also may be related to the conjectured S-duality between
the two models.

\section*{Acknowledgements}
We would like to thank the 2004 Simons Workshop on Mathematics and Physics, which was the motivation
for this review.
We are especially grateful to Martin Ro\v{c}ek for his hospitality during the workshop and
for preparing many of the figures in this paper, and to Lubo\v{s} Motl, Boris Pioline and Marcel
Vonk for helpful comments on earlier versions.

This work was supported in part by NSF grants PHY-0244821
 and DMS-0244464.

\renewcommand{\baselinestretch}{1}
\small\normalsize

\bibliography{physics}
\end{document}